\documentclass[12pt]{article}

\usepackage{pdfpages}

\usepackage{setspace}
\usepackage{amssymb}
\usepackage{amsfonts}
\usepackage{amsmath}
\usepackage{amsthm}
\usepackage[margin=1in]{geometry}

\usepackage{amssymb}
\usepackage{comment}

\usepackage{booktabs}
\usepackage{threeparttable}
\usepackage{dcolumn}
\usepackage{multirow}

\usepackage{etoolbox} 

\usepackage[usenames,dvipsnames]{pstricks}
\usepackage{epsfig}

\usepackage{tikz}
\usetikzlibrary{positioning, arrows}
\usetikzlibrary{shapes.geometric,calc}

\usepackage{pgfplots}
\pgfplotsset{compat=newest}

\usepackage{enumitem}

\usepackage{graphicx}
\usepackage{float}

\usepackage{epstopdf}
\usepackage[all]{xy}

\usepackage{natbib}

\usepackage{subcaption}
\usepackage{natbib}
\usepackage{booktabs}
\setcitestyle{authoryear,open={(},close={)},aysep={}}

\usepackage{color}

\newtheorem{definition}{Definition}

\usepackage{amsthm}

\usepackage{amsthm}
\usepackage{soul} 

\theoremstyle{plain}
\newtheorem{theorem}{Theorem}
\newtheorem{proposition}{Proposition}
\newtheorem{lemma}{Lemma}

\theoremstyle{remark}
\newtheorem{example}{\textit{Example}}
\newtheorem*{example*}{\textit{Example}}

\usepackage{caption}
\usepackage{subcaption}
\usepackage{bbding} 
\usepackage{rotating,xcolor}

\definecolor{BerkeleyBlue}{RGB}{0,50,98}  

\usepackage{hyperref}
\hypersetup{
    colorlinks=true,
    linkcolor=BerkeleyBlue,
    citecolor= BerkeleyBlue,
   filecolor=magenta,      
  urlcolor=cyan,
   pdftitle={Overleaf Example},
    pdfpagemode=FullScreen,
    }

\usepackage{url}

\UseRawInputEncoding

\newtheorem{assumption}{Assumption}
\newtheorem*{assumption*}{Assumption}

\usepackage[strict]{changepage}
\usepackage{framed}
\definecolor{formalshade}{rgb}{0.95,0.95,1}
\definecolor{darkblue}{rgb}{0.7,0.7,0.7}
\definecolor{formalshadeR}{rgb}{1,0.95,0.95}

\theoremstyle{definition}

\usepackage{multibib}
\newcites{App}{Appendix References}
\newcites{REd}{References}
\newcites{Rtwo}{References}
\newcites{Rthree}{References}

\renewcommand{\thesection}{\arabic{section}}
\renewcommand{\thesubsection}{\thesection.\arabic{subsection}}

\usepackage{titlesec}

\titleformat{\section}
  {\normalfont\Large\bfseries}{\thesection}{1em}{}

\titleformat{\subsection}[block]
  {\normalfont\large\bfseries}{\thesubsection}{1em}{}
\titlespacing*{\subsection}
  {0pt}{3.25ex plus 1ex minus .2ex}{1em}

\begin{document}
\newcommand{\cmmnt}[1]{\ignorespaces}
\sloppy
\title{Berk-Nash Rationalizability\thanks{We thank Attila Ambrus, Andreas Blume, Elliot Lipnowski, Laurent Mathevet, David Pearce, Kevin Reffet, and Yuichi Yamamoto for helpful comments. Esponda: iesponda@ucsb.edu. Pouzo: dpouzo@berkeley.edu}\medskip{}}
\author{%
\begin{tabular}{ccc}
Ignacio Esponda~~~~~~~~  & Demian Pouzo\tabularnewline
UC Santa Barbara~~~~~~~~ & UC Berkeley\tabularnewline
\end{tabular}}
\maketitle
\begin{abstract}
We introduce Berk--Nash rationalizability, a set-valued counterpart to Berk--Nash equilibrium for environments in which agents may hold misspecified models. A set of behaviors is self-justified if every behavior in the set is optimal under beliefs fitted to data generated within the set. Berk--Nash rationalizability is the largest self-justified set and reduces to standard rationalizability under correct specification and identification. We provide a learning foundation allowing Bayesian agents to learn from different, selectively sampled histories: almost surely, the limit set of play is self-justified. Conversely, under identification, any compact self-justified set can arise as the limit set with arbitrarily high probability. We illustrate how the concept yields informative predictions that are robust across learning procedures, without committing to a specific dynamic or assuming convergence.

\end{abstract}
\thispagestyle{empty}

\newpage{}

\setcounter{tocdepth}{1} 
\tableofcontents
\thispagestyle{empty}
\newpage{}

\setcounter{page}{1} 

\maketitle

\section{Introduction}

Equilibrium requires that players' actions be optimal given their beliefs and that those beliefs be consistent with the behavior they support. In Nash equilibrium, players correctly anticipate one another's behavior; in Berk--Nash equilibrium, their possibly misspecified models must also best fit the observations that behavior generates. Behavior and beliefs, however, need not settle on a single equilibrium configuration. This motivates a set-valued counterpart to equilibrium: a range of behavior that can justify itself. The consistency requirement then applies to the range as a whole, allowing it to contain action profiles that are not themselves equilibria. We call a set of action profiles \emph{self-justified} if every behavior in the set can be optimal under beliefs formed from observations generated within the set itself.

More concretely, consider a candidate set $S$ of action profiles. For each player and each profile in $S$, we ask whether that player's action can be optimal under a belief concentrated on the models that best fit the data generated by some distribution over behavior in $S$. Supporting distributions may differ across players and across profiles: self-justification does not require a single distribution to justify the entire set. What matters is that all supporting evidence is generated from within $S$. The set is self-justified if every player's action at every profile in $S$ can be justified in this way. A singleton self-justified set is a pure-strategy Berk--Nash equilibrium, or a Nash equilibrium when players know the environment.

\emph{Berk--Nash rationalizability} consists of the action profiles that belong to some self-justified set. Equivalently, it is the largest set $S$ such that every action profile in $S$ can be justified using data generated by some distribution over behavior in $S$. The set can also be constructed iteratively, by repeatedly eliminating actions that cannot be justified using data generated by the actions that remain. Under our maintained compactness and continuity assumptions, the resulting set is nonempty and compact.  When the environment is known, the best-fitting-model restriction disappears and Berk--Nash rationalizability reduces to standard rationalizability (\cite{bernheim1984rationalizable,pearce1984rationalizable}), in its correlated version (\cite{brandenburger1987rationalizability}). Our interpretation, however, is not epistemic: the concept is not meant to capture common knowledge of rationality,\footnote{The standard epistemic foundations of rationalizability are developed by \cite{brandenburger1987rationalizability} and \cite{tan1988bayesian}.} but rather the internal consistency of a range of behavior with the beliefs generated by that range. Under misspecification, the same self-justification logic applies, but beliefs must additionally be supported on models that best fit the data generated by behavior in the set.

The need for a set-valued notion is particularly natural from a learning perspective. A large literature gives equilibrium a dynamic foundation by specifying how agents learn and asking whether behavior converges to a steady state. In complete-information games, fictitious play and related adaptive procedures converge to Nash equilibrium in important classes of games, but convergence can fail and play may instead cycle or exhibit multiple limit points (\cite{Shapley1964,monderer1996potential,hofbauer2002global}). The same issue arises under misspecification. A growing literature studies learning dynamics and provides conditions under which behavior converges to Berk--Nash equilibrium or related steady states, as well as examples in which convergence fails or is more delicate (e.g., \cite{nyarko1991learning,fudenberg2017active,heidhues2018unrealistic,heidhues2021convergence,bohren2021learning,he2022mislearning,ba2023multi,murooka2025bayesian}). When behavior does not converge to a point, however, the relevant long-run object is its \emph{limit set}: the collection of action profiles that remain possible asymptotically. This suggests replacing the usual pointwise steady-state requirement by a set-valued one---the long-run range of behavior should be capable of generating beliefs that justify the behavior that continues to occur within that range.

Once the long-run object is a set rather than a point, the data that sustain behavior need not be summarized by a single empirical distribution. Different players may learn from different subsets of past observations, and even for a given player, different subsequences of the learning process may generate different limiting empirical distributions over the same long-run range. The researcher may have limited information about these sampling choices. We therefore seek restrictions on long-run behavior that remain valid across a broad class of sampling schemes. Allowing supporting distributions to differ across players and action profiles accommodates this uncertainty while retaining the central requirement of self-justification: each behavior in the set must be supported by beliefs fitted to data generated from within that set. Accordingly, our learning environment allows player-specific and selectively sampled histories, subject to restrictions ensuring that the selected data remain informative about the behavior that persists in the long run.

We provide a learning foundation for self-justified sets in a deliberately broad environment. Each player is Bayesian about the payoff-relevant parameters of a possibly misspecified model, but need not learn from the entire history. Instead, players may use different, selectively sampled collections of past observations. Sampling may be nonnested: an observation may be used at one date, omitted at another, or recalled again later. We impose two main restrictions on these sampling rules. First, whether an observation will be retained cannot depend on the consequence that is realized in that observation. Second, the number of selected observations must grow sufficiently quickly. Under these conditions, we extend Berk's asymptotic-belief argument to selected samples and compact action spaces: whenever a player's selected empirical distribution converges along a subsequence, posterior beliefs concentrate on the models that best fit the data generated by that limiting distribution. This statistical result is independent of the particular behavioral rule generating actions.

To connect beliefs to behavior, we impose two asymptotic requirements. Forecasts of opponents' play must become consistent with the empirical distribution in the player's own selected sample, and actions must become asymptotically optimal given the player's current beliefs and forecasts. Apart from these restrictions, behavior may be history dependent, randomized, and generated by very different learning procedures. Our main result is that, almost surely, the entire limit set of play is self-justified. Thus every action profile that is approached infinitely often is Berk--Nash rationalizable. The result does not require behavior, beliefs, forecasts, or empirical distributions to converge to a point. Instead, it says that whatever range of behavior survives asymptotically must be capable of generating the beliefs that justify the behavior within that range.

We also establish a partial converse. Under an identification condition, for any self-justified set and any arbitrarily small probability of failure, one can construct a learning process satisfying the sampling and behavioral requirements of our limit-set theorem whose limit set is exactly that set with the remaining probability. The construction uses the flexibility in sample selection to generate the different empirical distributions needed to sustain different parts of the set. We therefore do not interpret the converse as saying that every self-justified set will arise under a natural or canonical learning rule. Rather, it serves a more limited purpose. It shows that self-justification is not merely a necessary restriction produced by the proof of our limit theorem: within the broad class of learning environments covered by the theorem, the restriction cannot in general be tightened without imposing additional structure on how agents sample data or adjust behavior.


The set-valued perspective delivers useful predictions about long-run behavior, as we demonstrate in three examples. In the returns-to-effort setting of \cite{heidhues2018unrealistic}, Berk--Nash rationalizability yields both convergence results and economically informative bounds when available results do not establish convergence. Under overconfidence, the rationalizable set is the interval between the smallest and largest Berk--Nash equilibria. When the equilibrium is unique, this interval collapses, extending the convergence conclusion of \cite{heidhues2018unrealistic} to our broad class of learning procedures. With multiple equilibria, their convergence result no longer applies, but the rationalizable interval still bounds the behavior that can persist in the long run. Under underconfidence, the equilibrium is unique, but available results do not establish convergence to it. Berk--Nash rationalizability nevertheless bounds long-run behavior by an interval, and when this interval collapses to the equilibrium, our learning foundation implies convergence across the broad class of learning procedures we consider. 

Second, convergence to a unique equilibrium under a particular learning rule need not be robust to alternative ways of sampling experience. In an example with strategic feedback and misspecification, a natural full-history Bayesian learning process converges globally to the unique Berk--Nash equilibrium for some parameters, yet the Berk--Nash rationalizable set is large. Here the set-valued prediction reflects the scope of the robustness we seek: other ways of sampling experience within our learning framework can sustain a much wider range of behavior. Thus a global convergence result for a particular learning rule need not make the equilibrium a robust prediction across learning procedures.

Our baseline notion deliberately allows different players, and different behaviors within the same self-justified set, to be supported by different distributions over the long-run range. This is natural when agents may learn from different samples of experience. We also study a stronger \emph{common-sample} refinement, in which the beliefs used to justify players' actions must be fitted to a common empirical distribution of past play. The distinction can have substantial bite. In a linear-quadratic network example, baseline rationalizability is large even though the Nash equilibrium is unique, while imposing the common-sample restriction can collapse the rationalizable set to the unique equilibrium. The example shows that restrictions on how agents use experience can meaningfully sharpen the set of self-justified behaviors.

The approach also has a more traditional application to the analysis of particular learning dynamics. A general literature studies asymptotic behavior using tools such as stochastic approximation, differential inclusions, martingale methods, and related techniques (\cite{fudenberg2021limit,esponda2021asymptotic,frick2023belief,murooka2023convergence}). These methods can deliver substantially sharper predictions for a fully specified dynamic. Berk--Nash rationalizability provides a complementary route to long-run conclusions that does not require solving the dynamics in full. If the rationalizable set is a singleton, our limit-set theorem implies convergence for every learning process in the class we study; when it is larger, the set bounds the behavior that can persist asymptotically. These conclusions can be useful when direct convergence analysis is difficult or when available convergence results do not apply. Dynamic-specific analysis may of course sharpen these predictions, but it addresses a different question from the robustness embodied in our solution concept.


\cite{milgrom1991adaptive} establish a broad link between adaptive learning and correlated rationalizability. Roughly, their adaptive-learning condition requires current behavior to be justifiable by beliefs that eventually place weight only on actions that continue to appear in sufficiently long histories. In the correctly specified and identified benchmark, our learning procedures satisfy a more restrictive version of this condition, and both approaches imply that persistent behavior is correlated rationalizable.\footnote{A related connection is to \cite{milgrom1990rationalizability}, whose monotone-iteration argument bounds rationalizable actions in supermodular games by the smallest and largest Nash equilibria. The same logic yields our returns-to-effort characterization.} Our analysis further gives a set-valued interpretation of long-run behavior: under our learning rules, the actual limit set is self-justified, and our partial converse shows that any compact self-justified set can be sustained as a limit set with arbitrarily high probability.

The connection changes more fundamentally once agents' models may be misspecified. In our baseline formulation, the same selected history is used for two distinct purposes. First, agents use the observed consequences in that history to learn which subjective model best describes their environment. Second, they use the opponents' actions in that history to forecast how those opponents will behave. Bayesian updating disciplines the first inference, while forecast consistency ties the second to the same selected history. These two requirements can be separated. One can retain Bayesian learning about the subjective model while weakening forecast consistency to require only that current behavior be justified by an adaptive belief about opponents' play. Section~\ref{subsec:alternatives} develops this extension and shows how the limit-set conclusion changes as broader forms of adaptive behavior are allowed.


Our analysis builds on \citet[henceforth EP16]{esponda2016berk}, which introduces Berk--Nash equilibrium for environments in which agents learn using possibly misspecified models. Broadly, EP16 show that if behavior converges, the limiting action profile must be a Berk--Nash equilibrium.\footnote{EP16 also provide a learning foundation for mixed-strategy Berk--Nash equilibrium using the payoff-perturbation approach that \cite{fudenberg1993learning} use to justify mixed-strategy Nash equilibrium. More recent work, including \cite{esponda2021equilibrium} and \cite{esponda2021asymptotic}, dispenses with payoff perturbations and instead interprets a mixed action profile as the limiting empirical distribution of realized actions.} In general, however, behavior need not converge to a single action profile, mixed strategy, or empirical distribution. The present paper therefore replaces the point-valued steady state by a set-valued one: the long-run object may be a range of action profiles, with different points in that range justified by beliefs fitted to different limiting empirical distributions generated from within the same range. Our partial converse further adapts the construction of \cite{fudenberg1993learning}, which EP16 extend to Berk--Nash equilibrium under misspecification. The earlier arguments sustain a single equilibrium through behavior that becomes approximately optimal over time. Here we instead construct a learning process whose entire limit set is a given self-justified set, with different dates and players potentially relying on different selected samples from the same long-run range.

We discuss several extensions and related concepts in Section~5, including rationalizable distributions, mixed strategies, forward-looking behavior, stronger support restrictions on persistent behavior, alternative sampling and learning specifications, and other set-valued notions of long-run behavior. We also compare our approach with \cite{frick2023belief}, who develop a related iterative procedure for restricting long-run beliefs under misspecification.\footnote{Related iterative arguments have been used in more specific settings: \cite{heidhues2018unrealistic} obtain bounds that collapse to a singleton, and \cite{he2022mislearning} adapts their approach to a particular multidimensional environment.}

\section{Solution Concept} \label{sec:games}

We introduce the solution concept as a set-valued analogue of equilibrium. An equilibrium is a singleton steady state: if play remains at an action profile \(a\), then the evidence generated by \(a\) supports beliefs under which \(a\) is optimal. This is the traditional way of using equilibrium as a long-run prediction. However, learning and adjustment processes need not converge to an equilibrium action profile. The relevant long-run object may instead be a range of behavior: the action profiles that remain possible in the long run. We therefore allow for a steady-state range of behavior, a set \(S\), within which play may continue to vary. The set is self-justified if every profile in \(S\) can be optimal under beliefs generated by observations drawn from \(S\) itself. We then define Berk--Nash rationalizability by asking which action profiles can appear in some self-justified range.

\subsection{Primitives}

There is a finite set of players \(I\). For each \(i\in I\), the individual action space is \(\mathbb A^i\), and the joint action space is
\[
\mathbb A:=\prod_{i\in I}\mathbb A^i.
\]
Player \(i\)'s consequences take values in \(\mathbb Y^i\). The true environment is described by a mapping
\[
Q^i:\mathbb A\to \Delta \mathbb Y^i,
\]
which assigns to every action profile a distribution over player \(i\)'s consequences. Player \(i\) need not know \(Q^i\). Instead, she entertains a parametric model
\[
\{Q^i_{\theta^i}:\theta^i\in\Theta^i\},
\qquad
Q^i_{\theta^i}:\mathbb A\to \Delta \mathbb Y^i.
\]
An action profile is denoted \(a=(a^i)_{i\in I}\in\mathbb A\). The payoff function of player \(i\) is
\[
\pi^i:\mathbb A^i\times \mathbb Y^i\to \mathbb R,
\]
where the distribution of \(y^i\in\mathbb Y^i\) already incorporates the effect of all players' actions through \(Q^i\).

We impose the following regularity assumptions.\footnote{All Euclidean spaces, including \(\mathbb A\) and \(\Theta\), are endowed with their Borel \(\sigma\)-algebra. The spaces \(\Delta\mathbb A\) and \(\Delta\Theta\) denote the sets of Borel probability measures on \(\mathbb A\) and \(\Theta\), respectively, endowed with the topology of weak convergence.}

\begin{assumption}\label{ass:game}
For each \(i\in I\):

\begin{enumerate}[label=(\roman*)]
\item \label{ass:game:spaces} \emph{Spaces.} The sets \(\mathbb A^i\) and \(\Theta^i\) are nonempty compact subsets of Euclidean spaces, and \(\mathbb Y^i\) is a Borel subset of a Euclidean space.

\item \label{ass:game:densities} \emph{Densities and a.e.\ continuity.} There exists a Borel measure \(\nu^i\) on \(\mathbb Y^i\) such that, for every \(a\in\mathbb A\), \(Q^i(\cdot\mid a)\ll \nu^i\) with density \(q^i(\cdot\mid a)\), and, for every \(\theta^i\in\Theta^i\), \(Q^i_{\theta^i}(\cdot\mid a)\ll \nu^i\) with density \(q^i_{\theta^i}(\cdot\mid a)\). Moreover, for \(\nu^i\)-a.e.\ \(y\), the maps
\[
a\mapsto q^i(y\mid a),
\qquad
(\theta^i,a)\mapsto q^i_{\theta^i}(y\mid a),
\qquad
(a^i,y)\mapsto \pi^i(a^i,y)
\]
are continuous.

\item \label{ass:game:bounds} \emph{Likelihood-ratio bound, density envelope, and integrability.} There exist measurable functions \(M^i:\mathbb Y^i\to[1,\infty)\) and \(r^i:\mathbb Y^i\to[0,\infty)\) such that, for all \((\theta^i,a)\in\Theta^i\times\mathbb A\) and \(\nu^i\)-a.e.\ \(y\),
\[
q^i_{\theta^i}(y\mid a)\,M^i(y)^{-1}
\leq
q^i(y\mid a)
\leq
q^i_{\theta^i}(y\mid a)\,M^i(y),
\]
and
\[
q^i(y\mid a)\leq r^i(y),
\qquad
\int_{\mathbb Y^i} M^i(y)r^i(y)\,\nu^i(dy)<\infty.
\]

\item \label{ass:game:payoffs} \emph{Payoffs.} There exists a measurable function \(h^i_\pi:\mathbb Y^i\to[0,\infty)\) such that
\[
|\pi^i(a^i,y)|\leq h^i_\pi(y)
\quad
\text{for all }a^i\in\mathbb A^i
\]
and
\[
\int_{\mathbb Y^i} h^i_\pi(y)M^i(y)r^i(y)\,\nu^i(dy)<\infty.
\]
\end{enumerate}
\end{assumption}

These assumptions are regularity conditions. They ensure that expected payoffs are well defined and continuous, that optimal-action correspondences are nonempty, and that the set of Kullback--Leibler minimizing models is nonempty and upper hemicontinuous in the relevant distribution of action profiles.

\subsection{Optimality and Best-Fitting Models}

The definition uses two ingredients. First, given a belief over models and a forecast of opponents' actions, we define the set of optimal actions. Second, given a distribution over past action profiles, we define the models that best fit the data generated by those profiles.

Fix player \(i\) and a distribution over opponents' actions \(\beta^{-i}\in\Delta\mathbb A^{-i}\). For \(\theta^i\in\Theta^i\) and \(a^i\in\mathbb A^i\), player \(i\)'s expected payoff is
\begin{equation}\label{eq:EU}
U^i(a^i,\beta^{-i},\theta^i)
:=
\int_{\mathbb A^{-i}}\int_{\mathbb Y^i}
\pi^i(a^i,y^i)\,
Q^i_{\theta^i}(dy^i\mid a^i,a^{-i})\,
\beta^{-i}(da^{-i}).
\end{equation}
Given a belief \(\mu^i\in\Delta\Theta^i\), the optimal action correspondence is
\[
F^i(\mu^i,\beta^{-i})
:=
\arg\max_{a^i\in\mathbb A^i}
\int_{\Theta^i}
U^i(a^i,\beta^{-i},\theta^i)\,\mu^i(d\theta^i).
\]

Next, for model \(\theta^i\in\Theta^i\) and action profile \(a\in\mathbb A\), define the Kullback--Leibler divergence from the true consequence distribution to the model consequence distribution by
\[
K^i(\theta^i,a)
:=
\int_{\mathbb Y^i}
\ln\!\left(
\frac{q^i(y^i\mid a)}
     {q^i_{\theta^i}(y^i\mid a)}
\right)
q^i(y^i\mid a)\,\nu^i(dy^i).
\]
For a distribution \(\sigma\in\Delta\mathbb A\), define player \(i\)'s set of best-fitting models as
\[
\Theta^{m,i}(\sigma)
:=
\arg\min_{\theta^i\in\Theta^i}
\int_{\mathbb A} K^i(\theta^i,a)\,\sigma(da).
\]
Thus \(\Theta^{m,i}(\sigma)\) contains the models that best fit the data player \(i\) would observe if action profiles were distributed according to \(\sigma\).

\subsection{Self-Justification and Rationalizability}

Let \(S\subseteq\mathbb A\) be a candidate steady-state range of behavior. For player \(i\), a distribution \(\sigma^i\in\Delta S\) represents a selected or weighted sample of action profiles from this range. The same distribution \(\sigma^i\) plays two roles. First, it determines the set \(\Theta^{m,i}(\sigma^i)\) of models that best fit player \(i\)'s data. Second, its marginal \((\sigma^i)^{-i}\) over opponents' actions is player \(i\)'s forecast of opponents' behavior.

For every \(S\subseteq\mathbb A\), define the Berk--Nash self-justification operator
\begin{equation}\label{eq:Gamma_BN}
\Gamma(S)
:=
\left\{
a\in\mathbb A:
\forall i\in I,\ 
\exists\,\sigma^i\in\Delta S,\ 
\exists\,\mu^i\in\Delta\Theta^{m,i}(\sigma^i)
\text{ such that }
a^i\in F^i\bigl(\mu^i,(\sigma^i)^{-i}\bigr)
\right\}.
\end{equation}
Here \((\sigma^i)^{-i}\in\Delta S^{-i}\) denotes the marginal of \(\sigma^i\) over opponents' actions.\footnote{Formally, \(S^{-i}:=\{a^{-i}:\exists a^i \text{ such that }(a^i,a^{-i})\in S\}\). For \(\sigma^i\in\Delta S\), the marginal \((\sigma^i)^{-i}\) is the probability distribution on \(S^{-i}\) induced by \(\sigma^i\): for every measurable set \(B\subseteq S^{-i}\), \((\sigma^i)^{-i}(B)=\sigma^i(\{a\in S:a^{-i}\in B\})\).}

Equivalently, \(a\in\Gamma(S)\) if, for each player \(i\), there is a selected sample \(\sigma^i\) from \(S\), a belief \(\mu^i\) supported on the models that best fit that sample, and an optimality condition making \(a^i\) a best response to the resulting belief and forecast.

\begin{definition}
A set \(S\subseteq\mathbb A\) is \emph{self-justified} if it is nonempty and compact, and
\[
S\subseteq \Gamma(S).
\]
\end{definition}

A self-justified set is a set-valued steady state. If \(S=\{a\}\) is a singleton, then every selected sample supported on \(S\) is the degenerate distribution on \(a\). The condition \(\{a\}\subseteq\Gamma(\{a\})\) then says that \(a\) is optimal under beliefs fitted to data generated by \(a\) itself. This is the usual steady-state logic of equilibrium: if play remains at \(a\), then players have no incentive to change. Indeed, the condition \(\{a\}\subseteq\Gamma(\{a\})\) is exactly the definition of a pure-strategy Berk--Nash equilibrium.

The reason to move from points to sets is that long-run behavior need not converge to a single action profile. Even when point convergence fails, an infinite path has a long-run range: the profiles that continue to be approached, or, in a finite action space, the profiles played infinitely often. The definition extends the steady-state logic of equilibrium to such ranges. If the long-run range of behavior is \(S\), players may form beliefs from selected or weighted samples of observations generated within \(S\). The set is self-justified when every behavior in \(S\) can be supported by such internally generated beliefs. Thus the question is not whether play can settle permanently at a single profile, but whether the range of behavior can justify itself. Play may continue to vary within \(S\), but every behavior that remains possible is justified by the range's own evidence. The compactness requirement in teh definition reflects this long-run interpretation: in a compact action space, the set of limit points of any action path is nonempty and compact.

Let
\[
\mathcal S
:=
\{S\subseteq\mathbb A:S\text{ is nonempty, compact, and }S\subseteq\Gamma(S)\}.
\]
denote the collection of self-justified sets. These sets are the possible steady-state ranges of behavior. The singleton elements of \(\mathcal S\) are exactly the pure-strategy Berk--Nash equilibria. As with equilibrium, there may be many such ranges; indeed, allowing for set-valued steady states may enlarge the multiplicity of possible long-run predictions. The benefit is that the concept does not require long-run behavior to converge to a single action profile. Different learning or adjustment processes may select different ranges, or may fail to select a singleton, that is, an equilibrium, at all. Self-justification asks which ranges are internally consistent, without specifying the particular dynamics that select among them. We now use the collection \(\mathcal S\) to define the individual action profiles that are consistent with self-justification, in the sense that they belong to at least one self-justified range.

\begin{definition}
The set of \emph{Berk--Nash rationalizable} action profiles is
\[
\mathcal B := \bigcup_{S\in\mathcal S} S.
\]
\end{definition}

Requiring self-justified sets to be compact does not affect the set of rationalizable profiles. Indeed, if a nonempty set \(S\subseteq\mathbb A\) satisfies \(S\subseteq\Gamma(S)\), then its closure satisfies \(\overline S\subseteq\Gamma(\overline S)\).\footnote{By monotonicity, \(S\subseteq\overline S\) implies \(\Gamma(S)\subseteq\Gamma(\overline S)\), so \(S\subseteq\Gamma(\overline S)\). Since \(\Gamma(\overline S)\) is closed by Lemma~\ref{lem:Gamma-regular}, it contains \(\overline S\).} Because \(\mathbb A\) is compact, \(\overline S\) is compact and hence self-justified. Thus every profile that belongs to a noncompact set satisfying the inclusion also belongs to a compact self-justified set.

Although \(\mathcal B\) is defined by taking the union over all self-justified sets, the following result shows that this union is itself a self-justified set. Thus the profile-level prediction can be represented by a single largest self-justified set. The theorem also gives an iterative characterization of this set.

\begin{theorem}[Existence and characterization of rationalizable set] \label{theo:characterization}
The set of Berk--Nash rationalizable action profiles, $\mathcal{B}\subseteq \mathbb{A}$, is nonempty, compact, and is the largest fixed point of $\Gamma$. 
It can be obtained iteratively, as
\[
\mathcal{B} = \bigcap_{k=0}^\infty B^k \quad \text{with} \quad B^0 = \mathbb{A} \quad \text{and} \quad B^{k+1} = \Gamma(B^k).
\]
Moreover, \(\mathcal B\) is a product set: there exist nonempty compact sets \(\mathcal B^i\subseteq\mathbb A^i\) such that
\[
\mathcal B=\prod_{i\in I}\mathcal B^i.
\]
\end{theorem}
\begin{proof}
See the Appendix.
\end{proof}

The result that \(\mathcal B\) is the largest fixed point of \(\Gamma\) follows from monotonicity of \(\Gamma\) and the argument in the proof of Tarski's fixed point theorem. In general, the largest fixed point of a monotone operator may require a transfinite iteration to characterize (e.g., \cite{lipman1994note}, \cite{arieli2010rationalizability}). Here, countably many iterations suffice because \(\Gamma\) has the relevant compactness and closedness properties: Lemma \ref{lem:basic-properties} establishes the upper hemicontinuity and closed-graph properties of the best-response and best-fitting-model correspondences, and Lemma \ref{lem:Gamma-regular} implies that \(\Gamma\) maps nonempty closed sets to nonempty closed sets. These properties also imply that the rationalizable set is nonempty and compact. An analogous countable-iteration result was established by \cite{weinstein2017interim} for interim correlated rationalizability. Finally, although self-justified sets need not be product sets, the operator \(\Gamma\) is product-valued because each player's action is justified separately. Hence the largest fixed point \(\mathcal B\) is a product set.

The theorem gives a simple way to characterize the rationalizable set. Begin with all action profiles as the candidate long-run range, and repeatedly remove profiles that cannot be justified by beliefs generated from the remaining range. The profiles that survive every round are exactly those that can appear in some self-justified range. Thus the iterative procedure turns the collection of possible self-justified steady-state ranges into a single profile-level prediction, \(\mathcal B\).

\subsection{Special Cases}

\subsubsection{Single-agent case}
The definition of the solution concept is easiest to visualize in the single-agent case. There is one decision maker, with action space \(\mathbb A\), consequence space \(\mathbb Y\), true consequence kernel \(Q:\mathbb A\to\Delta\mathbb Y\), subjective model \(\{Q_\theta:\theta\in\Theta\}\), and payoff function \(\pi:\mathbb A\times\mathbb Y\to\mathbb R\).

Because there are no opponents, there is no strategic forecast. Thus the optimal-action correspondence has only one argument. For a belief \(\mu\in\Delta\Theta\), write
\[
F(\mu)
:=
\arg\max_{a\in\mathbb A}
\int_{\Theta}
\left[
\int_{\mathbb Y}\pi(a,y)\,Q_\theta(dy\mid a)
\right]
\mu(d\theta).
\]
This is the single-agent analogue of \(F^i(\mu^i,\beta^{-i})\): the belief over models remains, but the forecast of opponents' actions is omitted.

For \(\sigma\in\Delta\mathbb A\), let
\[
\Theta^m(\sigma)
:=
\arg\min_{\theta\in\Theta}
\int_{\mathbb A} K(\theta,a)\,\sigma(da)
\]
denote the set of models that best fit data generated by the action distribution \(\sigma\). Then, for any set \(S\subseteq\mathbb A\), the self-justification operator becomes
\[
\Gamma(S)
=
\left\{
a\in\mathbb A:
\exists\,\sigma\in\Delta S,\ 
\exists\,\mu\in\Delta\Theta^m(\sigma)
\text{ such that }
a\in F(\mu)
\right\}.
\]

Thus, in the single-agent case, a nonempty compact set \(S\subseteq\mathbb A\) is self-justified if every action in \(S\) can be optimal under beliefs fitted to some selected or weighted sample of observations generated by actions in \(S\). Different actions in \(S\) may be justified by different samples \(\sigma\). If \(S=\{a\}\), then the only possible sample is the degenerate distribution on \(a\), and self-justification reduces to the single-agent version of pure-strategy Berk--Nash equilibrium.

\subsubsection{Correct-model benchmark: Nash equilibrium and classical rationalizability}

We next consider the benchmark in which players' models are correctly specified and identified. Player \(i\)'s model is \emph{correctly specified} if there exists \(\theta^{i,*}\in\Theta^i\) such that
\[
Q^i_{\theta^{i,*}}(\cdot\mid a)=Q^i(\cdot\mid a)
\quad\text{for all }a\in\mathbb A.
\]
Thus the true consequence kernel belongs to player \(i\)'s model.

Player \(i\)'s model is \emph{identified} if, for every \(\sigma\in\Delta\mathbb A\), the set \(\Theta^{m,i}(\sigma)\) is a singleton. If the model is both correctly specified and identified, then
\[
\Theta^{m,i}(\sigma)=\{\theta^{i,*}\}
\quad\text{for every }\sigma\in\Delta\mathbb A,
\]
because \(\theta^{i,*}\) attains the minimum KL divergence. The \emph{game is correctly specified and identified} if both properties hold for every player.

Under correct specification and identification, the learning component of \(\Gamma\) disappears. For \(\beta^{-i}\in\Delta\mathbb A^{-i}\), let \(BR^i(\beta^{-i})\) denote player \(i\)'s best-response correspondence when payoffs are evaluated under the true consequence kernel \(Q^i\). In this benchmark, it is useful to write the operator as
\[
\Gamma^{BR}(S)
=
\left\{
a\in\mathbb A:
\forall i\in I,\ 
\exists\,\sigma^i\in\Delta S
\text{ such that }
a^i\in BR^i\bigl((\sigma^i)^{-i}\bigr)
\right\}.
\]
Thus, under correct specification and identification, \(\Gamma=\Gamma^{BR}\).\footnote{Indeed, correct specification and identification imply \(\Theta^{m,i}(\sigma^i)=\{\theta^{i,*}\}\) for every \(\sigma^i\). Hence any belief \(\mu^i\in\Delta\Theta^{m,i}(\sigma^i)\) is degenerate at the true model, and the condition \(a^i\in F^i(\mu^i,(\sigma^i)^{-i})\) is exactly \(a^i\in BR^i((\sigma^i)^{-i})\).}
In this benchmark, a nonempty compact set \(S\) is self-justified if every behavior in \(S\) is a best response to some belief over opponents' behavior generated from \(S\).

The singleton case gives Nash equilibrium. If \(S=\{a\}\), then the only distribution in \(\Delta S\) is the degenerate distribution on \(a\), and the induced belief about opponents' behavior assigns probability one to \(a^{-i}\). Hence \(\{a\}\) is self-justified if and only if
\[
a^i\in BR^i(a^{-i})
\quad\text{for every }i\in I,
\]
that is, if and only if \(a\) is a pure-strategy Nash equilibrium.

The set-valued case gives the best-response property underlying classical rationalizability. The condition \(S\subseteq\Gamma^{BR}(S)\) says that every behavior in \(S\) has a best-response justification using only beliefs generated from \(S\). Equivalently, in terms of the projections \(S^i\) of \(S\) onto individual action spaces, each action of player \(i\) that appears in \(S\) is a best response to some conjecture over opponents' actions that remain possible. \cite{pearce1984rationalizable} calls this the \emph{best-response property}. Thus, under correct specification and identification, self-justified sets are precisely the nonempty compact sets that have the best-response property.

Classical rationalizability is obtained by taking the union of all sets with the best-response property. Restricting attention to nonempty compact sets with this property is without loss by the closure argument above. Since Berk--Nash rationalizability is obtained by taking the union of all self-justified sets, the two notions coincide in the correctly specified and identified benchmark. In two-player games, this is the rationalizability notion of \cite{bernheim1984rationalizable} and \cite{pearce1984rationalizable}, since each player has only one opponent and there is no distinction between independent and correlated conjectures over opponents' actions. With three or more players, our formulation allows player \(i\)'s belief over \(a^{-i}\) to be an arbitrary joint distribution over opponents' actions. It therefore corresponds to the correlated version of rationalizability introduced by \cite{brandenburger1987rationalizability}, rather than to the original Bernheim--Pearce formulation with independent conjectures across opponents.


\subsection{Common sample refinement}

The baseline operator allows each player to justify her action using a different selected-sample distribution. A natural refinement requires all players to rely on the same sample.

For $S\subseteq\mathbb A$, define
\[
\Gamma^{CS}(S)
:=
\left\{
a\in\mathbb A:
\exists\,\sigma\in\Delta S
\text{ such that, for every }i\in I,\ 
\exists\,\mu^i\in\Delta\Theta^{m,i}(\sigma)
\text{ with }
a^i\in F^i(\mu^i,\sigma^{-i})
\right\}.
\]
A nonempty compact set $S$ is \emph{common-sample self-justified} if \(S\subseteq\Gamma^{CS}(S)\), and an action profile is \emph{common-sample Berk--Nash rationalizable} if it belongs to some such set. Let $\mathcal B^{CS}$ denote the set of common-sample Berk--Nash rationalizable profiles.

Thus, a profile must now be justified by a single distribution $\sigma$ over past action profiles. Players may still have different subjective models and limiting beliefs, but both their model fitting and their forecasts of opponents' behavior must be based on the same evidence.

The basic characterization of $\mathcal B$ carries over to $\Gamma^{CS}$. In particular, if
\[
C^0=\mathbb A,
\qquad
C^{k+1}=\Gamma^{CS}(C^k),
\]
then
\[
\mathcal B^{CS}
=
\bigcap_{k=0}^{\infty}C^k,
\]
and $\mathcal B^{CS}$ is nonempty, compact, and the largest fixed point of $\Gamma^{CS}$. Unlike $\Gamma$, however, $\Gamma^{CS}$ need not be product-valued: for each fixed $\sigma$ the profiles justified by $\sigma$ form a product set, but their union over $\sigma\in\Delta S$ need not.

\begin{proposition}[Refinement hierarchy]\label{prop:refinement-hierarchy}
For every $S\subseteq\mathbb A$, \(\Gamma^{CS}(S)\subseteq\Gamma(S)\). Consequently,
\(\mathcal B^{CS}\subseteq\mathcal B\).
\end{proposition}

\begin{proof}
If $a\in\Gamma^{CS}(S)$ is justified by $\sigma\in\Delta S$, then in the definition of $\Gamma(S)$ simply take \(\sigma^i=\sigma\) for every $i$. Thus $\Gamma^{CS}(S)\subseteq\Gamma(S)$. The inclusion \(\mathcal B^{CS}\subseteq\mathcal B\) follows immediately.
\end{proof}

The inclusion can be strict because $\Gamma$ allows different players to extract different sample distributions from the same long-run set. In two-player games with correct specification and identification, however, the common-sample restriction does not further refine the rationalizable set.

\begin{proposition}[Two-player correct-model benchmark]\label{prop:CS-two-player-equivalence}
Suppose the game is correctly specified and identified and $|I|=2$. Then \(
\mathcal B^{CS}=\mathcal B\).
\end{proposition}

\begin{proof}
We already know that $\mathcal B^{CS}\subseteq\mathcal B$. For the reverse inclusion, recall that under correct specification and identification, player $i$'s optimality condition depends only on the distribution of her opponent's action. Moreover, $\mathcal B=\mathcal B^1\times\mathcal B^2$ is a product fixed point of $\Gamma$. Fix $a\in\mathcal B$. Since \(a\in\Gamma(\mathcal B)\), there are distributions $\sigma^1,\sigma^2\in\Delta\mathcal B$ such that $(\sigma^1)^2$ justifies $a^1$ and $(\sigma^2)^1$ justifies $a^2$. Because $\mathcal B$ is a product set, 
\[
\bar\sigma
:=
(\sigma^2)^1\otimes(\sigma^1)^2
\in\Delta\mathcal B.
\]
This single distribution has exactly the two marginals needed to justify $a^1$ and $a^2$. Hence \(a\in\Gamma^{CS}(\mathcal B)\). Therefore \(\mathcal B\subseteq\Gamma^{CS}(\mathcal B)\), so $\mathcal B$ is common-sample self-justified and \(\mathcal B\subseteq\mathcal B^{CS}\). Combining the two inclusions yields \(\mathcal B^{CS}=\mathcal B\).
\end{proof}

\section{Learning Foundations for Self-Justification} \label{sec:foundations}

The solution concept introduced above can be understood on its own as a set-valued steady-state notion: a self-justified set is a range of behavior whose elements can be supported by beliefs generated from within the range itself. This section provides additional foundations for the concept. The first result gives a learning foundation: in a dynamic environment with Bayesian learning from selected samples and myopic optimization, every limit action is Berk--Nash rationalizable. The second result goes in the opposite direction: under permissive sampling and asymptotically vanishing optimization mistakes, any self-justified set can arise as the limiting range of behavior.

\subsection{A general learning environment}

We now describe a general class of stochastic processes underlying the learning results. Time is discrete, $t=1,2,\ldots$. Let $\mathbb Y=\times_{i\in I}\mathbb Y^i$, and let $Q(\cdot\mid a)\in\Delta\mathbb Y$ be a measurable joint consequence kernel whose $i$th marginal is $Q^i(\cdot\mid a)$ for every player $i$. No independence across players' consequences is required.

To describe sampling, for each player $i$ and dates $\tau,t\geq1$, let
\[
W_{\tau,t}^i\in\{0,1\}
\]
indicate whether player $i$ uses the period-$\tau$ observation in her sample at date $t$. Since only past observations can be used, set $W_{\tau,t}^i=0$ whenever $\tau\geq t$. Thus, for each player $i$, the sampling scheme can be represented by the triangular array
\[
W^i=
\begin{pmatrix}
0 & W_{1,2}^i & W_{1,3}^i & W_{1,4}^i & \cdots\\
0 & 0 & W_{2,3}^i & W_{2,4}^i & \cdots\\
0 & 0 & 0 & W_{3,4}^i & \cdots\\
0 & 0 & 0 & 0 & \ddots\\
\vdots & \vdots & \vdots & \vdots & \ddots
\end{pmatrix}.
\]
Row $\tau$ specifies the future dates at which observation $\tau$ is used, while column $t$ specifies the observations used at date $t$. The entries on and below the diagonal are zero because an observation cannot be used before, or at the same date as, its realization.

For each date $t$, let
\[
W_t:=\bigl(W_{t,s}^i\bigr)_{i\in I,\ s>t}
\]
denote the collection of future recall decisions concerning the period-$t$ observation. Thus $W_t$ is the $t$th row of the sampling arrays across players. The entire row $W_t$ is determined after $a_t$ is chosen but before the consequence $y_t$ is realized.

This sampling structure includes permanent sampling as a special case: if $J_t^i\in\{0,1\}$ and
\[
W_{t,s}^i=J_t^i
\qquad\text{for every }s>t,
\]
then observation $t$ is either retained forever or never used. It also allows deterministic ex ante resampling and randomized recall schedules, provided the future use of observation $t$ is determined before its consequence is realized.

Let
\[
\mathcal W_t
:=
\prod_{i\in I}\{0,1\}^{\{t+1,t+2,\ldots\}}
\]
denote the space of possible $t$th rows $W_t$, endowed with its product Borel $\sigma$-algebra. Define the path space
\[
\Omega
=
\prod_{t=1}^\infty
\bigl(\mathbb A\times\mathcal W_t\times\mathbb Y\bigr),
\]
endowed with the corresponding product $\sigma$-algebra $\mathcal F$. A realization is written
\[
\omega=(a_1,W_1,y_1,a_2,W_2,y_2,\ldots),
\]
and the cumulative history after period $t$ is
\[
h_t=(a_1,W_1,y_1,\ldots,a_t,W_t,y_t).
\]

An action rule is an arbitrary sequence of measurable stochastic kernels
\[
\alpha_t(\cdot\mid h_{t-1})\in\Delta\mathbb A.
\]
Thus actions may depend arbitrarily on the entire past and may be randomized. We impose no further restriction here because the statistical learning result below does not require one: it uses only that actions and sampling decisions concerning period $t$ are determined before the corresponding consequence is realized. Behavioral restrictions linking actions to selected samples, posterior beliefs, and optimality are imposed separately below.

After $a_t$ is chosen, but before $y_t$ is realized, the row $W_t$ is drawn according to a measurable stochastic kernel
\[
\lambda_t(\cdot\mid h_{t-1},a_t)\in\Delta\mathcal W_t.
\]
Thus the entire future sampling pattern for observation $t$ is determined before its consequence is realized. It may depend on the past, the current action profile, and additional randomization, but not on $y_t$ or on later endogenous variables.

Conditional on $(h_{t-1},a_t,W_t)$, the consequence vector is drawn according to $Q(\cdot\mid a_t)$. Hence the one-period transition kernel is
\[
\alpha_t(da_t\mid h_{t-1})
\lambda_t(dW_t\mid h_{t-1},a_t)
Q(dy_t\mid a_t).
\]
By the Ionescu--Tulcea theorem, these kernels induce a unique probability measure $\mathbf P$ on $(\Omega,\mathcal F)$. In particular, for every player $i$ and every Borel set $B\subseteq\mathbb Y^i$,
\[
\mathbf P(y_t^i\in B\mid h_{t-1},a_t,W_t)
=
Q^i(B\mid a_t).
\]

For each player $i$, the sample used at date $t$ is
\[
H_t^i
=
\{\tau<t:W_{\tau,t}^i=1\}.
\]
Thus samples need not be nested: an observation may be used at some dates, omitted at others, and used again later.

We impose a sample-growth condition. For every player $i$,
\[
\frac{|H_t^i|}{\log t}\longrightarrow\infty
\qquad \mathbf P\text{-almost surely}.
\]
Thus the number of observations used at date $t$ must grow faster than $\log t$, although it may still be a vanishing fraction of the full history. In particular, the condition is satisfied whenever there is a positive lower bound on the asymptotic fraction of past observations used,\footnote{For example, if the recall indicators $W_{\tau,t}^i$ are independent across $\tau$ and satisfy $\mathbf P(W_{\tau,t}^i=1)=p_{\tau,t}^i\geq \underline p>0$, then standard concentration and Borel--Cantelli imply a positive asymptotic sampling fraction almost surely.}
\[
\liminf_{t\to\infty}\frac{|H_t^i|}{t}>0
\qquad \mathbf P\text{-almost surely}.
\]

Player $i$ starts from a full-support prior $\mu_0^i\in\Delta\Theta^i$ and updates using only the observations in $H_t^i$. Thus, for every Borel set $B\subseteq\Theta^i$,
\[
\mu_t^i(B)
=
\frac{
\int_B
\prod_{\tau\in H_t^i}
q_\theta^i(y_\tau^i\mid a_\tau)
\,\mu_0^i(d\theta)
}{
\int_{\Theta^i}
\prod_{\tau\in H_t^i}
q_\theta^i(y_\tau^i\mid a_\tau)
\,\mu_0^i(d\theta)
},
\]
with $\mu_t^i=\mu_0^i$ when $H_t^i=\varnothing$.

We call any process satisfying the preceding timing, sampling, growth, and Bayesian-updating conditions an \emph{admissible learning process}.

Whenever $H_t^i\neq\varnothing$, let
\[
\hat\sigma_t^i
=
\frac{1}{|H_t^i|}
\sum_{\tau\in H_t^i}\delta_{a_\tau}
\]
denote player $i$'s empirical action distribution induced by the selected sample.

\begin{lemma}[Asymptotic beliefs]\label{lemm:Berk-selected}
For every admissible learning process, with $\mathbf P$-probability one, the following holds simultaneously for every player $i$. If, along some subsequence $(t_k)_k$, $\hat\sigma_{t_k}^i\Rightarrow\sigma^i\in\Delta\mathbb A$, then for every closed set $E\subseteq\Theta^i$ with $E\cap\Theta^{m,i}(\sigma^i)=\varnothing$, there exist constants $C>0$, $\rho>0$, and an integer $K$ such that
\begin{equation}\label{eq:exponential}
\mu_{t_k}^i(E)
\leq
C\exp\{-\rho |H_{t_k}^i|\}.
\end{equation}
for all $k\geq K$. In particular, if also $\mu_{t_k}^i\Rightarrow\mu^i$, then $\mu^i\in\Delta\Theta^{m,i}(\sigma^i)$.
\end{lemma}

Lemma~\ref{lemm:Berk-selected} says that along any subsequence for which player $i$'s selected empirical distribution converges, posterior probability assigned to any closed set of models separated from the best-fitting models for the limiting selected data vanishes exponentially fast.\footnote{The exponential rate is not needed for the sequel.} Consequently, every limiting posterior belief is supported on $\Theta^{m,i}(\sigma^i)$. The result builds on \citet{berk1966limiting}'s analysis of misspecified Bayesian learning and its extensions to endogenous data, including \citet{esponda2016berk} and \citet{esponda2021asymptotic}. Relative to those results, the lemma allows inference to be based on selected, possibly nonnested, samples of realized observations and allows for a compact action space. Related concentration arguments for stochastic and nonmonotone samples appear in \citet{fudenberg2024selective} in the context of selective memory. Their recall process may depend on realized outcomes, whereas our sampling rule requires the future use of an observation to be determined before its consequence is realized.

Lemma~\ref{lemm:Berk-selected} places no optimality restriction on the action rule. Actions may be generated by any history-dependent and randomized procedure. What matters for the statistical argument is that sampling is non-anticipating and that the selected sample grows sufficiently quickly. Non-anticipation preserves the conditional mean-zero structure of selected likelihood contributions, and a concentration argument then yields the uniform strong law needed for posterior concentration. The Supplemental Appendix provides the details.

\subsection{Behavioral restrictions}

We now impose additional restrictions on admissible learning processes that connect players' beliefs and forecasts to their actions.

At date $t$, player $i$ forms a forecast
\[
\tilde\sigma_t^{-i}\in\Delta\mathbb A^{-i}
\]
as a measurable function of the history $h_{t-1}$. For any
$\sigma\in\Delta\mathbb A$, let $\sigma^{-i}$ denote its marginal on
$\mathbb A^{-i}$. We require forecasts to be asymptotically consistent with
the selected sample: with $\mathbf P$-probability one, for every player $i$
and every subsequence $(t_k)_k$,
\begin{equation} \label{eq:forecast-consistency}
\hat\sigma_{t_k}^i\Rightarrow\sigma^i
\quad\Longrightarrow\quad
\tilde\sigma_{t_k}^{-i}\Rightarrow(\sigma^i)^{-i}.
\end{equation}
Thus, whenever player $i$'s selected empirical distribution converges along
a subsequence, her forecast converges to the opponents' marginal of the same
limiting distribution. The simplest example is
$\tilde\sigma_t^{-i}=(\hat\sigma_t^i)^{-i}$, so that player $i$ forecasts opponents' play using the empirical marginal distribution of opponents' actions in her own selected sample.\footnote{Condition \eqref{eq:forecast-consistency} also allows finite-date smoothing and initial conditions, provided their effect vanishes asymptotically. For example, forecasts may mix the selected empirical marginal with an initial full-support forecast using a weight that converges to zero. The separated treatment of model learning and forecasts can also be viewed as a reduced form of a joint Bayesian model over consequences and opponents' play when the latter is governed by flexible nuisance parameters that are variation independent of $\theta^i$. If the subjective model instead imposes cross-restrictions between $\theta^i$ and opponents' behavior, the relevant limiting object would generally involve a joint KL problem.}

For $\varepsilon\geq0$, define the $\varepsilon$-optimal action
correspondence
\[
F_\varepsilon^i(\mu^i,\beta^{-i})
:=
\left\{
a^i\in\mathbb A^i:
\sup_{x^i\in\mathbb A^i}
\int_{\Theta^i}U^i(x^i,\beta^{-i},\theta^i)\,\mu^i(d\theta^i)
-
\int_{\Theta^i}U^i(a^i,\beta^{-i},\theta^i)\,\mu^i(d\theta^i)
\leq\varepsilon
\right\}.
\]
Thus $F_0^i=F^i$.

We now impose a behavioral restriction on the action kernels of an admissible learning process. Following \cite{fudenberg1993learning}, we say that the action rule is \emph{asymptotically myopic} if there exists a deterministic sequence $\varepsilon_t\downarrow0$ such that, for almost every history $h_{t-1}$ generated by the process,
\[
\alpha_t\!\left(
\prod_{i\in I}
F_{\varepsilon_t}^i(\mu_t^i,\tilde\sigma_t^{-i})
\,\middle|\,h_{t-1}
\right)=1.
\]
Thus $\alpha_t$ may depend arbitrarily on the past and may randomize, but its support is restricted to action profiles that are $\varepsilon_t$-optimal for every player. In particular, any selection among multiple approximately optimal actions can depend only on the past history.

The formulation is intentionally permissive. The action rule may be history dependent and randomized, and no particular adjustment process, such as fictitious play or best-response dynamics, is required. Exact best responses are also unnecessary: it is enough that payoff losses vanish asymptotically. The selected-sample formulation can be extended to weighted observations, provided the effective sample size diverges and no single observation has asymptotically dominant weight, and forecasts may be smoothed or adjusted at finite dates provided these effects vanish asymptotically. The key restrictions are that the data used for inference grow sufficiently rich, forecasts become consistent with those data, and sample selection is non-anticipatory with respect to consequence realizations. The non-anticipation restriction prevents the sampling rule from selecting observations on the basis of unusually favorable or unfavorable realized consequences, directly or through later endogenous variables that reveal those consequences. Without it, the conditional distribution of selected consequences given actions need not coincide with the true kernel. This differs from selective-memory models such as \citet{fudenberg2024selective}, which allow recall to depend on realized outcomes and therefore permit systematic memory bias.

\subsection{Limit behavior and self-justification}

Let \(L(a^\infty)\) denote the set of limit action profiles of the realized action path \(a^\infty\). That is,
\[
L(a^\infty)
=
\left\{
a\in\mathbb A:
\exists\, t_k\to\infty
\text{ such that }
a_{t_k}\to a
\right\}.
\]
When \(\mathbb A\) is finite, \(a\in L(a^\infty)\) if and only if \(a\) is played infinitely often along the path.

The next lemma records a purely topological property of limit sets. It uses no learning assumptions and follows only from compactness of the action space. This also motivates requiring self-justified sets to be compact, since they are meant to represent long-run ranges of behavior.

\begin{lemma}\label{lem:limit-set-compact}
For every action path \(a^\infty\in\mathbb A^{\mathbb N}\), the set \(L(a^\infty)\) is nonempty and compact.
\end{lemma}


\begin{proof}
For each \(T\geq1\), let $K_T:=\overline{\{a_t:t\geq T\}}$. Then \(K_{T+1}\subseteq K_T\). Moreover, each \(K_T\) is nonempty and compact because it is a closed subset of the compact space \(\mathbb A\). Hence \(\bigcap_{T=1}^{\infty}K_T\) is nonempty and compact. This intersection is exactly \(L(a^\infty)\): a point belongs to every tail closure if and only if it is the limit of a subsequence of the path. Therefore \(L(a^\infty)\) is nonempty and compact.
\end{proof}

The next result shows that the learning environment just described generates only self-justified long-run behavior.

\begin{theorem}[Limit behavior is self-justified]\label{theo:limit-self-justified}
Consider an admissible learning process with consistent forecasts and asymptotically myopic action rules. With probability one, \(L(a^\infty)\) is self-justified; that is, \(L(a^\infty)\) is nonempty and compact, and $L(a^\infty)\subseteq \Gamma\bigl(L(a^\infty)\bigr)$. Consequently, every limit action profile is Berk--Nash rationalizable.
\end{theorem}
\begin{proof}
Fix a realization in the $\mathbf P$-probability-one event on which Lemma~\ref{lemm:Berk-selected}, forecast consistency, persistence, and asymptotic myopia hold. Let $L=L(a^\infty)$. By Lemma~\ref{lem:limit-set-compact}, $L$ is nonempty and compact. Pick $a\in L$ and a subsequence $(t_k)_k$ with $a_{t_k}\to a$. By compactness, passing to a further subsequence if needed, assume that, for every player $i$, $\hat\sigma_{t_k}^i\Rightarrow\sigma^i$ and $\mu_{t_k}^i\Rightarrow\mu^i$.

We first show that each $\sigma^i$ is supported on $L$. If $x\notin L$, some open neighborhood $U$ of $x$ is visited only finitely often. Since $|H_t^i|\to\infty$, we have $\hat\sigma_t^i(U)\to0$, and hence $\hat\sigma_{t_k}^i(U)\to0$. Since $\hat\sigma_{t_k}^i\Rightarrow\sigma^i$, the Portmanteau theorem gives $\sigma^i(U)\leq\liminf_k\hat\sigma_{t_k}^i(U)=0$. Thus $x\notin\operatorname{supp}(\sigma^i)$, and therefore $\operatorname{supp}(\sigma^i)\subseteq L$.

By Lemma \ref{lemm:Berk-selected}, $\mu^i\in\Delta\Theta^{m,i}(\sigma^i)$ for every $i$. 
By forecast consistency, for every player $i$, $\tilde\sigma_{t_k}^{-i}\Rightarrow(\sigma^i)^{-i}$. Asymptotic myopia implies $a_{t_k}^i\in F_{\varepsilon_{t_k}}^i(\mu_{t_k}^i,\tilde\sigma_{t_k}^{-i})$, with $\varepsilon_{t_k}\to0$. By the closed-graph property of $(\varepsilon,\mu^i,\beta^{-i})\mapsto F_\varepsilon^i(\mu^i,\beta^{-i})$ (Lemma \ref{lem:basic-properties}(ii) in the Appendix), it follows that $a^i\in F^i(\mu^i,(\sigma^i)^{-i})$ for every player $i$.

Thus, for each player \(i\), there exists \(\sigma^i\in\Delta L\) and \(\mu^i\in\Delta\Theta^{m,i}(\sigma^i)\) such that $a^i\in F^i(\mu^i,(\sigma^i)^{-i})$. By the definition of \(\Gamma\), \(a\in\Gamma(L)\). Since \(a\in L\) was arbitrary, \(L\subseteq\Gamma(L)\). Since \(L\) is nonempty and compact, \(L\in\mathcal S\), and hence every \(a\in L\) belongs to \(\mathcal B=\bigcup_{S\in\mathcal S}S\). Thus every limit action profile is Berk--Nash rationalizable.
\end{proof}

The intuition is as follows. Take a limit action profile $a$. Along a subsequence converging to $a$, each player's selected empirical distribution has a limit $\sigma^i$ supported on $L(a^\infty)$. Lemma~\ref{lemm:Berk-selected} implies that the limiting posterior is supported on the models that best fit $\sigma^i$, while forecast consistency makes the limiting forecast equal to $(\sigma^i)^{-i}$. Asymptotic myopia then implies that $a^i$ is optimal under these limiting beliefs. Thus every component of every limit action profile is justified by some distribution supported on the limit set, yielding $L(a^\infty)\subseteq\Gamma(L(a^\infty))$.

\subsection{Foundation for common-sample refinement}

The learning foundation above allows different players to rely on different selected samples from the realized history. This is why a limit profile may be justified by player-specific distributions $\sigma^i$. The common-sample refinement imposes the stronger requirement that all components of a profile be justified by a single distribution over action profiles. We now obtain the corresponding learning result by requiring players to use the same selected sample.

Consider an admissible learning process in which sampling is common across players: for every date $t$, $J_t^i=J_t^j$ for all $i,j\in I$. Hence $H_t^i=H_t$ for every player $i$. Write
\[
\hat\sigma_t=\frac{1}{|H_t|}\sum_{\tau\in H_t}\delta_{a_\tau}
\]
for the common selected empirical distribution. We also require forecasts to be consistent with this common sample: with $\mathbf P$-probability one, for every subsequence $(t_k)_k$,
\[
\hat\sigma_{t_k}\Rightarrow\sigma
\quad\Longrightarrow\quad
\tilde\sigma_{t_k}^{-i}\Rightarrow\sigma^{-i}
\quad\text{for every }i\in I.
\]

\begin{theorem}[Common-sample limit behavior]\label{theo:common-sample-foundation}
Consider an admissible learning process with a common selected sample, forecasts consistent with the common sample, and asymptotically myopic action rules. With $\mathbf P$-probability one, $L(a^\infty)\subseteq\Gamma^{CS}\bigl(L(a^\infty)\bigr)$.
Consequently, $L(a^\infty)$ is common-sample self-justified and every limit action profile is common-sample Berk--Nash rationalizable.
\end{theorem}

\begin{proof}
The proof is identical to that of Theorem~\ref{theo:limit-self-justified}, except that the common selected sample yields a single subsequential limit $\sigma$ for all players, in place of the player-specific limits $\sigma^i$. Hence the same $\sigma$ justifies every component of the limit profile.
\end{proof}

\subsection{Sustaining Self-Justified Sets}

We next establish a partial converse to the limit result above. The previous subsection showed that the long-run limit set of any admissible learning process with consistent forecasts and asymptotically myopic behavior must be self-justified. We now show that, under identification, every nonempty compact self-justified set can in turn be sustained as the long-run range of behavior by some admissible learning process satisfying the same asymptotic behavioral requirements.

Throughout this subsection, assume that the game is identified: for every player $i$ and every $\sigma\in\Delta\mathbb A$, the set $\Theta^{m,i}(\sigma)$ is a singleton. We write its unique element as
\[
\theta^i(\sigma),
\qquad
\Theta^{m,i}(\sigma)=\{\theta^i(\sigma)\}.
\]

\begin{theorem}[Sustaining self-justified sets]\label{theo:sustain-self-justified}
Suppose the game is identified. Let $S\subseteq\mathbb A$ be a nonempty compact self-justified set. Then, for every $\alpha>0$, there exists an admissible learning process with consistent forecasts and an asymptotically myopic action rule such that
\[
\mathbf P\bigl(L(a^\infty)=S\bigr)>1-\alpha.
\]
\end{theorem}

\begin{proof}
See the Appendix.
\end{proof}

The result follows the logic of \cite{fudenberg1993learning}'s learning foundation for Nash equilibrium. Their key insight is that equilibrium behavior can be sustained by a process that is not exactly optimal after every finite history, but becomes arbitrarily close to optimal over time. \cite{esponda2016berk} extend this argument to Berk--Nash equilibrium under misspecification. Theorem~\ref{theo:sustain-self-justified} uses the same basic idea, but the object being sustained is now a compact set $S$ rather than a singleton steady state. The construction therefore specifies an admissible learning process whose action path has limit set $S$, while different dates and players may use different selected samples from that same long-run range.

The compactness restriction comes from the conclusion of the theorem. Since $\mathbb A$ is compact, the limit set of any action path is nonempty and compact. Thus only compact sets can literally be sustained as $L(a^\infty)$. This restriction does not affect the set of Berk--Nash rationalizable profiles: if a profile belongs to some self-justified set $S$, then it also belongs to the compact self-justified set $\overline S$.\footnote{By monotonicity, $S\subseteq\overline S$ implies $\Gamma(S)\subseteq\Gamma(\overline S)$. Thus, if $S\subseteq\Gamma(S)$, then $S\subseteq\Gamma(\overline S)$. Since $\Gamma(\overline S)$ is closed, it contains $\overline S$.}

The idea of the proof is as follows. Since $S$ is self-justified, each profile $a\in S$ can be justified by suitable selected empirical distributions supported on $S$. The construction chooses a target path whose limit set is $S$ and, at dates when the target path calls for action profile $a$, selects past observations so that each player's empirical distribution is close to one that justifies $a^i$. The selected samples are constructed to satisfy the sample-growth condition required for admissibility, while still allowing different dates and players to use different empirical distributions over the same long-run range $S$.

Identification ensures that the empirical distributions selected by the construction pin down the relevant limiting beliefs. If player $i$'s selected empirical distribution is close to $\sigma$, Bayesian learning places most posterior mass near the unique best-fitting parameter $\theta^i(\sigma)$, so actions justified by $\sigma$ become approximately optimal. The target path is followed whenever the prescribed action is $\varepsilon_t$-optimal, and otherwise the process switches to an exact best response. The sequence $\varepsilon_t\downarrow0$ is chosen slowly enough that the probability of ever departing from the target path is smaller than $\alpha$. On the event that no departure occurs, the realized path coincides with the target path and therefore has limit set exactly $S$.

The conclusion is not almost sure because the posterior is based on realized consequences. Even when the selected empirical distribution is close to a distribution that justifies the prescribed action, finite samples can generate atypical consequence realizations that make the prescribed action fail to be $\varepsilon_t$-optimal. The construction makes the probability of such deviations arbitrarily small, but does not rule them out entirely.

The construction therefore has a selected-history interpretation. The action path generates the long-run range $S$, while beliefs at different dates may be based on different selected samples drawn from past play in that range. This allows different actions in $S$ to be supported by different empirical distributions, even though all observations come from the same realized history. The theorem is a possibility result rather than a selection result: it does not say which self-justified set will arise under a given learning environment, but rather that under identification no nonempty compact self-justified set is ruled out on learning grounds alone.

\section{Examples}

\subsection{Returns to effort and misguided learning}

Following the returns-to-effort application in \cite{heidhues2018unrealistic}, consider a single agent choosing effort \(a\in\mathbb A=[0,\bar a]\), where \(\bar a\) is large enough not to bind. Output is

$$
y=(\alpha^*+a)\theta^*+\varepsilon,
\qquad
\varepsilon\sim\mathcal N(0,1),
$$
where \(\alpha^*\geq0\) is true ability and \(\theta^*>0\) is the true return to effort. Payoffs are $\pi(a,y)=y-c(a)$, where \(c\) is differentiable and strictly convex, \(c(0)=c'(0)=0\), and \(c'(a)\to\infty\).

The agent is misspecified about ability. He believes ability is \(\alpha>0\), which he holds fixed, and learns only the return parameter \(\theta\in[0,\bar\theta]\):

$$
y=(\alpha+a)\theta+\varepsilon,
\qquad
\varepsilon\sim\mathcal N(0,1).
$$

Overconfidence corresponds to \(\alpha>\alpha^*\), and underconfidence to \(\alpha<\alpha^*\).

As a benchmark, if the agent knew the true return \(\theta^*\), the mistaken belief about ability would affect the perceived level of output but not its marginal return to effort. Hence the correct-prior, no-learning action is $a^{opt}=(c')^{-1}(\theta^*)$.

For a selected-sample distribution \(\sigma\in\Delta\mathbb A\), the KL-minimizing return solves

$$
\theta^m(\sigma)
\in
\arg\min_{\theta\in[0,\bar\theta]}
\frac12
\int_{\mathbb A}
\left(
(\alpha+a)\theta-(\alpha^*+a)\theta^*
\right)^2
\sigma(da).
$$

The objective is strictly convex, so the minimizer is unique. For a degenerate action \(a\),

$$
\theta^m(\delta_a)
=
\theta^*
\frac{\alpha^*+a}{\alpha+a}.
$$

More generally, \(\theta^m(\sigma)\) is a weighted average of the values \(\theta^m(\delta_a)\) in the support of \(\sigma\).

Given belief \(\theta\), the agent chooses effort satisfying \(c'(a)=\theta\). Define the justification map

$$
T(a)
:=
(c')^{-1}\!\left(\theta^m(\delta_a)\right).
$$

Berk--Nash equilibrium actions are exactly the fixed points of \(T\). Moreover, for an interval \(A=[L,H]\),

$$
\Gamma(A)
=
\left\{
(c')^{-1}(\theta^m(\sigma)):
\sigma\in\Delta A
\right\}.
$$

Since \(\theta^m(\sigma)\) ranges between the extreme values of \(\theta^m(\delta_a)\) on \(A\),

$$
\Gamma([L,H])
=
\begin{cases}
[T(L),T(H)], & \text{if }T\text{ is increasing},\\[3pt]
[T(H),T(L)], & \text{if }T\text{ is decreasing}.
\end{cases}
$$

Thus Berk--Nash rationalizability is characterized by iterating this interval map.

\emph{Overconfidence.}
Suppose \(\alpha>\alpha^*\). Then $\theta^m(\delta_a)<\theta^*$ for every \(a\), and
$$
\frac{d}{da}\theta^m(\delta_a)
=
\theta^*
\frac{\alpha-\alpha^*}{(\alpha+a)^2}
>0.
$$
Hence \(T\) is strictly increasing. Overconfidence makes the agent attribute disappointing output to a low return to effort; this bias is strongest at low effort and becomes weaker as effort rises.

The left panel of Figure~\ref{fig:returns_effort} illustrates the resulting self-defeating learning. Since \(\theta^m(\delta_a)<\theta^*\), every Berk--Nash equilibrium lies below $a^{opt}=(c')^{-1}(\theta^*)$. 

Starting from \(A^0=[0,\bar a]\) and defining \(A^{k+1}=\Gamma(A^k)\), write $A^k=[a_{\min}^k,a_{\max}^k]$. Because \(T\) is increasing,
$$
a_{\min}^{k+1}=T(a_{\min}^k),
\qquad
a_{\max}^{k+1}=T(a_{\max}^k).
$$

The endpoints converge to the smallest and largest fixed points of \(T\). Hence Berk--Nash rationalizability is the interval $[a_{\min}^\infty,a_{\max}^\infty]$ between the smallest and largest Berk--Nash equilibrium actions. If the fixed point is unique, the interval collapses to that equilibrium.

Thus the self-defeating mechanism extends beyond equilibrium selection: throughout the rationalizable range, the agent can justify low effort with pessimistic beliefs about the return to effort, and every rationalizable action lies below the correct-prior benchmark.

\begin{figure}[htbp]
\hspace{-1cm}
\centering
\begin{subfigure}[b]{0.5\textwidth}
\centering

\begin{tikzpicture}
\begin{axis}[
width=8cm,
height=5cm,
xmin=0, xmax=3,
ymin=0, ymax=1.5,
samples=100,
domain=0:3,
axis lines=left,
xtick=\empty,
ytick={1},
yticklabels={$\theta^*$},
xlabel={$a$},
xlabel style={anchor=north east, at={(1,0)}, xshift=5pt},
ylabel style={at={(0,1)}, anchor=north, rotate=-90, xshift=-15pt, yshift=10pt}
]

\addplot[thick, red] {1 + (1)*(1 - 3)/(3 + x)};
\node[red] at (axis cs:2.35,0.5) {\scriptsize overconfident: $\theta^m(\delta_\cdot)$};
\node[black] at (axis cs:2,1.4) {\scriptsize mg cost: $c'(\cdot)$};

\addplot[only marks, mark=*, mark size=1.25pt, red] coordinates {
(.27,.38)
(.65,.45)
(1.4,.54)
};

\addplot[only marks, mark=*, mark size=1.25pt, black] coordinates {
(2.04,1)
};

\node at (axis cs:2,1.1) {\tiny $a^{opt}$};
\node at (axis cs:.23,.47) {\tiny $a_S$};
\node at (axis cs:.65,.4) {\tiny $a_M$};
\node at (axis cs:1.4,.62) {\tiny $a_L$};

\addplot[thick, black, domain=0:3, samples=300]
{0.45 / (1 + exp(-25 * (x - 0.2))) + ln(1 + exp(5 * (x - 1.5))) / 5};

\addplot[dashed, black] coordinates {(0,1) (3,1)};

\end{axis}
\end{tikzpicture}

\caption{Overconfident agent}

\end{subfigure}
\begin{subfigure}[b]{0.5\textwidth}
\centering

\begin{tikzpicture}
\begin{axis}[
width=8cm,
height=5cm,
xmin=0, xmax=3,
ymin=0, ymax=2.1,
samples=100,
domain=0:3,
axis lines=left,
xtick=\empty,
ytick={1},
yticklabels={$\theta^*$},
xlabel={$a$},
xlabel style={anchor=north east, at={(1,0)}, xshift=5pt},
ylabel style={at={(0,1)}, anchor=north, rotate=-90, xshift=-15pt, yshift=10pt},
xtick={0.22, 0.94},
xticklabels={{\tiny\textcolor{blue}{$a^{\infty}_{\min}$}}, {\tiny\textcolor{blue}{$a^{\infty}_{\max}$}}},
]

\addplot[thick, blue] {1 + (1)*(1 - 0.5)/(0.5 + x)};
\node[blue] at (axis cs:2.2,1.35) {\scriptsize underconfident: $\theta^m(\delta_\cdot)$};

\addplot[thick, black, domain=0:3, samples=300]
{1.5 * (1 - exp(-10*x)) + 0.5 * ln(1 + exp(10*(x - 1.0)))};

\node[black] at (axis cs:1.5,1.8) {\scriptsize mg cost: $c'(\cdot)$};

\addplot[dashed, black] coordinates {(0,1) (3,1)};

\node at (axis cs:.2,.9) {\tiny $a^{opt}$};
\addplot[only marks, mark=*, mark size=1.25pt, black] coordinates {
(.11,1)
};

\addplot[only marks, mark=*, mark size=1.25pt, black] coordinates {
(0.511,1.495)
};

\node at (axis cs:0.511,1.608) {\tiny $a^*$};

\draw[dashed, green!70] (axis cs:.22,1.34) rectangle (axis cs:.94,1.7);
\draw[dashed, black!40] (axis cs:0.22,0) -- (axis cs:0.22,1.34);
\draw[dashed, black!40] (axis cs:0.94,0) -- (axis cs:0.94,1.34);

\end{axis}
\end{tikzpicture}

\caption{Underconfident agent}

\end{subfigure}
\caption{
		Returns to effort example
{\footnotesize
The optimal action \(a^{opt}\) is where the marginal cost curve \(c'\) intersects the true return \(\theta^*\). Berk--Nash equilibrium actions lie at intersections of \(c'\) and the KL-minimizing return curve \(\theta^m(\delta_\cdot)\). In the overconfident case, there are three Berk-Nah equilibria, $a_S$, $a_M$, and $a_L$, and Berk--Nash rationalizability is the interval between the smallest and largest equilibrium actions, $[a_S,a_L]$. In the underconfident case, the Berk--Nash equilibrium is unique at $a^*$, but the rationalizable set is the interval $[a^\infty_{min},a^\infty_{max}]$ whose endpoints form a two-cycle of \(T\).
}
}
\label{fig:returns_effort}
\end{figure}

\emph{Underconfidence.}
Suppose instead that \(\alpha<\alpha^*\). Then $\theta^m(\delta_a)>\theta^*$ for every \(a\), and \(\theta^m(\delta_a)\) is strictly decreasing, so \(T\) is strictly decreasing. The agent now understates ability and therefore attributes unexpectedly high output to a high return to effort. Low sampled effort generates especially optimistic beliefs about returns and therefore justifies high effort, whereas high sampled effort weakens the ability mistake and justifies lower effort. Thus the feedback is self-correcting rather than self-reinforcing.

Because \(T\) is decreasing, it has a unique fixed point, so the Berk--Nash equilibrium is unique. Since \(\theta^m(\delta_a)>\theta^*\), this equilibrium lies above \(a^{opt}\): the agent overestimates the return to effort and chooses too much effort relative to the correct-prior benchmark.

For rationalizability, if $A^k=[a_{\min}^k,a_{\max}^k],$ then $a_{\min}^{k+1}=T(a_{\max}^k)$ and $a_{\max}^{k+1}=T(a_{\min}^k)$. The limiting endpoints satisfy $T(a_{\min}^\infty)=a_{\max}^\infty$ and $T(a_{\max}^\infty)=a_{\min}^\infty$. Hence they form a two-cycle of \(T\), or equivalently fixed points of \(T^2\).

The underconfidence case therefore shows that a unique Berk--Nash equilibrium need not imply a singleton rationalizable prediction. If \(T^2\) has a unique fixed point, then $
a_{\min}^\infty=a_{\max}^\infty$ and Berk--Nash rationalizability collapses to the equilibrium. Otherwise, the rationalizable set is the nontrivial interval $[a_{\min}^\infty,a_{\max}^\infty]$, whose endpoints form a two-cycle of \(T\), as depicted in the right panel of Figure~\ref{fig:returns_effort}.


This characterization is closely related to the belief-space iteration in
\cite{frick2023belief} (see Section \ref{subsec:othersolutions} for further
discussion). Their one-dimensional active-learning result applies here because
the optimal action is uniquely given by $a(\mu)=(c')^{-1}\left(\int\theta\,\mu(d\theta)\right)$ 
and is continuous in beliefs, while the KL criterion is continuous and has a
unique minimizer at each induced action. Let
\(b(\theta)=(c')^{-1}(\theta)\) and
\(M(\theta)=\theta^m(\delta_{b(\theta)})\). Their belief-space iteration is
generated by \(M\), whereas our action-space iteration is generated by \(T\).
The two maps are conjugate, $M=c'\circ T\circ b$, 
and therefore the two procedures deliver the same prediction in this example.

The scope and interpretation of the two approaches nevertheless differ. With
a finite effort set and nonconstant optimal behavior, the optimal-action rule
is necessarily discontinuous at some beliefs, so the continuity condition in
\cite{frick2023belief} fails. Our rationalizability analysis does not require a
continuous single-valued policy and therefore continues to apply. Moreover,
our learning result gives the rationalizable action set a robustness
interpretation: it bounds the long-run behavior generated by the broader class
of learning and sampling procedures considered here.

\paragraph{Relation to \citet{heidhues2018unrealistic}.}
This example is a special case of the returns-to-effort environment in \citet{heidhues2018unrealistic}, but we use it to study a different object. In the overconfidence case, HKS analyze a particular Bayesian learning process and, under uniqueness, show convergence to the Berk--Nash equilibrium. In our notation, this is the case in which the increasing justification map $T$ has a unique fixed point. Berk--Nash rationalizability recovers the same point prediction: the iterated rationalizable interval collapses to that fixed point. Our learning foundation strengthens this conclusion by showing convergence for the broader class of learning procedures considered here, including selected samples and asymptotically optimal behavior; because the model is identified, forward-looking behavior is also covered as a special case (see Section~\ref{subsec:forward}).

When the overconfidence case admits multiple Berk--Nash equilibria, the HKS argument no longer yields a convergence result. Berk--Nash rationalizability nevertheless gives a sharp set-valued prediction: since $T$ is increasing, the rationalizable set is the interval between the smallest and largest Berk--Nash equilibrium actions. Throughout this interval, the agent underestimates the return to effort and therefore chooses inefficiently low effort relative to the correct-prior benchmark. The learning foundation implies that this interval bounds the long-run behavior of the learning procedures considered here, including the HKS process, while the converse result shows that such self-justified long-run ranges can in fact be sustained.

The underconfidence case illustrates a different role for the concept. Here $T$ is decreasing, so the Berk--Nash equilibrium is unique, but uniqueness alone need not imply robust convergence across learning procedures. If $T^2$ has a unique fixed point, Berk--Nash rationalizability collapses to the equilibrium and the learning foundation yields almost-sure convergence. If $T^2$ has multiple fixed points, the rationalizable set is instead the interval $[a_{\min}^\infty,a_{\max}^\infty]$, whose endpoints form a two-cycle of $T$. Thus the set-valued concept identifies when the unique equilibrium is a robust long-run prediction and when robustness instead requires a range.

Thus Berk--Nash rationalizability is not merely a technique for analyzing one fully specified dynamic. For a given dynamic, it can be useful because it gives simple conditions under which convergence follows and, when convergence to a point is not guaranteed, provides bounds on the possible long-run range. Dynamic-specific analysis may of course sharpen these predictions. But the solution concept also stands on its own: it is a set-valued analogue of Berk--Nash equilibrium, describing which ranges of behavior are internally justified by the evidence generated within those ranges. The learning foundation explains why these self-justified sets are relevant across a broad class of learning procedures, rather than being artifacts of any particular dynamic.

\subsection{Strategic feedback and self-justification}

This example studies learning with strategic feedback, where each player's action affects conditions faced by others and those conditions then guide future behavior. Such feedback arises in supply-chain interactions, monetary and fiscal policy, organizational and regulatory decision making, and firms' adjustments of strategic choices over time. A natural misspecification is to treat these observed conditions as draws from an underlying environment while neglecting their endogenous dependence on behavior. The example shows how this can generate large self-justified sets even when Berk--Nash equilibrium is unique, and how a common-sample requirement can sharply refine the prediction.\footnote{Supply-chain environments provide a well-documented example. \citet{lee1997information} emphasize that upstream firms use downstream orders as information about demand, while \citet{sterman1989modeling} provides experimental evidence that decision makers in inventory systems systematically underappreciate the endogenous feedback generated by their own actions.}

Consider two players, \(i=1,2\), whose actions \(a^i\in[-1,1]\) interact through a feedback system. Player \(i\) observes a target \(y^i\) and has payoff $u^i(a^i,y^i)=-(a^i-y^i)^2$. The true targets are
$$
y^1=\gamma a^1-\omega a^2+\varepsilon^1,
\qquad
y^2=\omega a^1+\gamma a^2+\varepsilon^2,
$$
where \(\varepsilon^1,\varepsilon^2\) are independent standard normal shocks, \(\gamma\geq0\), and \(\omega>0\). The parameter \(\gamma\) captures own-action feedback, while the opposite signs of the cross-effects generate a rotational component.

Players neglect this endogeneity. Player \(i\) believes instead that \(y^i=\theta^i+\varepsilon^i\), with \(\theta^i\) unknown, and therefore chooses \(a^i=\Pi(\theta^i)\), where \(\Pi\) denotes projection onto \([-1,1]\). This captures situations in which agents treat endogenous conditions as draws from a stationary environment rather than as outcomes partly generated by past actions.

For a fixed action profile \(a\), the KL-minimizing parameters are

$$
\theta^{m,1}(a)=\gamma a^1-\omega a^2,
\qquad
\theta^{m,2}(a)=\omega a^1+\gamma a^2.
$$

Hence a Berk--Nash equilibrium satisfies \(a=\Pi(Ba)\), where

$$
B=
\begin{pmatrix}
\gamma & -\omega\\
\omega & \gamma
\end{pmatrix}.
$$

Throughout, assume $\gamma<1+\omega$. Under this restriction the unique Berk--Nash equilibrium is \(a^*=(0,0)\).\footnote{For \(\gamma<1\), the projection inequality and \(a'Ba=\gamma\|a\|^2\) rule out any nonzero fixed point. For \(\gamma>1\), any nonzero fixed point must lie on the boundary; the projection conditions then require \(\omega\leq\gamma-1\), contrary to \(\gamma<1+\omega\).}

Berk--Nash rationalizability can be very different. If player \(i\) selects a distribution \(\sigma^i\) over a set \(S\), let \(\bar a^{\,i}\) denote its mean. Then
$$
\theta^{m,1}(\sigma^1)=\gamma\bar a^{\,1}_1-\omega\bar a^{\,1}_2,
\qquad
\theta^{m,2}(\sigma^2)=\omega\bar a^{\,2}_1+\gamma\bar a^{\,2}_2.
$$
Because players may use different selected samples, the two coordinates of a candidate profile can be justified by different mean action profiles.

The largest self-justified set is
$$
\mathcal B=
\begin{cases}
\{(0,0)\}, & \gamma+\omega<1,\\[3pt]
[-1,1]^2, & \gamma+\omega\geq1.
\end{cases}
$$
To see this, let \(M\) be the largest absolute coordinate attained in a self-justified set \(S\). Any sample supported on \(S\) has mean coordinates bounded by \(M\), so any fitted parameter, and hence any justified action, has absolute value at most \((\gamma+\omega)M\). Self-justification therefore requires \(M\leq(\gamma+\omega)M\), implying \(M=0\) when \(\gamma+\omega<1\). Conversely, if \(\gamma+\omega\geq1\), taking \(S=[-1,1]^2\), each player's fitted parameter can range over \([-(\gamma+\omega),\gamma+\omega]\), which contains every action in \([-1,1]\). Since the players may use different samples, every profile can be justified.\footnote{Here rationalizability is characterized directly, without iteration; we thank David Pearce for emphasizing to us that such direct arguments can sometimes be useful.}

Thus the threshold \(\gamma+\omega=1\) measures whether feedback is strong enough for the data generated within a bounded range of behavior to reproduce that same range. Importantly, the unique Berk--Nash equilibrium remains \((0,0)\) on both sides of this threshold.

The common-sample refinement can sharply reduce this set. If both players use the same sample with mean \(\bar a\), their fitted parameters are \(B\bar a\), so

$$
\Gamma^{CS}(S)
=
\{\Pi(B\bar a):\bar a\in\operatorname{co}(S)\}.
$$

Because \(\|Bx\|=\sqrt{\gamma^2+\omega^2}\|x\|\), any common-sample self-justified set collapses to \(\{(0,0)\}\) whenever \(\gamma^2+\omega^2<1\). 
Thus, when \(\gamma+\omega\geq 1\) but \(\gamma^2+\omega^2<1\), requiring a common sample reduces the rationalizable set from the entire square to the unique equilibrium:
\[
\mathcal B=[-1,1]^2,
\qquad
\mathcal B^{CS}=\{(0,0)\}.
\]
The restriction need not sharpen the prediction, however. If \(\gamma^2+\omega^2\geq\gamma+\omega\), then \(B^{-1}\) maps the square into itself, so every profile \(a\) is justified by a common sample with mean \(B^{-1}a\). Hence \(\mathcal B^{CS}=\mathcal B=[-1,1]^2\).

\paragraph{Comparison with a benchmark learning dynamic.}
To compare our predictions with those of a specific learning rule, suppose players use their full histories and have normal priors over \(\theta^i\). Standard stochastic-approximation arguments imply that play converges almost surely to \((0,0)\) when \(\gamma<1\), whereas for \(\gamma>1\) the equilibrium is unstable and, in this Gaussian environment, the probability of convergence to it is zero.\footnote{With normal priors, posterior means satisfy \(m_{t+1}=m_t+\alpha_{t+1}[B\Pi(m_t)-m_t+\varepsilon_{t+1}]\), with \(\alpha_t\sim t^{-1}\). The associated ODE is \(\dot m=B\Pi(m)-m\). For \(\gamma<1\), a Lyapunov argument gives global asymptotic stability of zero. For \(\gamma>1\), the local eigenvalues are \((\gamma-1)\pm i\omega\), so zero is a repeller; the exact linear recursion and nondegenerate Gaussian innovations imply avoidance of the equilibrium. See, for example, \citet{heidhues2021convergence} for the use of stochastic approximation in misspecified learning.}

The contrast is especially clear in the region
$$
\gamma<1
\qquad\text{and}\qquad
\gamma+\omega\geq1.
$$
Here full-history Bayesian learning converges almost surely to the unique Berk--Nash equilibrium \((0,0)\), yet $\mathcal B=[-1,1]^2$. Thus the large rationalizable set is not a consequence of equilibrium instability. The two objects answer different questions: the benchmark fixes a particular way of using past observations, whereas Berk--Nash rationalizability asks what can be justified when sampling from experience is left unspecified. The common-sample refinement can narrow this prediction for some parameters, but for others even \(\mathcal B^{CS}\) remains the whole square.


\subsection{A linear-quadratic network example}

The common-sample restriction can have substantial bite even when the game is correctly known. Network games provide a natural setting for this possibility because several players may respond to the same neighbors: classical rationalizability allows them to justify their actions using different beliefs about those neighbors, whereas a common sample forces these beliefs to come from the same empirical distribution. The following three-player linear-quadratic example makes this contrast stark. Linear-quadratic network games are widely used to study peer effects, local public goods, crime, and strategic complementarities or substitutes; see, for example, \cite{ballester2006whos}, \cite{bramoulle2007public}, and \cite{bramoulle2014strategic}.

There are three players, with actions \(a^i\in[-1,1]\). Payoffs are

$$
u^1(a)=-\frac12(a^1-\lambda a^3)^2,\qquad
u^2(a)=-\frac12(a^2-\lambda a^3)^2,
$$

and

$$
u^3(a)
=
-\frac12
\left(
a^3+\mu\frac{a^1+a^2}{2}
-\kappa\frac{a^1-a^2}{2}
\right)^2.
$$

Assume \(0<\lambda\leq1\), \(\mu>0\), \(\lambda\mu<1\), and \(\kappa\lambda\geq1\). Players 1 and 2 track player 3 with strength \(\lambda\). Player 3 responds negatively to their average action and positively to the imbalance between them. The imbalance term will be the source of permissiveness under classical rationalizability.

Because the game is correctly specified and identified, beliefs matter only through opponents' empirical action distributions. If \(m_j(\sigma)\) denotes the mean action of player \(j\) under \(\sigma\), then

$$
BR^1(\sigma)=\{\lambda m_3(\sigma)\},\qquad
BR^2(\sigma)=\{\lambda m_3(\sigma)\},
$$

while

$$
BR^3(\sigma)
=
\left\{
\Pi\!\left(
-\mu\frac{m_1(\sigma)+m_2(\sigma)}{2}
+\kappa\frac{m_1(\sigma)-m_2(\sigma)}{2}
\right)
\right\}.
$$

\paragraph{Nash equilibrium.}
At equilibrium, \(a^1=a^2=\lambda a^3\), so the imbalance term vanishes and player 3's condition becomes \(a^3=-\mu\lambda a^3\). Hence the unique Nash equilibrium is $a^*=(0,0,0)$.

\paragraph{Rationalizability.}
Under the baseline operator, players may use different empirical distributions. Starting from \([-1,1]^3\), players 1 and 2 can justify any action in \([-\lambda,\lambda]\). Player 3 can justify any action in \([-1,1]\): over the set \([-\lambda,\lambda]^2\), the imbalance term alone can span an interval containing \([-1,1]\) because \(\kappa\lambda\geq1\). Hence

$$
\mathcal B
=
[-\lambda,\lambda]\times[-\lambda,\lambda]\times[-1,1].
$$

This set is self-justified, so no further iteration is needed. The key source of permissiveness is that players 1 and 2 may justify different actions using different empirical distributions over player 3's behavior.\footnote{Only marginal means matter, so the same set obtains under independent-beliefs rationalizability.}

\paragraph{Common-sample rationalizability.}
Now require all players to use the same distribution \(\sigma\). Since players 1 and 2 respond to the same mean \(m_3(\sigma)\), any common-sample justified profile must satisfy \(a^1=a^2\). Thus after one round the candidate set has the form

$$
\{(q,q,r):q\in[-\lambda,\lambda],\ r\in[-1,1]\}.
$$

Once \(a^1=a^2\), every common sample also satisfies \(m_1(\sigma)=m_2(\sigma)\), so player 3's imbalance term disappears. The remaining feedback is therefore

$$
q=\lambda m_3(\sigma),
\qquad
r=-\mu m_1(\sigma).
$$

Two applications of the common-sample operator multiply the relevant ranges by \(\lambda\mu\). Since \(\lambda\mu<1\), repeated application contracts both ranges to zero. Hence $\mathcal B^{CS}=\{(0,0,0)\}$. Thus

$$
NE=\mathcal B^{CS}=\{(0,0,0)\},
\qquad
\mathcal B
=
[-\lambda,\lambda]\times[-\lambda,\lambda]\times[-1,1].
$$

The common-sample restriction works through two simple steps. First, players 1 and 2 must respond to the same empirical distribution of player 3, which forces \(a^1=a^2\). Second, this eliminates the imbalance term in player 3's payoff, leaving only a feedback loop of strength \(\lambda\mu<1\), which contracts to the unique Nash equilibrium.

\section{Extensions and Related Concepts}

\subsection{Rationalizable Distributions}

We defined rationalizability over actions. There is an equivalent definition in terms of distributions over actions, and this alternative definition captures the limit points of empirical-action distributions.

For a set $\Sigma\subseteq\Delta\mathbb A$, define
\[
\Phi(\Sigma)
:=
\left\{
\sigma\in\Delta\mathbb A:
\begin{array}{l}
\text{for every }a\in\operatorname{supp}(\sigma)\text{ and every }i\in I,\\
\text{there exist }\sigma_a^i\in\Sigma
\text{ and }\mu_a^i\in\Delta\Theta^{m,i}(\sigma_a^i)\\
\text{such that }a^i\in F^i\!\left(\mu_a^i,(\sigma_a^i)^{-i}\right)
\end{array}
\right\}.
\]
Thus every action profile in the support of $\sigma$ must be justified, for each player, by beliefs fitted to some reference distribution in $\Sigma$.

\begin{definition}
A distribution $\sigma\in\Delta\mathbb A$ is \emph{Berk--Nash rationalizable} if there exists a set $\Sigma\subseteq\Delta\mathbb A$ such that
$\sigma\in\Sigma\subseteq\Phi(\Sigma)$.
\end{definition}

Moreover, a distribution $\sigma\in\Delta\mathbb A$ is a \emph{generalized Berk--Nash equilibrium} if $\sigma\in\Phi(\{\sigma\})$; see also \cite{esponda2021asymptotic} and \cite{murooka2023convergence}. This generalizes the mixed-strategy Berk--Nash equilibrium notion in EP16 in two respects. First, $\sigma$ may be correlated across players. Second, different action profiles in $\operatorname{supp}(\sigma)$ may be justified by different beliefs. Because the reference set is the singleton $\{\sigma\}$, however, all such beliefs are fitted to the same empirical distribution $\sigma$. In EP16's original definition of Berk-Nash equilibrium, play is represented by a product mixed-strategy profile, and each player's mixed strategy is supported on actions that are optimal under a single belief for that player. Their learning foundation justifies this more restrictive structure by introducing independent Harsanyi-style payoff perturbations.

As with the action-based operator, $\Phi$ is monotone. Hence the union of all sets $\Sigma$ satisfying $\Sigma\subseteq\Phi(\Sigma)$ also satisfies this inclusion. We denote this largest distributionally rationalizable set by $\mathcal M \subseteq\Delta\mathbb A$. As before, $\mathcal B\subseteq\mathbb A$ is  set of Berk--Nash rationalizable action profiles.

\begin{proposition}[Equivalence of action and distributional approaches]
\label{prop:EquivalenceGammaPhi}
\[
\mathcal M=\Delta\mathcal B.
\]
\end{proposition}

\begin{proof}
We first show that $\Delta\mathcal B\subseteq\mathcal M$. Since
$\mathcal B=\Gamma(\mathcal B)$, for every $a\in\mathcal B$ and every player
$i$ there exist $\sigma_a^i\in\Delta\mathcal B$ and
$\mu_a^i\in\Delta\Theta^{m,i}(\sigma_a^i)$ such that
$a^i\in F^i(\mu_a^i,(\sigma_a^i)^{-i})$. Hence every
$\sigma\in\Delta\mathcal B$ belongs to $\Phi(\Delta\mathcal B)$. Since
$\mathcal M$ is the largest set satisfying $\Sigma\subseteq\Phi(\Sigma)$,
it follows that $\Delta\mathcal B\subseteq\mathcal M$.

Conversely, let $B:=\bigcup_{\sigma\in\mathcal M}\operatorname{supp}(\sigma)$. Fix $a\in B$. Then $a\in\operatorname{supp}(\sigma)$ for some
$\sigma\in\mathcal M$. Since $\mathcal M\subseteq\Phi(\mathcal M)$, for every
player $i$ there exist $\sigma_a^i\in\mathcal M$ and
$\mu_a^i\in\Delta\Theta^{m,i}(\sigma_a^i)$ such that
$a^i\in F^i(\mu_a^i,(\sigma_a^i)^{-i})$. By construction,
$\operatorname{supp}(\sigma_a^i)\subseteq B$, so $\sigma_a^i\in\Delta B$.
Therefore \(a\in\Gamma(B)\), so \(B\subseteq\Gamma(B)\). By the closure argument following the definition of self-justification, this implies that \(\overline B\) is self-justified. Hence
\(B\subseteq\overline B\subseteq\mathcal B\). Finally, every \(\sigma\in\mathcal M\) satisfies
\(\operatorname{supp}(\sigma)\subseteq B\subseteq\mathcal B\), so
\(\sigma\in\Delta\mathcal B\). Thus
\(\mathcal M\subseteq\Delta\mathcal B\).
\end{proof}

The proposition shows that, under our approach, the distributional and action-based formulations are equivalent. Thus working directly with distributions yields no additional restrictions on long-run frequencies or correlations beyond those already implied by the rationalizable action set. For our purposes, it is therefore more convenient to work with action profiles.

Finally, the analogous equivalence also holds for the common-sample refinement: the corresponding rationalizable distributions are exactly the distributions supported on the common-sample rationalizable action set.

\subsection{Mixed-Strategy Rationalizability} \label{sec:mixedR}

Empirical play can be correlated because players respond to a common history. Following \cite{fudenberg1993learning} and EP16, one can instead describe intended mixed strategies by introducing idiosyncratic payoff perturbations that are independent across players and time. This separates correlation in realized histories from correlation in players' intended choices.

For simplicity, suppose that each $\mathbb A^i$ is finite. Let $\varepsilon^i=(\varepsilon_a^i)_{a\in\mathbb A^i}$ be an idiosyncratic payoff shock with an absolutely continuous distribution, independent across players. Given belief $\mu^i$ and forecast $\beta^{-i}$, define the induced stochastic best response by
\[
\kappa^i(a^i\mid\mu^i,\beta^{-i})
:=
P\!\left(
a^i\in\arg\max_{x^i\in\mathbb A^i}
\{U^i(x^i,\beta^{-i},\mu^i)+\varepsilon_{x^i}^i\}
\right).
\]
Thus, conditional on beliefs and forecasts, intended choices are independent across players.

Let $\Sigma\subseteq\prod_i\Delta\mathbb A^i$, and abuse notation by identifying each mixed-strategy profile with its associated product distribution on $\mathbb A$. Define
\[
\Phi_M(\Sigma)
:=
\left\{
\sigma\in\prod_i\Delta\mathbb A^i:
\text{ for every }i,\ \exists\,\hat\sigma^i\in\Sigma,\
\exists\,\mu^i\in\Delta\Theta^{m,i}(\hat\sigma^i)
\text{ such that }
\sigma^i=\kappa^i(\cdot\mid\mu^i,(\hat\sigma^i)^{-i})
\right\}.
\]
A mixed-strategy profile $\sigma$ is \emph{Berk--Nash rationalizable} if there exists $\Sigma\subseteq\prod_i\Delta\mathbb A^i$ such that
$\sigma\in\Sigma\subseteq\Phi_M(\Sigma)$. Under the same regularity conditions as above, the usual monotonicity and learning arguments carry over: there is a largest rationalizable set of intended mixed-strategy profiles, and every limit intended mixed-strategy profile generated by the learning process belongs to it.

If players use a common selected sample, the corresponding refinement requires the same reference profile $\hat\sigma\in\Sigma$ for every player. Independent payoff perturbations therefore eliminate correlation in intended strategies conditional on the history, but they do not eliminate the common-sample restriction on beliefs. The two restrictions have different sources: product-form intended strategies follow from independent idiosyncratic shocks, whereas a common reference distribution follows from learning from the same sample.

\subsection{Forward-Looking Behavior} \label{subsec:forward}

The behavioral assumption above also accommodates forward-looking Bayesian agents under identification. Following EP16 (Appendix C), suppose player $i$ evaluates actions according to the discounted Bayesian value function
\[
V^i(\mu^i,\beta^{-i})
=
\max_{a^i\in\mathbb A^i}
\left\{
U^i(a^i,\beta^{-i},\mu^i)
+
\delta^i
\int
V^i\!\left(
B^i(\mu^i,a^i,\beta^{-i},y^i),
\beta^{-i}
\right)
\bar Q_{\mu^i}^i(dy^i\mid a^i,\beta^{-i})
\right\},
\]
where
\[
\bar Q_{\theta^i}^i(\cdot\mid a^i,\beta^{-i})
:=
\int_{\mathbb A^{-i}}
Q_{\theta^i}^i(\cdot\mid a^i,a^{-i})\,\beta^{-i}(da^{-i})
\]
is the consequence distribution predicted by parameter $\theta^i$, $\bar Q_{\mu^i}^i$ is the corresponding mixture under belief $\mu^i$, and $B^i$ is the Bayesian posterior following the observed consequence. Let $\Phi^i(\mu^i,\beta^{-i})$ denote the set of actions attaining the maximum. We say that player $i$ is \emph{forward looking} if
$a_t^i\in\Phi^i(\mu_t^i,\tilde\sigma_t^{-i})$ at every date $t$.

Under identification---that is, when $\Theta^{m,i}(\sigma)$ is a singleton for every $\sigma$---forward-looking behavior is a special case of the asymptotically myopic behavior assumed above. Indeed, suppose that $\Theta^{m,i}(\sigma)$ is a singleton for every $\sigma$. Along any subsequence for which $\hat\sigma_{t_k}^i\Rightarrow\sigma^i$, Lemma \ref{lemm:Berk-selected} implies that $\mu_{t_k}^i$ converges to a degenerate belief on the unique element of $\Theta^{m,i}(\sigma^i)$. At a degenerate belief, Bayesian updating leaves beliefs unchanged after every action and consequence, so there is no value of experimentation and dynamic optimality coincides with myopic optimality. Under the usual continuity conditions, it follows that forward-looking behavior satisfies the asymptotic-myopia condition imposed above.

\subsection{A support refinement} \label{subsec:supportref}

The solution concept does not require an action profile to belong to the support of the distribution used to justify it. The reason is that recurrence of an action and representation of that action in the empirical distribution need not be synchronized. Along a subsequence with $a_{t_k}\to a$, the selected empirical distributions may converge to some $\sigma$ with $a\notin\operatorname{supp}(\sigma)$: the action may recur precisely at dates at which it receives asymptotically negligible weight in the accumulated sample. Since the empirical distributions need not converge along the full path, different recurrent subsequences may be associated with different limiting empirical regimes. Requiring $a\in\operatorname{supp}(\sigma)$ therefore imposes the additional condition that recurrence of the action be synchronized with the empirical distribution that justifies it.

When $\mathbb A$ is finite and the full history is used, this distinction has a particularly sharp interpretation. Membership in the limit set means that an action is played infinitely often, whereas the support refinement corresponds to requiring a nonvanishing empirical frequency along some subsequence. Thus the refinement distinguishes mere recurrence from recurrence that remains quantitatively represented in the data.

Motivated by this distinction, we consider a refinement that requires the recurrent action to remain represented in the empirical distribution that justifies it. For a nonempty set $S\subseteq\mathbb A$, define
\[
\Gamma^{+}(S)
=
\left\{
a\in\mathbb A:
\begin{array}{l}
\text{for every }i\in I\text{ there exist }\sigma^i\in\Delta S
\text{ and }\mu^i\in\Delta\Theta^{m,i}(\sigma^i)\\
\text{such that }a\in\operatorname{supp}(\sigma^i)
\text{ and }a^i\in F^i(\mu^i,(\sigma^i)^{-i})
\end{array}
\right\}.
\]
The common-sample refinement $\Gamma^{CS+}$ is defined analogously by requiring a single $\sigma\in\Delta S$, with $a\in\operatorname{supp}(\sigma)$, to justify all players. We call $S$ \emph{$+$-self-justified} if $S\subseteq\Gamma^+(S)$, and \emph{common-sample $+$-self-justified} if $S\subseteq\Gamma^{CS+}(S)$.

When $\mathbb A$ is finite, the support refinement has a particularly sharp learning interpretation. Under a common selected sample, the Supplemental Appendix shows, as part of a more general analysis of \emph{self-synchronized behavior} that the set of action profiles played with non-vanishing frequency, i.e., $\{
a\in\mathbb A:
\limsup_{t\to\infty}\hat\sigma_t(a)>0
\}$, 
is $+$-self-justified. Under full-history learning, this is precisely the set of action profiles played with nonvanishing empirical frequency. By contrast, the ordinary limit set consists of the action profiles played infinitely often. Thus, in the finite-action case, the baseline and support-refined concepts correspond naturally to two different notions of long-run persistence: recurrence and nonvanishing frequency.

The distinction can matter even in a simple single-agent problem.

\begin{example}[Vanishing-frequency experimentation]
Let $\mathbb A=\{a,b\}$ and $\mathbb Y=\{0,1\}$. The true consequence distribution satisfies $Q(1\mid a)=q$ and $Q(1\mid b)=q+d$, where $0<q<q+d<1$. The agent's misspecified model ignores the effect of the action and assumes
\[
Q_\theta(1\mid a)=Q_\theta(1\mid b)=\theta.
\]
Payoffs are $\pi(a,y)=0$ and $\pi(b,y)=q-y$.

For a distribution $\sigma$ with $\sigma(b)=p$, the unique KL-minimizing parameter is
\[
\theta(\sigma)=q+pd.
\]
The subjective expected payoff from $a$ is therefore $0$, while that from $b$ is $-pd$.

Under the baseline operator $\Gamma$, both actions survive. Action $a$ is optimal for every $p\geq0$, while $b$ is optimal when $p=0$: under $\sigma=\delta_a$, the agent is indifferent between $a$ and $b$. Hence the Berk--Nash rationalizable set is $\{a,b\}$.

The support refinement eliminates $b$. Any distribution that can justify $b$ under $\Gamma^+$ must satisfy $\sigma(b)=p>0$. But then the subjective payoff from $b$ is $-pd<0$, while the payoff from $a$ is zero. Thus the $+$-rationalizable set is $\{a\}$.

Accordingly, the baseline concept permits $b$ to remain part of the long-run range even when it is played with vanishing frequency. The support refinement rules out this possibility: under full-history learning, $b$ may still be played infinitely often, but its empirical frequency must vanish.
\end{example}

The example also illustrates when the refinement has additional bite. Action $b$ survives the baseline concept only because it is weakly optimal under the rationalizing distribution $\delta_a$. More generally, suppose an action is strictly optimal under a rationalizing distribution that excludes it from its support. Under the usual continuity conditions, the distribution can be perturbed to place an arbitrarily small positive probability on that action while preserving strict optimality. The action then also survives the support refinement. Hence the distinction between the two concepts arises primarily when justification relies on an indifference or, more generally, is not locally robust to placing positive probability on the action itself.

\subsection{Alternative learning and sampling rules} \label{subsec:alternatives}

The particular form of the operator $\Gamma$ reflects the Bayesian-updating and sampling assumptions maintained above, but the limit-set argument is more general. Its key input is an asymptotic characterization of beliefs. Suppose that, for each player $i$, there is a correspondence
\[
\mathcal M^i:\Delta\mathbb A\rightrightarrows\Delta\Theta^i
\]
such that whenever $\hat\sigma_{t_k}^i\Rightarrow\sigma^i$ and
$\mu_{t_k}^i\Rightarrow\mu^i$, the learning process implies
$\mu^i\in\mathcal M^i(\sigma^i)$. In our baseline model, Lemma~\ref{lemm:Berk-selected} gives
\[
\mathcal M^i(\sigma^i)=\Delta\Theta^{m,i}(\sigma^i),
\]
where $\Theta^{m,i}(\sigma^i)$ is defined by KL minimization. For another updating rule, one could instead use the corresponding characterization of asymptotic beliefs. Replacing $\Delta\Theta^{m,i}(\sigma^i)$ by $\mathcal M^i(\sigma^i)$ in the definition of $\Gamma$ leaves the logic of the limit-set result unchanged, subject to the analogous regularity conditions.

The same approach can accommodate systematically biased sampling. For example, \citet{fudenberg2024selective} study selective-memory equilibrium, where the probability that an experience is recalled may depend on its realized outcome and agents update as if the recalled experiences were the only ones that occurred. Their long-run beliefs are characterized by memory-weighted likelihood maximizers rather than the usual KL minimizers. Thus one could extend our framework to their setting by replacing the baseline asymptotic-belief correspondence with the one induced by the memory distortion, and then defining the corresponding version of $\Gamma$. Our maintained sampling assumption rules out this form of memory bias by requiring the future use of an observation to be determined before its consequence is realized.\footnote{\citet{fudenberg2026limited} consider a different sampling regime in which the expected number of recalled observations remains bounded. Sampling noise therefore does not vanish: even when action frequencies converge, agents continue to face a nondegenerate distribution of recalled databases and posterior beliefs. Their stochastic memory equilibrium is consequently a fixed point in action distributions. Incorporating this case here would require keeping track of the distribution of long-run beliefs, rather than only the set of beliefs that can arise asymptotically.}

Another interesting alternative, inspired by the adaptive-learning approach of \citet{milgrom1991adaptive}, is to relax forecast consistency while leaving Bayesian learning about the subjective model unchanged. Under this rule, player $i$ continues to update beliefs about $\theta^i$ using the selected sample, exactly as in the baseline model. However, the forecast used to evaluate current actions need not coincide asymptotically with the empirical distribution of that sample. Instead, current behavior need only be approximately optimal against some belief supported on opponent profiles that have occurred in a sufficiently long recent history. For now, we impose this adaptive requirement without allowing recombinations of opponents' actions.

This separates the two roles played by $\sigma^i$ in our baseline operator. The distribution $\sigma^i$ continues to determine which models fit the selected data, while a potentially different distribution $\beta^{-i}$ governs beliefs about opponents. Accordingly, for $S\subseteq\mathbb A$, let $S^{-i}$ denote its projection onto the joint action space of player $i$'s opponents and define
\[
\Gamma^A(S)
=
\left\{
a\in\mathbb A:
\begin{array}{l}
\text{for every } i,\text{ there exist }
\sigma^i\in\Delta S,\ 
\mu^i\in\Delta\Theta^{m,i}(\sigma^i),\\[2pt]
\beta^{-i}\in\Delta S^{-i}
\text{ such that }
a^i\in F^i(\mu^i,\beta^{-i})
\end{array}
\right\}.
\]
Thus $\sigma^i$ disciplines learning about the model, whereas $\beta^{-i}$ disciplines forecasts of opponents. The baseline operator $\Gamma$ imposes the additional restriction $\beta^{-i}=(\sigma^i)^{-i}$.

The proof of the limit-set result extends directly to this adaptive rule. Along any subsequence converging to a point in the limit set $L$, the asymptotic-beliefs argument still implies that limiting posterior beliefs are supported on $\Theta^{m,i}(\sigma^i)$ for some $\sigma^i\in\Delta L$. Adaptiveness, in turn, yields a limiting forecast $\beta^{-i}$ supported on $L^{-i}$. Passing optimality to the limit therefore gives $L\subseteq\Gamma^A(L)$. Hence forecast consistency is not essential for the limit-set argument: it can be replaced by this weaker adaptive restriction, provided that the beliefs used to justify behavior remain supported on opponent profiles that persist in the history.

In the correctly specified and identified benchmark, the distribution used to learn the model no longer affects behavior, and $\Gamma^A$ reduces to the best-response operator $\Gamma^{BR}$. Thus $L\subseteq\Gamma^{BR}(L)$, so replacing forecast consistency by this adaptive rule leaves the self-justification conclusion unchanged. \citet{milgrom1991adaptive} use a still weaker notion of adaptiveness, allowing beliefs to place weight on recombinations of opponents' actions that have appeared in the history even if those opponent profiles were never played jointly. Under this rule, the actual limit set $L$ need not be self-justified. Instead, letting $L^i$ denote player $i$'s marginal limit set and $R(L)=\prod_i L^i$, the same argument gives $R(L)\subseteq\Gamma^{BR}(R(L))$. Thus the rectangular hull of the limit set is self-justified, and every action that persists in the long run remains rationalizable.\footnote{\citet{milgrom1991adaptive} also study sophisticated learning, which allows further rounds of reasoning about opponents' behavior. Extending this notion to misspecification would require assumptions about what players know about others' subjective models and beliefs, so we leave it for future work.}

\subsection{Other solution concepts} \label{subsec:othersolutions}

\citeauthor{aumann1974subjectivity}'s (\citeyear{aumann1974subjectivity}) correlated equilibrium is a distributional rather than a set-valued solution concept. It has two defining features: (i) it permits correlated play; and (ii) it imposes conditional best responses---whenever an action is used with positive probability, it must be optimal given the conditional distribution of opponents' actions in those instances. Its learning foundations come from calibrated best responses \citep{foster1997calibrated}, conditional universal consistency \citep{fudenberg1999conditional}, and regret matching \citep{hart2000simple}, all of which implement this conditional best-response logic. Our framework can also accommodate correlated empirical play, but the optimality requirement is different: players best respond to marginal rather than action-contingent conditional distributions of opponents' play. A natural extension would be to allow conditional forecasts and best responses, yielding a set-valued analogue of correlated equilibrium within our framework.

We now turn to a separate comparison based on fixed-point properties of best-response operators. Following \cite{basu1991strategy}, a set of action profiles $S$ is \emph{closed under rational behavior (curb)} if $\Gamma(S)\subseteq S$. Thus, if beliefs or conjectures are restricted to behavior in $S$, every rational response remains inside $S$. Curb sets therefore capture a strategic closure property: once behavior is confined to the set, best-response reasoning does not lead outside it. 

This is different from the role played by self-justified sets in our analysis. A self-justified set satisfies $S\subseteq\Gamma(S)$. The interpretation is that if $S$ is the long-run range of behavior, then every action profile that continues to occur must be justifiable by beliefs and forecasts formed from data generated within $S$. The condition does not require every other action profile that could also be justified from such data to appear in the realized long-run range.

The two concepts therefore impose opposite inclusion requirements and are designed to capture different objects. Curb sets identify strategically closed regions of the game, while self-justified sets are intended to describe possible long-run ranges generated by learning. A curb set may contain actions that are never recurrent and need not themselves be justified by behavior within the set, so curb closure by itself does not provide the long-run-range interpretation we seek.

The two notions coincide when $S=\Gamma(S)$. In particular, the largest self-justified set $\mathcal B$ is a fixed point of $\Gamma$ and is therefore also a curb set. This is a property of the rationalizable set rather than the defining interpretation of self-justification. Other curb sets need not have the long-run interpretation that motivates our concept. Our learning results therefore naturally deliver self-justification rather than curb closure: recurrent behavior must be supported by evidence generated along the long-run path, but the dynamics need not generate every rational response to that evidence.

A closer connection to our long-run interpretation is provided by \cite{gilboa1991social}, who introduce \emph{cyclically stable sets} to describe persistent behavior that need not converge to a point. Their notion is tied to a particular continuous best-response adjustment process, a connection developed further in subsequent work such as \cite{matsui1992best}. A cyclically stable set is internally accessible and cannot be exited through the relevant adjustment paths. Thus, as in our approach, the long-run object may be a set rather than a point. The difference is that cyclical stability inherits the transition structure of a specific best-response dynamic, whereas our notion abstracts from the precise adjustment mechanism and imposes only the requirement that every behavior that persists be supportable by beliefs and evidence generated within the same long-run range.

A different literature comes closer to our interpretation of sets as long-run ranges of behavior. In the stochastic-approximation approach to learning, the object of interest need not be a point limit: the limit set of a learning process is often characterized using the dynamics of an associated differential equation or differential inclusion, and in particular as an internally chain transitive set \citep{benaim1996dynamical,benaim1999dynamics,benaim2005stochastic}. This approach has been applied to a variety of learning procedures in games, including fictitious play \citep{benaim1999mixed}. It also underlies the analysis of misspecified learning in \cite{esponda2021asymptotic}, where the long-run evolution of action frequencies is characterized through a differential inclusion. In that sense, the idea that persistent long-run behavior may be naturally described by a set rather than by a single equilibrium is well established.

The difference is that internally chain transitive sets are defined relative to a particular dynamic. In principle, once a learning rule is fixed, one can study the associated differential inclusion and characterize its internally chain transitive sets, thereby obtaining a detailed description of its possible long-run behavior. In practice, however, such sets can be difficult to characterize except in special classes of games and dynamics, and the conclusions are tied to the particular adjustment process under study. Our approach takes a different route. Rather than fixing a specific dynamic and characterizing its limit sets directly, we ask for a dynamic-independent restriction that any long-run range generated by a broad class of learning processes must satisfy. Self-justification provides such a restriction: every behavior that remains in the long-run range must be supportable by beliefs and forecasts generated from data drawn from that same range.

Thus the two approaches are complementary. Internally chain transitive sets provide a potentially sharper characterization once a specific dynamic is imposed, whereas self-justification deliberately gives up some of that sharpness in exchange for robustness across learning and adjustment processes. Our results identify a common restriction on long-run ranges without requiring the researcher to specify the exact mechanism that selects among them.

Finally, \cite{frick2023belief} provide a related iterative characterization of long-run beliefs. Their framework is formulated directly in terms of a Markov belief process and is sufficiently abstract to cover, among other applications, active learning and some social-learning environments. They iteratively eliminate parameter states that are uniformly KL-dominated on the simplex generated by the surviving states and show that beliefs are globally attracted to the simplex over the states that remain. Their proof relies on a different idea from ours: a $q$-dominance condition turns powers of posterior ratios into supermartingales, which are then used to establish global attraction.

Our analysis is complementary. We study a broader class of learning paths and seek restrictions on long-run behavior that do not depend on a particular law of motion for beliefs. At the same time, our environment is more specialized in other respects, since it is built around repeated decision and strategic problems rather than the abstract belief processes considered by \cite{frick2023belief}. Their global result also requires continuity of the induced signal process throughout the belief simplex; in active-learning applications this can be restrictive for finite action spaces or discontinuous best responses. Our self-justification approach instead always allows finite actions and multivalued optimal-choice correspondences and extends naturally to games with player-specific beliefs, samples, and forecasts.


\subsection{Asymmetric information}

We focused attention to games of complete information. Extending the analysis to asymmetric information raises both conceptual and technical challenges. Conceptually, one must first decide what the relevant long-run object is: a limit profile of realized states, signals, and actions, or a limit strategy profile mapping signals into actions. The latter is often more informative---for instance, it can capture monotonicity in private signals, which the former cannot. Moreover, in environments with private information it may be no longer natural to assume that players observe each other's signal-contingent strategies at the end of each period, so the feedback structure may need to be weakened. Technically, forecasts of others' play must now be conditioned on private signals; with a continuum of signals, this requires signal-contingent learning rules and, effectively, consistent nonparametric estimation of opponents' conditional behavior. A full treatment of games with asymmetric information is left for future work.
\appendix
\section{Appendix}

\subsection{Preliminary results}

\begin{lemma}\label{lem:basic-properties}
For each player \(i\in I\), define \(\bar U^i(a^i,\theta^i,a^{-i})=\int_{\mathbb Y^i}\pi^i(a^i,y^i)\,q^i_{\theta^i}(y^i\mid a^i,a^{-i})\,\nu^i(dy^i)\), \(V^i(a^i,\mu^i,\sigma^{-i})=\int_{\Theta^i}\int_{\mathbb A^{-i}}\bar U^i(a^i,\theta^i,a^{-i})\,\sigma^{-i}(da^{-i})\,\mu^i(d\theta^i)\), and \(\bar V^i(\mu^i,\sigma^{-i})=\max_{a^i\in\mathbb A^i}V^i(a^i,\mu^i,\sigma^{-i})\). Then:
\begin{enumerate}[label=(\roman*)]
    \item \(V^i\) is continuous in \((a^i,\mu^i,\sigma^{-i})\), and \(\bar V^i\) is continuous in \((\mu^i,\sigma^{-i})\).
    \item For \(\varepsilon\geq0\), let \(F^i_\varepsilon(\mu^i,\sigma^{-i})=\{a^i\in\mathbb A^i:\bar V^i(\mu^i,\sigma^{-i})-V^i(a^i,\mu^i,\sigma^{-i})\leq\varepsilon\}\). The correspondence \((\varepsilon,\mu^i,\sigma^{-i})\mapsto F^i_\varepsilon(\mu^i,\sigma^{-i})\) is nonempty, compact-valued, upper hemicontinuous, and has a closed graph. In particular, since \(F^i_0=F^i\), the best-response correspondence \(F^i\) is nonempty, compact-valued, upper hemicontinuous in \((\mu^i,\sigma^{-i})\), and has a closed graph.
    \item The set of KL minimizers $\Theta^{m,i}(\sigma)\subseteq\Theta^i$ is nonempty, compact-valued, and upper hemicontinuous in $\sigma$, and hence has a closed graph. The induced correspondence $\sigma\mapsto\Delta\Theta^{m,i}(\sigma)$ is likewise nonempty, compact-valued, and upper hemicontinuous, and therefore also has a closed graph.
    \item If the game is identified, so that \(\Theta^{m,i}(\sigma)=\{\theta^i(\sigma)\}\) for every \(\sigma\in\Delta\mathbb A\), then the map \(\sigma\mapsto\theta^i(\sigma)\) is continuous.
\end{enumerate}
\end{lemma}

\begin{proof}
The arguments are standard; details are provided in Supplemental Appendix~\ref{supp:basicproperties}. By Assumptions~\ref{ass:game}\ref{ass:game:densities}--%
\ref{ass:game}\ref{ass:game:payoffs} and dominated convergence,
\(\bar U^i\) is bounded and continuous. Hence \(V^i\) is continuous, and
compactness of \(\mathbb A^i\) from
Assumption~\ref{ass:game}\ref{ass:game:spaces}, together with the Maximum
Theorem \citep[Theorem~17.31]{aliprantis2006infinite}, gives continuity of
\(\bar V^i\), proving (i). For (ii), continuity of the payoff-loss function and compactness of
\(\mathbb A^i\) imply that \(F^i_\varepsilon\) is nonempty,
compact-valued, and has a closed graph; upper hemicontinuity follows from
\citet[Theorem~17.11]{aliprantis2006infinite}. For (iii), Assumptions~\ref{ass:game}\ref{ass:game:densities} and
\ref{ass:game}\ref{ass:game:bounds} imply, again by dominated convergence,
that \(K^i(\theta^i,a)\) is bounded and continuous. The Maximum Theorem then
implies that \(\Theta^{m,i}\) is nonempty, compact-valued, and upper
hemicontinuous, and Theorem~17.11 gives a closed graph. The corresponding
properties of \(\sigma\mapsto\Delta\Theta^{m,i}(\sigma)\) follow from
\cite{aliprantis2006infinite}, Theorems~17.13 and 17.11. Finally, under identification \(\Theta^{m,i}\) is singleton-valued, so (iv)
follows from \citet[Lemma~17.6]{aliprantis2006infinite}.
\end{proof}

\begin{lemma}\label{lem:Gamma-regular}
The correspondence \(\Gamma\) is nonempty-valued on nonempty subsets of \(\mathbb A\). Moreover, if \(A\subseteq\mathbb A\) is nonempty and closed, then \(\Gamma(A)\) is closed.
\end{lemma}

\begin{proof}
Nonemptiness follows directly from Lemma~\ref{lem:basic-properties}: If
\(A\neq\varnothing\), choose \(\sigma^i\in\Delta A\) for each \(i\), then
\(\mu^i\in\Delta\Theta^{m,i}(\sigma^i)\) by part (iii), and finally
\(a^i\in F^i(\mu^i,(\sigma^i)^{-i})\) by part (ii). Hence
\(a=(a^i)_{i\in I}\in\Gamma(A)\).

Now suppose that \(A\) is closed and let \(a_n\in\Gamma(A)\) with
\(a_n\to a\). Since \(\mathbb A\) is compact by
Assumption~\ref{ass:game}\ref{ass:game:spaces}, \(A\) is compact, and hence
\(\Delta A\) is compact. For each \(n\) and \(i\), choose
\(\sigma_n^i\in\Delta A\) and
\(\mu_n^i\in\Delta\Theta^{m,i}(\sigma_n^i)\) such that
\(a_n^i\in F^i(\mu_n^i,(\sigma_n^i)^{-i})\). Passing to a subsequence, and
using that \(I\) is finite, we may assume
\(\sigma_n^i\Rightarrow\sigma^i\in\Delta A\) and
\(\mu_n^i\Rightarrow\mu^i\in\Delta\Theta^i\) for every \(i\). By the closed-graph property in Lemma~\ref{lem:basic-properties}(iii),
\(\mu^i\in\Delta\Theta^{m,i}(\sigma^i)\). Since marginalization is continuous
under weak convergence,
\((\sigma_n^i)^{-i}\Rightarrow(\sigma^i)^{-i}\). The closed-graph property in
Lemma~\ref{lem:basic-properties}(ii) then implies
\(a^i\in F^i(\mu^i,(\sigma^i)^{-i})\) for every \(i\). Hence
\(a\in\Gamma(A)\), so \(\Gamma(A)\) is closed.
\end{proof}

\begin{lemma}[Deterministic tolerance]\label{lem:deterministic-tolerance}
Let \((\ell_t)_{t\geq1}\) be a sequence of nonnegative random variables on a probability space \((\Omega,\mathcal F,P)\). Suppose that \(\ell_t\to0\), \(P\)-almost surely, and that \(\sup_t \ell_t\leq M\) for some finite constant \(M\). Then, for every \(\alpha>0\), there exists a deterministic sequence \(\varepsilon_t\downarrow0\) such that
\[
P\left(\ell_t\leq\varepsilon_t\text{ for every }t\right)>1-\alpha.
\]
\end{lemma}
\begin{proof}
By Egorov's theorem, for every \(\alpha>0\) there exists an event
\(G\) with \(P(G)>1-\alpha\) such that \(\ell_t\to0\) uniformly on \(G\).
Let \(C:=\max\{M,1\}\). For each \(k\geq1\), uniform convergence yields
\(N_k\) such that $\ell_t\leq C/k$ on $G$ for every \(t\geq N_k\). Choose the sequence \((N_k)\) strictly increasing and define
\[
\varepsilon_t
:=
\begin{cases}
C, & t<N_1,\\
C/k, & N_k\leq t<N_{k+1}.
\end{cases}
\]
Then \(\varepsilon_t\downarrow0\). If \(t<N_1\), then
\(\ell_t\leq M\leq C=\varepsilon_t\). If
\(N_k\leq t<N_{k+1}\), then \(t\geq N_k\), so
\(\ell_t\leq C/k=\varepsilon_t\) on \(G\). Hence \(\ell_t\leq\varepsilon_t\) for every \(t\) on \(G\). Since
\(P(G)>1-\alpha\), the result follows.
\end{proof}

\subsection{Proof of Theorem \ref{theo:characterization}}

\begin{proof}
Let \(\mathcal J:=\{S\subseteq\mathbb A:S\text{ is nonempty and compact, and }S\subseteq\Gamma(S)\}\) be the collection of self-justified sets. By definition, \(\mathcal B=\bigcup_{S\in\mathcal J}S\). We first establish the iterative characterization and then show that this set is the largest fixed point of \(\Gamma\).

Define \(B^0=\mathbb A\) and \(B^{k+1}=\Gamma(B^k)\) for \(k\geq0\), and let \(\mathcal B^\infty:=\bigcap_{k=0}^\infty B^k\). The correspondence $\Gamma$ is monotone: if $S\subseteq S'$, then $\Delta S\subseteq\Delta S'$, and therefore $\Gamma(S)\subseteq\Gamma(S')$. Moreover, $B^1=\Gamma(B^0)\subseteq\mathbb A=B^0$. Hence, by induction, $B^{k+1}\subseteq B^k$ for every $k$, so $(B^k)_k$ is a decreasing sequence of sets. By Lemma \ref{lem:Gamma-regular}, \(\Gamma\) is nonempty-valued on nonempty sets and maps nonempty closed sets to closed sets. Since \(B^0=\mathbb A\) is nonempty and compact by Assumption \ref{ass:game}\ref{ass:game:spaces}, it follows by induction that each \(B^k\) is nonempty and compact. Therefore \(\mathcal B^\infty\), as the intersection of a decreasing sequence of nonempty compact sets, is nonempty and compact.

We next show that \(\mathcal B^\infty\) is self-justified. Let \(a\in\mathcal B^\infty\). Then, for every \(k\geq1\), \(a\in B^k=\Gamma(B^{k-1})\). Hence, for every player \(i\), there exist \(\sigma_k^i\in\Delta B^{k-1}\) and \(\mu_k^i\in\Delta\Theta^{m,i}(\sigma_k^i)\) such that \(a^i\in F^i\bigl(\mu_k^i,(\sigma_k^i)^{-i}\bigr)\). By compactness of \(\Delta\mathbb A\) and \(\Delta\Theta^i\), and since the set of players is finite, we may pass to a common subsequence such that, for every player \(i\), \(\sigma_k^i\Rightarrow\sigma^i\) and \(\mu_k^i\Rightarrow\mu^i\). We claim that \(\sigma^i\in\Delta\mathcal B^\infty\). Fix \(m\geq0\). For all \(k\geq m+1\), we have \(B^{k-1}\subseteq B^m\), and hence \(\sigma_k^i(B^m)=1\). Since \(B^m\) is closed, weak convergence implies \(\sigma^i(B^m)=1\). Since this holds for every \(m\), and \((B^m)_m\) is decreasing, continuity from above gives \(\sigma^i(\mathcal B^\infty)=\sigma^i\left(\bigcap_{m=0}^\infty B^m\right)=1\). Thus \(\sigma^i\in\Delta\mathcal B^\infty\). By Lemma~\ref{lem:basic-properties}(iii), the correspondence \(\sigma\mapsto\Delta\Theta^{m,i}(\sigma)\) has a closed graph. Hence \(\mu^i\in\Delta\Theta^{m,i}(\sigma^i)\). Also, \((\sigma_k^i)^{-i}\Rightarrow(\sigma^i)^{-i}\). Since \(F^i\) has a closed graph by Lemma~\ref{lem:basic-properties}(ii), taking limits in \(a^i\in F^i\bigl(\mu_k^i,(\sigma_k^i)^{-i}\bigr)\) gives \(a^i\in F^i\bigl(\mu^i,(\sigma^i)^{-i}\bigr)\). This holds for every player \(i\). Therefore \(a\in\Gamma(\mathcal B^\infty)\), and hence \(\mathcal B^\infty\subseteq\Gamma(\mathcal B^\infty)\).

The reverse inclusion follows from monotonicity. Since \(\mathcal B^\infty\subseteq B^k\) for every \(k\), monotonicity implies \(\Gamma(\mathcal B^\infty)\subseteq\Gamma(B^k)=B^{k+1}\) for every \(k\). Hence \(\Gamma(\mathcal B^\infty)\subseteq\bigcap_{k=0}^\infty B^k=\mathcal B^\infty\). Thus \(\mathcal B^\infty=\Gamma(\mathcal B^\infty)\).

We now show that \(\mathcal B^\infty\) contains every self-justified set. Let \(S\in\mathcal J\), so \(S\) is nonempty and compact, and \(S\subseteq\Gamma(S)\). Since \(S\subseteq\mathbb A=B^0\), suppose inductively that \(S\subseteq B^k\). By monotonicity, \(\Gamma(S)\subseteq\Gamma(B^k)=B^{k+1}\). Since \(S\subseteq\Gamma(S)\), it follows that \(S\subseteq B^{k+1}\). Hence \(S\subseteq B^k\) for every \(k\), and therefore \(S\subseteq\mathcal B^\infty\). Taking the union over all \(S\in\mathcal J\), we obtain \(\mathcal B\subseteq\mathcal B^\infty\). Since \(\mathcal B^\infty\) is nonempty, compact, and satisfies \(\mathcal B^\infty\subseteq\Gamma(\mathcal B^\infty)\), it belongs to \(\mathcal J\). Hence \(\mathcal B^\infty\subseteq\mathcal B\). Therefore \(\mathcal B=\mathcal B^\infty=\bigcap_{k=0}^\infty B^k\). In particular, \(\mathcal B\) is nonempty, compact, and a fixed point of \(\Gamma\). Moreover, any fixed point \(A=\Gamma(A)\) satisfies \(A\subseteq\Gamma(A)\). By the closure argument following the definition of self-justification, \(\overline A\subseteq\Gamma(\overline A)\). Since \(\mathbb A\) is compact, \(\overline A\) is compact, so \(\overline A\in\mathcal J\). Hence \(A\subseteq\mathcal B\). Since \(\mathcal B\) itself is a fixed point, it is the largest fixed point of \(\Gamma\).

It remains to prove that \(\mathcal B\) is a product set. For any \(S\subseteq\mathbb A\), define
\[
\Gamma^i(S)
:=
\left\{
a^i\in\mathbb A^i:
\exists\,\sigma^i\in\Delta S,\ 
\exists\,\mu^i\in\Delta\Theta^{m,i}(\sigma^i)
\text{ such that }
a^i\in F^i\bigl(\mu^i,(\sigma^i)^{-i}\bigr)
\right\}.
\]
Then \(\Gamma(S)=\prod_{i\in I}\Gamma^i(S)\). Indeed, membership in \(\Gamma(S)\) requires that each player \(i\)'s component be justified by some \(\sigma^i\in\Delta S\) and some \(\mu^i\in\Delta\Theta^{m,i}(\sigma^i)\), and these witnesses may differ across players. Since \(\mathcal B\) is a fixed point, \(\mathcal B=\Gamma(\mathcal B)=\prod_{i\in I}\Gamma^i(\mathcal B)\). Let \(\mathcal B^i:=\Gamma^i(\mathcal B)\). Then \(\mathcal B=\prod_{i\in I}\mathcal B^i\). Since \(\mathcal B\) is nonempty and compact, each factor \(\mathcal B^i=\operatorname{proj}_{\mathbb A^i}\mathcal B\) is nonempty and compact. Hence \(\mathcal B\) is a product set with nonempty compact components.
\end{proof}

\subsection{Proof of Theorem \ref{theo:sustain-self-justified}}

\begin{proof}
We proceed in five steps.

\medskip
\noindent
\emph{Step 1: Construct a target path with sufficiently large sampling reservoirs.} The key ingredient is a sequence of finite approximating sets whose mesh vanishes while their cardinalities remain small relative to the block index. Specifically, we first show that there exist finite sets $E_n\subseteq S$ such that $\operatorname{mesh}(E_n)\to0$
and $|E_n|\leq\sqrt n $ for all sufficiently large $n$, where, for a finite set $E\subseteq S$, $\operatorname{mesh}(E)
:=\sup_{a\in S}\min_{e\in E}d(a,e)$. To establish this fact, let $N(S,\delta)$ denote the minimum cardinality of a finite $\delta$-net of $S$. Since $S$ is compact, $N(S,1/m)<\infty$ for every integer $m\geq1$. Thus, for each fixed accuracy level $1/m$, only finitely many points are needed to approximate $S$ within that accuracy. At block $n$, we allow ourselves at most $\sqrt n$ net points. Since $\sqrt n\to\infty$, this cardinality bound eventually exceeds $N(S,1/m)$ for every fixed $m$. We can therefore let the desired accuracy improve sufficiently slowly with $n$: choose an integer sequence $m_n\to\infty$ such that $N(S,1/m_n)\leq\sqrt n$ for all sufficiently large $n$. In other words, we move to a finer net only once the growing cardinality bound $\sqrt n$ is large enough to accommodate it. For each such $n$, let $E_n$ be a $1/m_n$-net of minimum cardinality. For the finitely many remaining values of $n$, choose any nonempty finite subset $E_n\subseteq S$. Then $
\operatorname{mesh}(E_n) \leq \frac{1}{m_n} \to0$. Writing $K_n:=|E_n|$, we also have $K_n=N(S,1/m_n)\leq\sqrt n$ for all sufficiently large $n$.

We now use these sets to construct the target path. For each $n\geq1$, let $B_n$ be a finite block that lists every element of $E_n$ exactly $n$ times, in any order. Thus $|B_n|=nK_n$. The order within each block is immaterial; only the number of copies of each point will matter below. The two properties of $(E_n)_n$ established above immediately give the two features of the blocks that we will need. First, $\frac{K_n}{n}\leq\frac{1}{\sqrt n}\to0$. This will allow selected observations from block $B_n$ to approximate arbitrary distributions on $E_n$. Second, the blocks do not become too long relative to their index. Let $T_n:=\sum_{m=1}^n |B_m|$ denote the last date of block $B_n$. Since $K_m\leq\sqrt m$ for all sufficiently large $m$, there exists a constant $C_0<\infty$ such that $|B_m| \leq C_0m^{3/2}$ for every $m\geq1$. Hence $T_n
\leq C_0\sum_{m=1}^n m^{3/2} \leq C_0n^{5/2}$. It follows that if $t$ belongs to block $B_{n+1}$, then $t\leq T_{n+1} \leq C_0(n+1)^{5/2}$,
and therefore $\log t \leq \log C_0+\frac52\log(n+1)$. Consequently,
\[
\frac{n}{\log t}
\geq
\frac{n}{
\log C_0+\frac52\log(n+1)
}
\to\infty.
\]
Thus, when a date in block $B_{n+1}$ uses a sample of size $n$ drawn from block $B_n$, the required sample-growth condition $\frac{|H_t^i|}{\log t}\to\infty$
will hold automatically.

Finally, let $z^\infty=(z_t)_{t\geq1}$ be the deterministic path obtained by concatenating $B_1,B_2,\ldots$. Then $z_t\in S$ for every $t$, and $L(z^\infty)=S$. Indeed, since $S$ is compact and therefore closed, every limit point of $z^\infty$ belongs to $S$. Conversely, fix $a\in S$. For each $n$, choose $e_n\in E_n$ such that $d(e_n,a) \leq \operatorname{mesh}(E_n)$. Since $\operatorname{mesh}(E_n)\to0$, we have $e_n\to a$. Each $e_n$ appears in block $B_n$, so there exists a date $t_n$ in block $B_n$ such that $z_{t_n}=e_n$. Since $t_n\to\infty$, it follows that $z_{t_n}\to a$. Hence $a\in L(z^\infty)$. Since $a\in S$ was arbitrary, $L(z^\infty)=S$.

\medskip
\noindent
\emph{Step 2: Choose supporting distributions.} Fix a date $t$ and a player $i$. Since $z_t\in S$ and $S\subseteq\Gamma(S)$, the definition of $\Gamma$ implies that there exists a distribution $\rho_t^i\in\Delta S$ such that $z_t^i
\in F^i(\delta_{\theta^i(\rho_t^i)},(\rho_t^i)^{-i})$. Thus $\rho_t^i$ is a distribution over $S$ that justifies the target action $z_t^i$ for player $i$. Fix one such distribution $\rho_t^i$ for every pair $(t,i)$.

\medskip
\noindent
\emph{Step 3: Construct selected samples that approximate the supporting distributions.} We now construct the entire sampling array $W$ deterministically. Since the target path $(z_t)_t$ and the supporting distributions $(\rho_t^i)_{t,i}$ were fixed in Steps 1 and 2, all selected dates used below can be chosen ex ante. In particular, the resulting sampling scheme satisfies the timing requirement that the future use of observation $\tau$ be determined before its consequence is realized.

Fix $n\geq1$, a date $t$ in block $B_{n+1}$, and a player $i$. We use block $B_n$ as the reservoir from which player $i$'s date-$t$ sample is selected. Let
$\delta_n := \operatorname{mesh}(E_n)$, so that $\delta_n\to0$. Choose a measurable nearest-net projection $p_n:S\to E_n$ satisfying $d(a,p_n(a)) \leq \delta_n$ for every $a\in S$. Let $\bar\rho_t^i\in\Delta E_n$ be the distribution obtained by moving the mass of $\rho_t^i$ to the corresponding points of $E_n$:
\[
\bar\rho_t^i(e)
=
\rho_t^i
\left(
\{a\in S:p_n(a)=e\}
\right),
\qquad
e\in E_n.
\]
For each $e\in E_n$, write $w_e:=\bar\rho_t^i(e)$. Since $\sum_{e\in E_n}w_e=1$,
there exist nonnegative integers $(N_e)_{e\in E_n}$ such that $\sum_{e\in E_n}N_e=n$ and $\left| \frac{N_e}{n}-w_e \right| \leq \frac1n$ for every $e\in E_n$. For example, one may take $N_e=\lfloor nw_e\rfloor$ and allocate the remaining observations according to the largest fractional remainders. By construction, block $B_n$ contains exactly $n$ dates at which the target action is equal to each $e\in E_n$. Hence, for each $e\in E_n$, we can choose exactly $N_e$ of these dates and let $H_t^i$ be their union. Since $\sum_e N_e=n$, this gives $|H_t^i|=n$. Along the target path, the selected empirical distribution is therefore $\hat\sigma_t^i=\sum_{e\in E_n}\frac{N_e}{n}\delta_e$.

We next show that this selected empirical distribution is close to the supporting distribution $\rho_t^i$. Let $D:=\operatorname{diam}(S)$. Since both $\hat\sigma_t^i$ and $\bar\rho_t^i$ are supported on $E_n$, $\|\hat\sigma_t^i-\bar\rho_t^i\|_{\mathrm{TV}} \leq \frac{|E_n|}{2n}$. Hence their $1$-Wasserstein distance satisfies $d_{\Delta} \left(\hat\sigma_t^i,\bar\rho_t^i
\right) \leq \frac{D|E_n|}{2n}$, where $d_{\Delta}$ denotes the $1$-Wasserstein metric on $\Delta S$. Moreover, because the projection $p_n$ moves every point of $S$ by at most $\delta_n$, $d_{\Delta}\left(\bar\rho_t^i,\rho_t^i\right) \leq
\delta_n$. Therefore, by the triangle inequality, $d_{\Delta}\left(\hat\sigma_t^i,\rho_t^i\right) \leq \delta_n+\frac{D|E_n|}{2n}$. By Step 1, $\delta_n\to0$ and, for all sufficiently large $n$, $|E_n|\leq\sqrt n$. Hence $\frac{|E_n|}{n} \leq \frac1{\sqrt n} \to0$. It follows that, uniformly over dates $t$ in block $B_{n+1}$ and players $i$, $d_{\Delta} \left(\hat\sigma_t^i,\rho_t^i\right)\to0$.

It remains to verify the sample-growth condition. If $t$ belongs to block $B_{n+1}$, then the construction above gives $|H_t^i|=n$. Step 1 also showed that $t\leq T_{n+1} \leq C_0(n+1)^{5/2}$. Therefore, $\log t \leq \log C_0+\frac52\log(n+1)$, and hence
\[
\frac{|H_t^i|}{\log t}
=
\frac{n}{\log t}
\geq
\frac{n}{
\log C_0+\frac52\log(n+1)
}.
\]
The right-hand side converges to infinity as $n\to\infty$. Thus $\frac{|H_t^i|}{\log t}\to\infty$ for every player $i$ as $t\to\infty$. Hence the sampling scheme satisfies the sample-growth condition required for admissibility.

For dates in the first block, choose $H_t^i$ arbitrarily. Having chosen all sets $H_t^i$, define the sampling array by
\[
W_{\tau,t}^i
=
\mathbf 1\{\tau\in H_t^i\},
\qquad
\tau<t.
\]
Since all the sets $H_t^i$ were chosen deterministically from the target path and the supporting distributions fixed in advance, the entire array $W$ is deterministic ex ante. Hence it satisfies the non-anticipation requirement not only along the target path but after every possible realized history.

Finally, set $\tilde\sigma_t^{-i}=(\hat\sigma_t^i)^{-i}$. Thus forecast consistency holds automatically.

\medskip
\noindent
\emph{Step 4: Show that the target actions are asymptotically optimal along the target path.} Let $\mathbf P^z$ denote the probability law generated when the action rule is degenerate at the target path $z^\infty$ and the sampling array is the deterministic array constructed in Step 3. As shown in Step 3, this array satisfies $\frac{|H_t^i|}{\log t}\to\infty$ for every player $i$, and, being deterministic ex ante, it also satisfies the sampling-timing requirement. Hence the target-path process is an admissible learning process.

For each player $i$, let $\mu_t^{i,z}$, $\hat\sigma_t^{i,z}$, and $\tilde\sigma_t^{-i,z}$ denote player $i$'s posterior, selected empirical distribution, and forecast under this process. Since the target action path and the sampling array are deterministic, $\hat\sigma_t^{i,z}$ and $\tilde\sigma_t^{-i,z}$ are deterministic. The posterior $\mu_t^{i,z}$ depends only on the selected consequences $(y_\tau)_{\tau\in H_t^i}$, all of which are realized strictly before date $t$.

For each player $i$, define the payoff loss from playing the target action $z_t^i$ at date $t$ by
\[
\ell_t^i
:=
\sup_{x^i\in\mathbb A^i}
\int_{\Theta^i}
U^i(x^i,\tilde\sigma_t^{-i,z},\theta^i)\,
\mu_t^{i,z}(d\theta^i)
-
\int_{\Theta^i}
U^i(z_t^i,\tilde\sigma_t^{-i,z},\theta^i)\,
\mu_t^{i,z}(d\theta^i),
\]
and let $\ell_t:=\max_{i\in I}\ell_t^i$. Thus $\ell_t^i$, and therefore $\ell_t$, is measurable with respect to the consequence history through date $t-1$. This timing observation will be useful in Step 5.

We claim that $\ell_t\to0$ $\mathbf P^z$-almost surely: Let $\Omega^\ast$ be the probability-one event on which Lemma~\ref{lemm:Berk-selected} holds simultaneously for every player under the target-path process. Suppose, toward a contradiction, that $\ell_t\not\to0$ on a set of positive $\mathbf P^z$-probability. Since the player set is finite, there exist a player $i$, a number $\eta>0$, and an event $A\subseteq\Omega^\ast$ with $\mathbf P^z(A)>0$ such that
$\limsup_{t\to\infty}\ell_t^i(\omega)>\eta$ for every $\omega\in A$. Fix $\omega\in A$. Then there exists a subsequence $(t_m)_m$ such that $\ell_{t_m}^i(\omega)\geq\eta$ for every $m$. By compactness of $S$, the sequence $(z_{t_m})_m$ has a convergent subsequence. Since $\Delta S$ is compact under weak convergence, the sequence $(\rho_{t_m}^i)_m$ also has a convergent subsequence. Finally, since $\Theta^i$ is compact, $\Delta\Theta^i$ is compact under weak convergence, so the sequence $(\mu_{t_m}^{i,z}(\omega))_m$ has a convergent subsequence. Passing to a common further subsequence, we may therefore assume $z_{t_m}\to a\in S$, $\rho_{t_m}^i\Rightarrow\rho^i\in\Delta S$, and $\mu_{t_m}^{i,z}(\omega)\Rightarrow\mu^i\in\Delta\Theta^i$. By Step 3, $d_\Delta
\left(\hat\sigma_{t_m}^{i,z},\rho_{t_m}^i\right)\to0$. Together with $\rho_{t_m}^i\Rightarrow\rho^i$, this implies $\hat\sigma_{t_m}^{i,z}
\Rightarrow \rho^i$. Since $\tilde\sigma_t^{-i,z}=(\hat\sigma_t^{i,z})^{-i}$,
and taking marginals is continuous under weak convergence, we also have $\tilde\sigma_{t_m}^{-i,z} \Rightarrow (\rho^i)^{-i}$. Because $\omega\in\Omega^\ast$, Lemma~\ref{lemm:Berk-selected} applies to the subsequence $(t_m)_m$. Hence $\mu^i\in\Delta\Theta^{m,i}(\rho^i)$. By identification, $\Theta^{m,i}(\rho^i)=\{\theta^i(\rho^i)\}$,
so necessarily $\mu^i=\delta_{\theta^i(\rho^i)}$. On the other hand, by the choice of the supporting distributions in Step 2, $z_{t_m}^i \in F^i\left(\delta_{\theta^i(\rho_{t_m}^i)},(\rho_{t_m}^i)^{-i}\right)$ for every $m$. Identification and Lemma~\ref{lem:basic-properties}(iv) imply that the map
$\rho\mapsto\theta^i(\rho)$ is continuous. Hence $\theta^i(\rho_{t_m}^i)\to\theta^i(\rho^i)$, and therefore $\delta_{\theta^i(\rho_{t_m}^i)}\Rightarrow\delta_{\theta^i(\rho^i)}$. Together with $z_{t_m}^i\to a^i$ and $(\rho_{t_m}^i)^{-i}\Rightarrow(\rho^i)^{-i}$, the closed-graph property of $F^i$ implies $a^i\in F^i \left(\delta_{\theta^i(\rho^i)},(\rho^i)^{-i}\right)$. Now define the payoff-loss function
\[
\Lambda^i(a^i,\mu,\beta^{-i})
:=
\sup_{x^i\in\mathbb A^i}
\int_{\Theta^i}
U^i(x^i,\beta^{-i},\theta^i)\,
\mu(d\theta^i)
-
\int_{\Theta^i}
U^i(a^i,\beta^{-i},\theta^i)\,
\mu(d\theta^i).
\]
By continuity of expected payoffs and the value function, $\Lambda^i$ is continuous. Since $z_{t_m}^i\to a^i$, $\mu_{t_m}^{i,z}(\omega)\Rightarrow
\delta_{\theta^i(\rho^i)}$, and $\tilde\sigma_{t_m}^{-i,z}\Rightarrow(\rho^i)^{-i}$, we obtain
\[
\ell_{t_m}^i(\omega)
=
\Lambda^i
\left(
z_{t_m}^i,
\mu_{t_m}^{i,z}(\omega),
\tilde\sigma_{t_m}^{-i,z}
\right)
\to
\Lambda^i
\left(
a^i,
\delta_{\theta^i(\rho^i)},
(\rho^i)^{-i}
\right).
\]
The limit on the right-hand side is zero because $a^i \in F^i \left(\delta_{\theta^i(\rho^i)},(\rho^i)^{-i}\right)$. Therefore $\ell_{t_m}^i(\omega)\to0$, contradicting $\ell_{t_m}^i(\omega)\geq\eta$ for every $m$. Hence $\ell_t\to0$ $\mathbf P^z$-almost surely.

\medskip
\noindent
\emph{Step 5: Choose tolerances and define the actual process.} Since $\ell_t\to0$ $\mathbf P^z$-almost surely, Lemma~\ref{lem:deterministic-tolerance} implies that there exists a deterministic sequence $\varepsilon_t\downarrow0$ such that $\mathbf P^z\left(\ell_t\leq\varepsilon_t
\text{ for every }t\right)>1-\alpha$. Let $G:=\left\{\ell_t\leq\varepsilon_t
\text{ for every }t\right\}$, so that $\mathbf P^z(G)>1-\alpha$.

We now define the action rule of the actual process. At date $t$, given the posterior $\mu_t^i$ and forecast $\tilde\sigma_t^{-i}$ induced by the deterministic sampling array constructed in Step 3, player $i$ chooses the target action $z_t^i$ whenever $z_t^i\in F_{\varepsilon_t}^i\left(\mu_t^i,
\tilde\sigma_t^{-i}\right)$. Otherwise, she chooses an arbitrary exact best response in $F^i\left(\mu_t^i,\tilde\sigma_t^{-i}\right)$, which is nonempty by Lemma~\ref{lem:basic-properties}(ii), using any fixed measurable selection. Hence, at every history, $a_t^i \in F_{\varepsilon_t}^i\left(\mu_t^i,\tilde\sigma_t^{-i}\right)$, so the resulting action rule is asymptotically myopic. Together with the full-support priors, the deterministic sampling array constructed in Step 3, Bayesian updating, and the empirical forecasts $\tilde\sigma_t^{-i}=(\hat\sigma_t^i)^{-i}$, this action rule defines an admissible learning process with consistent forecasts. Let $\mathbf P$ denote the probability law generated by this actual process.

For each $T\geq1$, let
\[
F_T
:=
\left\{
a_t=z_t
\text{ for every }t\leq T
\right\}
\]
and
\[
G_T
:=
\left\{
\ell_t\leq\varepsilon_t
\text{ for every }t\leq T
\right\}.
\]
By the timing observation in Step 4, for each $t$ the target-path loss $\ell_t$ is a measurable function of the consequence history through date $t-1$. Thus the event $\{\ell_t\leq\varepsilon_t\}$ is determined by $y^{t-1}$ and is well defined under both probability laws.

To prove $\mathbf P(F_T)=\mathbf P^z(G_T)$ by induction, equality of the probabilities at date $T-1$ is not enough. The date-$T$ action depends on the realized consequence history $y^{T-1}$ through the event $\{ \ell_T\leq\varepsilon_T\}$. We therefore need the actual and target-path processes to induce the same distribution of $y^{T-1}$ on the corresponding events. We establish this distributional equality through the following equivalent test-function formulation: we next show that, for every $T\geq1$ and every bounded measurable $\varphi:\mathbb Y^T\to\mathbb R$,
\begin{equation}\label{eq:finite-horizon-coupling}
\mathbf E
\left[
\mathbf 1_{F_T}\varphi(y^T)
\right]
=
\mathbf E^z
\left[
\mathbf 1_{G_T}\varphi(y^T)
\right].
\end{equation}

We prove \eqref{eq:finite-horizon-coupling} by induction on $T$. For $T=1$, there is no preceding consequence history. Hence the posterior and forecast at date $1$ are the same in the actual process and in the target-path process. By construction of the action rule, $a_1=z_1\iff\ell_1\leq\varepsilon_1$. Moreover, whenever $a_1=z_1$, the consequence $y_1$ is drawn according to $Q(\cdot\mid z_1)$, which is also its distribution under the target-path process. Therefore, for every bounded measurable $\varphi:\mathbb Y\to\mathbb R$, $\mathbf E\left[\mathbf 1_{F_1}\varphi(y_1)\right]=\mathbf E^z\left[\mathbf 1_{G_1}\varphi(y_1)\right]$. Now suppose that \eqref{eq:finite-horizon-coupling} holds for some $T-1\geq1$. Thus, for every bounded measurable function $\psi:\mathbb Y^{T-1}\to\mathbb R$, $\mathbf E\left[\mathbf 1_{F_{T-}}\psi(y^{T-1})\right]=\mathbf E^z\left[\mathbf 1_{G_{T-1}}\psi(y^{T-1})
\right]$. On the event $F_{T-1}$, the actual action history through date $T-1$ coincides with the target history: $a_s=z_s$ for every $s<T$. Because the sampling array $W$ is deterministic ex ante, the actual process and the target-path process use the same selected dates at date $T$. Therefore, for any given consequence history $y^{T-1}$, the selected observations, posterior beliefs, and forecasts computed by the actual process on $F_{T-1}$ coincide with those computed by the target-path process from the same consequence history. It follows that, on $F_{T-1}$,
\[
a_T=z_T
\quad\Longleftrightarrow\quad
z_T^i
\in
F_{\varepsilon_T}^i
\left(
\mu_T^{i,z},
\tilde\sigma_T^{-i,z}
\right)
\text{ for every }i
\quad\Longleftrightarrow\quad
\ell_T\leq\varepsilon_T.
\]
Hence $\mathbf 1_{F_T}=\mathbf 1_{F_{T-1}}\mathbf 1\{\ell_T\leq\varepsilon_T\}$. 
Let $\varphi:\mathbb Y^T\to\mathbb R$ be bounded and measurable. Conditional on the history through date $T-1$, whenever $a_T=z_T$ the date-$T$ consequence is drawn according to $Q(\cdot\mid z_T)$. Hence
\begin{align*}
\mathbf E
\left[
\mathbf 1_{F_T}\varphi(y^T)
\right]
&=
\mathbf E
\left[
\mathbf 1_{F_{T-1}}
\mathbf 1\{\ell_T\leq\varepsilon_T\}
\int_{\mathbb Y}
\varphi(y^{T-1},y_T)
Q(dy_T\mid z_T)
\right].
\end{align*}

Define the bounded measurable function
\[
\psi(y^{T-1})
:=
\mathbf 1\{\ell_T\leq\varepsilon_T\}
\int_{\mathbb Y}
\varphi(y^{T-1},y_T)
Q(dy_T\mid z_T).
\]
Applying the induction hypothesis to this $\psi$ gives
\begin{align*}
\mathbf E
\left[
\mathbf 1_{F_T}\varphi(y^T)
\right]
&=
\mathbf E^z
\left[
\mathbf 1_{G_{T-1}}
\mathbf 1\{\ell_T\leq\varepsilon_T\}
\int_{\mathbb Y}
\varphi(y^{T-1},y_T)
Q(dy_T\mid z_T)
\right].
\end{align*}
Since
\[
G_T
=
G_{T-1}
\cap
\{\ell_T\leq\varepsilon_T\},
\]
and since under $\mathbf P^z$ the conditional distribution of $y_T$ given $y^{T-1}$ is $Q(\cdot\mid z_T)$, the right-hand side equals $\mathbf E^z\left[\mathbf 1_{G_T}\varphi(y^T)\right]$. This proves \eqref{eq:finite-horizon-coupling} for date $T$ and completes the induction.

Taking $\varphi\equiv1$ in \eqref{eq:finite-horizon-coupling}, we obtain $\mathbf P(F_T)=\mathbf P^z(G_T)$ for every $T\geq1$. The events $(F_T)_T$ are decreasing, with
\[
\bigcap_{T=1}^{\infty}F_T
=
\left\{
a_t=z_t
\text{ for every }t
\right\},
\]
and $(G_T)_T$ decreases to $G$. Hence, by continuity of probability from above,
\begin{align*}
\mathbf P
\left(
\bigcap_{T=1}^{\infty}F_T
\right)
&=
\lim_{T\to\infty}\mathbf P(F_T)
\\
&=
\lim_{T\to\infty}\mathbf P^z(G_T)
\\
&=
\mathbf P^z(G)
\\
&>
1-\alpha.
\end{align*}

On the event $\bigcap_{T=1}^{\infty}F_T$, the realized action path coincides with the target path $z^\infty$. Therefore, $L(a^\infty)=L(z^\infty)=S$. Hence $\mathbf P\bigl(L(a^\infty)=S\bigr)>1-\alpha$.
\end{proof}

\addcontentsline{toc}{section}{References}
\bibliography{references.bib}

\clearpage 
\clearpage
\pagenumbering{arabic}
\setcounter{page}{1}

\section*{Supplemental Appendix for ``Berk-Nash Rationalizability"}

\setcounter{section}{0}
\setcounter{subsection}{0}
\setcounter{equation}{0}
\renewcommand{\thesection}{S\arabic{section}}
\renewcommand{\thesubsection}{S\arabic{section}.\arabic{subsection}}
\renewcommand{\theequation}{S\arabic{equation}}

\makeatletter
\renewcommand{\section}[1]{%
  \refstepcounter{section}%
  \addcontentsline{dummy}{section}{}
  \noindent\par\vspace{1.5ex}%
  {\normalfont\Large\bfseries S\arabic{section}\quad#1}%
  \par\nobreak\vspace{1.0ex}\@afterheading%
}

\renewcommand{\subsection}[1]{%
  \refstepcounter{subsection}%
  \noindent\par\vspace{1.0ex}%
  {\normalfont\large\bfseries S\arabic{section}.\arabic{subsection}\quad#1}%
  \par\nobreak\vspace{0.8ex}\@afterheading%
}
\makeatother

\setcounter{proposition}{0}
\setcounter{definition}{0}
\renewcommand{\theproposition}{SM\arabic{proposition}}
\renewcommand{\thedefinition}{SM\arabic{definition}}

\section{Proof of Lemma~\ref{lem:basic-properties}} \label{supp:basicproperties}

\begin{proof}
Fix \(i\in I\). By Assumption~\ref{ass:game}\ref{ass:game:densities}, the integrand defining \(\bar U^i\) is \(\nu^i\)-a.e. continuous in \((a^i,\theta^i,a^{-i})\). By Assumption~\ref{ass:game}\ref{ass:game:bounds}, \(q^i_{\theta^i}(y^i\mid a)\leq M^i(y^i)q^i(y^i\mid a)\leq M^i(y^i)r^i(y^i)\), and by Assumption~\ref{ass:game}\ref{ass:game:payoffs}, \(|\pi^i(a^i,y^i)|\leq h_\pi^i(y^i)\), with \(\int h_\pi^i M^i r^i\,d\nu^i<\infty\). Hence the integrand is dominated by an integrable function. By the Dominated Convergence Theorem, \(\bar U^i\) is jointly continuous. The same domination implies that \(\bar U^i\) is bounded.

We first prove (i). Since \(\bar U^i\) is bounded and continuous, and since
\(\mathbb A^i\), \(\Theta^i\), and \(\mathbb A^{-i}\) are compact by
Assumption~\ref{ass:game}\ref{ass:game:spaces}, the map
\((a^i,\mu^i,\sigma^{-i})\mapsto V^i(a^i,\mu^i,\sigma^{-i})\) is continuous
under the weak topologies on \(\Delta(\Theta^i)\) and
\(\Delta(\mathbb A^{-i})\). Since the feasible set \(\mathbb A^i\) is compact
and does not depend on \((\mu^i,\sigma^{-i})\), the Maximum Theorem
\citep[Theorem~17.31]{aliprantis2006infinite} implies that
\(\bar V^i\) is continuous in \((\mu^i,\sigma^{-i})\).

We next prove (ii). By part (i), \(V^i\) and \(\bar V^i\) are continuous. Hence the payoff-loss function
\(D^i(a^i,\mu^i,\sigma^{-i})=\bar V^i(\mu^i,\sigma^{-i})-V^i(a^i,\mu^i,\sigma^{-i})\)
is continuous. Since \(\mathbb A^i\) is nonempty and compact by
Assumption~\ref{ass:game}\ref{ass:game:spaces}, Weierstrass' theorem implies
that the maximum defining \(\bar V^i(\mu^i,\sigma^{-i})\) is attained.
Therefore \(F^i_0(\mu^i,\sigma^{-i})\) is nonempty, and since
\(F^i_0(\mu^i,\sigma^{-i})\subseteq F^i_\varepsilon(\mu^i,\sigma^{-i})\)
for every \(\varepsilon\geq0\), \(F^i_\varepsilon(\mu^i,\sigma^{-i})\) is
nonempty. Moreover, \(F^i_\varepsilon(\mu^i,\sigma^{-i})\) is closed in
the compact set \(\mathbb A^i\), and hence is compact.

To see that the correspondence
\((\varepsilon,\mu^i,\sigma^{-i})\mapsto F^i_\varepsilon(\mu^i,\sigma^{-i})\)
has a closed graph, suppose \(\varepsilon_n\to\varepsilon\),
\((\mu_n^i,\sigma_n^{-i})\Rightarrow(\mu^i,\sigma^{-i})\),
\(a_n^i\to a^i\), and
\(a_n^i\in F^i_{\varepsilon_n}(\mu_n^i,\sigma_n^{-i})\) for every \(n\).
Then \(D^i(a_n^i,\mu_n^i,\sigma_n^{-i})\leq\varepsilon_n\) for every \(n\).
Taking limits and using the continuity of \(D^i\), we obtain
\(D^i(a^i,\mu^i,\sigma^{-i})\leq\varepsilon\). Hence
\(a^i\in F^i_\varepsilon(\mu^i,\sigma^{-i})\), proving that the graph is
closed.

Finally, since the range space \(\mathbb A^i\) is compact, the closed-graph
property implies upper hemicontinuity
\citep[Theorem~17.11]{aliprantis2006infinite}. Since \(F^i_0=F^i\), the stated
properties of the exact best-response correspondence \(F^i\) follow by
restricting the preceding correspondence to \(\varepsilon=0\).

We now prove (iii). For \((\theta^i,a)\in\Theta^i\times\mathbb A\), define
\[
K^i(\theta^i,a)
:=
\int_{\mathbb Y^i}
q^i(y^i\mid a)
\log\frac{q^i(y^i\mid a)}{q^i_{\theta^i}(y^i\mid a)}
\,\nu^i(dy^i),
\]
with the convention that \(0\log(0/0)=0\). By Assumption~\ref{ass:game}\ref{ass:game:bounds}, the likelihood-ratio bound implies \(\left|\log(q^i/q^i_{\theta^i})\right|\leq \log M^i\leq M^i\) whenever the ratio is well defined, and \(q^i\leq r^i\). Hence the integrand is dominated by \(M^i r^i\), which is integrable by Assumption~\ref{ass:game}\ref{ass:game:bounds}. By Assumption~\ref{ass:game}\ref{ass:game:densities}, if \((\theta_n^i,a_n)\to(\theta^i,a)\), then \(q^i(\cdot\mid a_n)\to q^i(\cdot\mid a)\) and \(q^i_{\theta_n^i}(\cdot\mid a_n)\to q^i_{\theta^i}(\cdot\mid a)\), \(\nu^i\)-a.e. If \(q^i_{\theta^i}(\cdot\mid a)>0\), continuity of \((u,v)\mapsto u\log(u/v)\) on \(\{u\geq0,v>0\}\) gives pointwise convergence of the integrand. If \(q^i_{\theta^i}(\cdot\mid a)=0\), then \(q^i(\cdot\mid a)=0\) by the likelihood-ratio bound, and the same bound implies that the integrand converges to \(0\). Therefore, by the Dominated Convergence Theorem, \(K^i(\theta_n^i,a_n)\to K^i(\theta^i,a)\). Thus \(K^i\) is bounded and continuous.

For \(\sigma\in\Delta\mathbb A\), write
\(K^i(\theta^i,\sigma)=\int_{\mathbb A}K^i(\theta^i,a)\,\sigma(da)\).
Since \(K^i(\theta^i,a)\) is bounded and continuous, the map
\((\theta^i,\sigma)\mapsto K^i(\theta^i,\sigma)\) is continuous under the
weak topology on \(\Delta\mathbb A\). Because \(\Theta^i\) is nonempty and
compact by Assumption~\ref{ass:game}\ref{ass:game:spaces}, the Maximum
Theorem \citep[Theorem~17.31]{aliprantis2006infinite}, applied to
\(-K^i\), implies that \(\sigma\mapsto\Theta^{m,i}(\sigma)\) is nonempty,
compact-valued, and upper hemicontinuous. Since the range space \(\Theta^i\)
is compact by Assumption~\ref{ass:game}\ref{ass:game:spaces},
\(\sigma\mapsto\Theta^{m,i}(\sigma)\) also has a closed graph by
\citet[Theorem~17.11]{aliprantis2006infinite}. Moreover, by
Assumption~\ref{ass:game}\ref{ass:game:spaces}, \(\Delta\mathbb A\) and
\(\Delta\Theta^i\) are compact metrizable under the weak topology. Hence
\citet[Theorem~17.13]{aliprantis2006infinite} implies that
\(\sigma\mapsto\Delta\Theta^{m,i}(\sigma)\) is nonempty, compact-valued,
and upper hemicontinuous, and
\citet[Theorem~17.11]{aliprantis2006infinite} then implies that it has a
closed graph.

Finally, we prove (iv). Under identification,
\(\Theta^{m,i}(\sigma)=\{\theta^i(\sigma)\}\) for every
\(\sigma\in\Delta\mathbb A\). By part (iii),
\(\sigma\mapsto\Theta^{m,i}(\sigma)\) is upper hemicontinuous. Since this
correspondence is singleton-valued,
\citet[Lemma~17.6]{aliprantis2006infinite} implies that
\(\sigma\mapsto\theta^i(\sigma)\) is continuous.

\end{proof}

\section{Proof of Lemma \ref{lemm:Berk-selected} (Asymptotic beliefs)} \label{sub:lemm:Berk-selected}

In this Supplemental Appendix, we prove Lemma~\ref{lemm:Berk-selected}. The argument extends standard results on misspecified Bayesian learning (e.g. EP16 and \cite{esponda2021asymptotic}) to the admissible learning processes defined in this paper and compact action spaces.

Because the player set $I$ is finite, it is enough to prove the result for a fixed player $i$ and then intersect the corresponding probability-one events. Henceforth, we suppress the superscript $i$.

For every \(t\) such that \(H_t\neq\varnothing\), let \(n_t:=|H_t|\) and define
\[
L_t(\theta):=\frac{1}{n_t}\sum_{\tau\in H_t}\ln\frac{q(y_\tau\mid a_\tau)}{q_\theta(y_\tau\mid a_\tau)}.
\]
Recall that the selected empirical distribution is
\[
\hat\sigma_t=\frac{1}{n_t}\sum_{\tau\in H_t}\delta_{a_\tau}.
\]

We will use the following fact, which we prove below in Section \ref{subsec:uLLN}:

\emph{Uniform SLLN.} For the fixed player, under any admissible learning process,
\begin{equation}\label{eq:uLLN}
\sup_{\theta\in\Theta}
\left|
L_t(\theta)
-
\int K(\theta,a)\hat\sigma_t(da)
\right|
\longrightarrow 0
\qquad \mathbf P\text{-a.s.}
\end{equation}
as $t\to\infty$ along dates for which $H_t\neq\varnothing$. Since $|H_t|\to\infty$ $\mathbf P$-a.s., this condition is eventually satisfied almost surely.

From this point on, fix a realization in the $\mathbf P$-probability-one event on which \eqref{eq:uLLN} holds and $|H_t|\to\infty$. Let $(t_k)_k$ be any subsequence such that $\hat\sigma_{t_k}\Rightarrow\sigma$ for some $\sigma\in\Delta\mathbb A$, and let $E\subseteq\Theta$ be closed with $E\cap\Theta^m(\sigma)=\varnothing$.

We rely on the following two observations:

\emph{Approximate $(\hat\sigma_{t_k})_k$ with its limit $\sigma$.} Since $K$ is continuous on the compact set $\Theta\times\mathbb A$ and
$\hat\sigma_{t_k}\Rightarrow\sigma$,
\[
\sup_{\theta\in\Theta}
\left|
\int K(\theta,a)\hat\sigma_{t_k}(da)
-
\int K(\theta,a)\sigma(da)
\right|
\longrightarrow0.
\]
Hence, for every $\varepsilon>0$, the displayed expression is smaller than
$\varepsilon$ for all sufficiently large $k$.

\emph{$E$ is well separated.} Since $E\subseteq\Theta$ is closed, it is compact. Because $E\cap\Theta^m(\sigma)=\varnothing$ and $\theta\mapsto\int K(\theta,a)\sigma(da)$ is continuous, there exists $\delta>0$ such that
\[
\int K(\theta,a)\sigma(da)\geq K^*(\sigma)+\delta
\]
for every $\theta\in E$, where
\[
K^*(\sigma):=\min_{\theta\in\Theta}\int K(\theta,a)\sigma(da).
\]

We now prove \eqref{eq:exponential}. For any $\xi>0$, using
$n_{t_k}=|H_{t_k}|$ and the definition of $L_{t_k}$,
\[
\mu_{t_k}(E)
=
\frac{
\int_E e^{-n_{t_k}L_{t_k}(\theta)}\mu_0(d\theta)
}{
\int_\Theta e^{-n_{t_k}L_{t_k}(\theta)}\mu_0(d\theta)
}
\leq
\frac{
\int_E e^{-n_{t_k}L_{t_k}(\theta)}\mu_0(d\theta)
}{
\int_{\Theta_\xi(\sigma)}
e^{-n_{t_k}L_{t_k}(\theta)}\mu_0(d\theta)
},
\]
where
\[
\Theta_\xi(\sigma)
:=
\left\{
\theta\in\Theta:
\int K(\theta,a)\sigma(da)
\leq K^*(\sigma)+\xi
\right\}.
\]

For the numerator, fix $\varepsilon>0$. By \eqref{eq:uLLN}, the uniform convergence of the empirical KL criterion to its limit, and the separation of $E$, for all sufficiently large $k$ and every $\theta\in E$,
\[
L_{t_k}(\theta)
\geq
K^*(\sigma)+\delta-2\varepsilon.
\]
Hence
\[
\int_E e^{-n_{t_k}L_{t_k}(\theta)}\mu_0(d\theta)
\leq
\mu_0(E)
e^{-n_{t_k}(K^*(\sigma)+\delta-2\varepsilon)}.
\]

For the denominator, by \eqref{eq:uLLN} and the same uniform approximation, for all sufficiently large $k$ and every $\theta\in\Theta_\xi(\sigma)$,
\[
L_{t_k}(\theta)
\leq
K^*(\sigma)+\xi+2\varepsilon.
\]
Therefore
\[
\int_{\Theta_\xi(\sigma)}
e^{-n_{t_k}L_{t_k}(\theta)}\mu_0(d\theta)
\geq
\mu_0(\Theta_\xi(\sigma))
e^{-n_{t_k}(K^*(\sigma)+\xi+2\varepsilon)}.
\]

Moreover, $\mu_0(\Theta_\xi(\sigma))>0$. Indeed, a minimizer
$\theta^*\in\Theta^m(\sigma)$ exists by compactness and continuity, and for
$\xi>0$ continuity implies that $\Theta_\xi(\sigma)$ contains a neighborhood
of $\theta^*$. Since $\mu_0$ has full support, this neighborhood has positive
$\mu_0$-measure.

Combining the numerator and denominator bounds gives, for all sufficiently large $k$,
\[
\mu_{t_k}(E)
\leq
\frac{\mu_0(E)}{\mu_0(\Theta_\xi(\sigma))}
\exp\{-n_{t_k}(\delta-\xi-4\varepsilon)\}.
\]
Choose $\varepsilon=\xi<\delta/10$. Then
$\delta-\xi-4\varepsilon>\delta/2$, and hence
\[
\mu_{t_k}(E)
\leq
\frac{\mu_0(E)}{\mu_0(\Theta_\xi(\sigma))}
e^{-n_{t_k}\delta/2}.
\]
Thus \eqref{eq:exponential} holds with
$\rho=\delta/2$ and
$C=\mu_0(E)/\mu_0(\Theta_\xi(\sigma))$.

Finally, suppose that $\mu_{t_k}\Rightarrow\mu$. Let
$\theta\notin\Theta^m(\sigma)$. Since $\Theta^m(\sigma)$ is closed, there exists an open neighborhood $U$ of $\theta$ whose closure $\overline U$ is disjoint from $\Theta^m(\sigma)$. By \eqref{eq:exponential},
\[
\mu_{t_k}(\overline U)\to0.
\]
Since $U$ is open, the Portmanteau theorem gives
\[
\mu(U)\leq \liminf_{k\to\infty}\mu_{t_k}(U)
\leq
\liminf_{k\to\infty}\mu_{t_k}(\overline U)
=0.
\]
Hence $\theta\notin\operatorname{supp}(\mu)$. Therefore
$\operatorname{supp}(\mu)\subseteq\Theta^m(\sigma)$, or equivalently
$\mu\in\Delta\Theta^m(\sigma)$.

\subsection{Uniform SLLN} \label{subsec:uLLN}

\cite{esponda2021asymptotic} (henceforth, EPY 2021) proves the uniform SLLN in the full-history case with a finite action space. We adapt their argument to admissible learning processes in three steps. First, we establish a strong law for selected observations using a conditional exponential-moment bound and a Borel--Cantelli argument. Second, we verify the required exponential-moment and continuity bounds uniformly over the compact action space. Third, we apply a finite-cover argument over the parameter space.

Fix a player $i$ and suppress the superscript $i$. Let
\[
\mathcal F_t:=\sigma(h_t)
\qquad\text{and}\qquad
\mathcal G_t:=\sigma(h_{t-1},a_t,W_t).
\]
Thus
\[
\mathcal F_{t-1}\subseteq\mathcal G_t\subseteq\mathcal F_t,
\]
and $\mathcal G_t$ contains all information determined before the period-$t$ consequence is realized. By admissibility,
\[
\mathbf P(y_t\in B\mid\mathcal G_t)
=
Q(B\mid a_t)
\]
for every Borel set $B\subseteq\mathbb Y$.

Define
\[
g(\theta,y,a)
:=
\log\frac{q(y\mid a)}{q_\theta(y\mid a)}.
\]
Recall that
\[
H_t=\{\tau<t:W_{\tau,t}=1\},
\qquad
N_t:=|H_t|
=
\sum_{\tau<t}W_{\tau,t}.
\]
We impose the sample-growth condition
\begin{equation}\label{eq:sample-growth-log}
\frac{N_t}{\log t}\longrightarrow\infty
\qquad
\mathbf P\text{-a.s.}
\end{equation}

\paragraph{Step 1: Selected-sample convergence.}

Fix $\theta\in\Theta$ and $\varepsilon>0$, and let
\[
O(\theta,\varepsilon)
=
\{\theta'\in\Theta:\|\theta'-\theta\|<\varepsilon\}.
\]
For each period $\tau$, define
\[
Z_\tau
:=
\sup_{\theta'\in O(\theta,\varepsilon)}
g(\theta',y_\tau,a_\tau)
\]
and its centered version
\[
\zeta_\tau
:=
Z_\tau
-
\mathbf E[Z_\tau\mid\mathcal G_\tau].
\]
By construction,
\[
\mathbf E[\zeta_\tau\mid\mathcal G_\tau]=0.
\]
Moreover, because conditional on $\mathcal G_\tau$ the consequence
$y_\tau$ is distributed according to $Q(\cdot\mid a_\tau)$,
\[
\mathbf E[Z_\tau\mid\mathcal G_\tau]
=
\mathbf E_{Q(\cdot\mid a_\tau)}
\left[
\sup_{\theta'\in O(\theta,\varepsilon)}
g(\theta',Y,a_\tau)
\right].
\]

Our objective in this step is to show that
\begin{equation}\label{eq:selected-centered-SLLN}
\frac{1}{N_t}
\sum_{\tau\in H_t}\zeta_\tau
\longrightarrow0
\qquad
\mathbf P\text{-a.s.}
\end{equation}

The main issue is that the sample $H_t$ may vary arbitrarily with $t$:
samples need not be nested, and an observation may be used at one date,
omitted at another, and used again later. The non-anticipation restriction
on sampling is what allows us to handle this flexibility. For every
$\tau<t$, the indicator $W_{\tau,t}$ is a component of the row $W_\tau$,
which is determined before $y_\tau$ is realized. Hence
$W_{\tau,t}$ is $\mathcal G_\tau$-measurable. It follows that
\[
\mathbf E[W_{\tau,t}\zeta_\tau\mid\mathcal G_\tau]
=
W_{\tau,t}
\mathbf E[\zeta_\tau\mid\mathcal G_\tau]
=
0.
\]
Thus selecting observation $\tau$ according to $W_{\tau,t}$ preserves
the conditional mean-zero property of its centered likelihood contribution.

For each date $t$, define the selected sum
\[
S_t
:=
\sum_{\tau<t}W_{\tau,t}\zeta_\tau
=
\sum_{\tau\in H_t}\zeta_\tau.
\]
Then \eqref{eq:selected-centered-SLLN} is equivalent to showing that
\[
\frac{S_t}{N_t}\longrightarrow0
\qquad
\mathbf P\text{-a.s.}
\]

As shown in Step 2 (Claim 1) below, the likelihood-ratio and envelope assumptions
imply that there exists a constant $C\in (0,\infty)$, independent of $\tau$,
such that
\begin{equation}\label{eq:conditional-mgf-zeta}
\mathbf E
\left[
e^{\lambda\zeta_\tau}
\mid\mathcal G_\tau
\right]
\leq
e^{C\lambda^2}
\qquad
\mathbf P\text{-a.s.}
\end{equation}
for every $|\lambda|\leq1/2$.

We first use \eqref{eq:conditional-mgf-zeta} to obtain an exponential
bound for the entire selected sum. Fix $t$ and $|\lambda|\leq1/2$.
Since $W_{\tau,t}\in\{0,1\}$ and is $\mathcal G_\tau$-measurable,
\begin{align}
&\mathbf E
\left[
\left.
\exp\left\{
\lambda W_{\tau,t}\zeta_\tau
-
C\lambda^2W_{\tau,t}
\right\}
\right|
\mathcal G_\tau
\right]
\nonumber\\
&\qquad
=
(1-W_{\tau,t})
+
W_{\tau,t}e^{-C\lambda^2}
\mathbf E
\left[
e^{\lambda\zeta_\tau}
\mid\mathcal G_\tau
\right]
\nonumber\\
&\qquad
\leq1.
\label{eq:selected-one-period-mgf}
\end{align}
Thus a selected observation contributes an exponential factor whose
conditional expectation is at most one, while an unselected observation
contributes exactly one.

We now iterate this inequality across periods. Since
\[
\lambda S_t-C\lambda^2N_t
=
\sum_{\tau<t}
\left(
\lambda W_{\tau,t}\zeta_\tau
-
C\lambda^2W_{\tau,t}
\right),
\]
we have
\[
\exp\{\lambda S_t-C\lambda^2N_t\}
=
\prod_{\tau<t}
\exp\left\{
\lambda W_{\tau,t}\zeta_\tau
-
C\lambda^2W_{\tau,t}
\right\}.
\]
Starting with the last factor and repeatedly conditioning on the corresponding
$\mathcal G_\tau$, equation \eqref{eq:selected-one-period-mgf} yields the desired
bound for the entire product. To see this, define
\[
X_{\tau,t}
:=
\exp\left\{
\lambda W_{\tau,t}\zeta_\tau
-
C\lambda^2W_{\tau,t}
\right\}.
\]
Then
\[
\exp\{\lambda S_t-C\lambda^2N_t\}
=
\prod_{\tau<t}X_{\tau,t}.
\]
Moreover, for every $\tau<t$,
\[
\mathbf E[X_{\tau,t}\mid\mathcal G_\tau]\leq1
\]
by \eqref{eq:selected-one-period-mgf}.

Consider first the last factor, $X_{t-1,t}$. The product
\[
\prod_{\tau<t-1}X_{\tau,t}
\]
depends only on information realized through period $t-2$, and is therefore
$\mathcal F_{t-2}$-measurable. Since
\[
\mathcal F_{t-2}\subseteq\mathcal G_{t-1},
\]
this product is also $\mathcal G_{t-1}$-measurable. Hence, by the tower
property,
\begin{align*}
\mathbf E\left[\prod_{\tau<t}X_{\tau,t}\right]
&=
\mathbf E\left[
\mathbf E\left[
\left.
\prod_{\tau<t}X_{\tau,t}
\right|
\mathcal G_{t-1}
\right]
\right]
\\
&=
\mathbf E\left[
\left(\prod_{\tau<t-1}X_{\tau,t}\right)
\mathbf E[X_{t-1,t}\mid\mathcal G_{t-1}]
\right]
\\
&\leq
\mathbf E\left[
\prod_{\tau<t-1}X_{\tau,t}
\right].
\end{align*}
Applying the same argument successively to $X_{t-2,t},X_{t-3,t},\ldots,X_{1,t}$
gives
\[
\mathbf E\left[
\prod_{\tau<t}X_{\tau,t}
\right]
\leq
\mathbf E\left[
\prod_{\tau<t-1}X_{\tau,t}
\right]
\leq
\cdots
\leq
\mathbf E[X_{1,t}]
\leq1.
\]
Therefore,
\begin{equation}\label{eq:selected-mgf}
\mathbf E
\left[
\exp\{\lambda S_t-C\lambda^2N_t\}
\right]
\leq1.
\end{equation}
The same inequality holds for negative $\lambda$, since
\eqref{eq:conditional-mgf-zeta} holds for every $|\lambda|\leq1/2$.

We next convert \eqref{eq:selected-mgf} into an almost-sure bound.
Fix $m\in\mathbb N$. Choose $\lambda_m\in(0,1/2]$ sufficiently small that
\[
\kappa_m
:=
\frac{\lambda_m}{m}
-
C\lambda_m^2
>0.
\]
For example, one may take
\[
\lambda_m
=
\min\left\{
\frac{1}{2Cm},
\frac14
\right\}.
\]
Define
\[
B_{t,m}
:=
\left\{
|S_t|\geq\frac{1}{m}N_t
\right\}.
\]

For any deterministic $b>0$, consider first the upper-tail event

$$
\left\{
S_t\geq\frac{1}{m}N_t
\right\}
\cap
\{N_t\geq b\}.
$$

This event says that the average selected centered contribution,
$S_t/N_t$, is at least $1/m$, and that the selected sample contains at
least $b$ observations.

On this event, since $\lambda_m>0$,

$$
\lambda_m S_t
\geq
\frac{\lambda_m}{m}N_t.
$$

Subtracting $C\lambda_m^2N_t$ from both sides gives

$$
\lambda_m S_t-C\lambda_m^2N_t
\geq
\left(
\frac{\lambda_m}{m}
-
C\lambda_m^2
\right)N_t
=
\kappa_m N_t.
$$

Since the event also requires $N_t\geq b$, and $\kappa_m>0$, it follows that

$$
\lambda_m S_t-C\lambda_m^2N_t
\geq
\kappa_m b.
$$

Exponentiating both sides,

$$
\exp\left\{
\lambda_m S_t-C\lambda_m^2N_t
\right\}
\geq
e^{\kappa_m b}.
$$

Therefore,

$$
\left\{
S_t\geq\frac{1}{m}N_t,\,
N_t\geq b
\right\}
\subseteq
\left\{
\exp\left\{
\lambda_m S_t-C\lambda_m^2N_t
\right\}
\geq
e^{\kappa_m b}
\right\}.
$$

Consequently,
\begin{align*}
\mathbf P
\left(
S_t\geq\frac{1}{m}N_t,
N_t\geq b
\right)
&\leq
\mathbf P
\left(
\exp \left\{
\lambda_m S_t-C\lambda_m^2N_t
 \right\}
\geq
e^{\kappa_m b}
\right).
\end{align*}
The random variable inside the probability on the right-hand side is
nonnegative. Hence, by Markov's inequality,
\begin{align*}
\mathbf P
\left(
S_t\geq\frac{1}{m}N_t,
N_t\geq b
\right)
&\leq
e^{-\kappa_m b}
\mathbf E
\left[
\exp\left\{
\lambda_m S_t-C\lambda_m^2N_t
\right\}
\right].
\end{align*}
Finally, applying \eqref{eq:selected-mgf} yields

$$
\mathbf P
\left(
S_t\geq\frac{1}{m}N_t,\,
N_t\geq b
\right)
\leq
e^{-\kappa_m b}.
$$

Because the exponential-moment bound also holds for $-\lambda_m$, the same
argument applied to $-S_t$ gives

$$
\mathbf P
\left(
S_t\leq-\frac{1}{m}N_t,\,
N_t\geq b
\right)
\leq
e^{-\kappa_m b}.
$$

Thus, defining

$$
B_{t,m}
:=
\left\{
|S_t|\geq\frac{1}{m}N_t
\right\},
$$

we can write

$$
B_{t,m}
=
\left\{
S_t\geq\frac{1}{m}N_t
\right\}
\cup
\left\{
S_t\leq-\frac{1}{m}N_t
\right\}.
$$

Therefore, using the upper- and lower-tail bounds above and applying the union bound,
\begin{equation}\label{eq:selected-tail}
\mathbf P
\left(
B_{t,m}\cap\{N_t\geq b\}
\right)
\leq
2e^{-\kappa_m b}.
\end{equation}

Now choose a deterministic constant $d_m>0$ such that
\[
\kappa_m d_m>1,
\]
and set
\[
b_t:=d_m\log t.
\]
Equation \eqref{eq:selected-tail} implies
\[
\mathbf P
\left(
B_{t,m}
\cap
\{N_t\geq d_m\log t\}
\right)
\leq
2t^{-\kappa_m d_m}.
\]
Since $\kappa_m d_m>1$,
\[
\sum_{t=1}^{\infty}
\mathbf P
\left(
B_{t,m}
\cap
\{N_t\geq d_m\log t\}
\right)
<\infty.
\]
By the Borel--Cantelli lemma,
\begin{equation}\label{eq:BC-selected}
B_{t,m}
\cap
{N_t\geq d_m\log t}
\qquad
\text{occurs only finitely often, a.s.}
\end{equation}
Thus, with probability one, there exists a finite random date $T_1$ such that, for every $t\geq T_1$,

$$
B_{t,m}
\cap
\{N_t\geq d_m\log t\}
$$

does not occur.

On the other hand, the sample-growth condition
\eqref{eq:sample-growth-log} implies that, for this fixed $d_m$, with probability one there exists a finite random date $T_2$ such that, for every $t\geq T_2$,

$$
N_t\geq d_m\log t.
$$

Therefore, on the intersection of these two probability-one events, for every

$$
t\geq \max\{T_1,T_2\},
$$

the event ${N_t\geq d_m\log t}$ holds while

$$
B_{t,m}\cap\{N_t\geq d_m\log t\}
$$

does not. It follows that $B_{t,m}$ itself cannot occur. Hence, for every fixed $m\in\mathbb N$,

$$
B_{t,m}
\qquad
\text{occurs only finitely often, a.s.}
$$

Equivalently,

$$
\frac{|S_t|}{N_t}<\frac1m
$$

for all sufficiently large $t$, almost surely.

Finally, intersecting these probability-one events over
$m=1,2,\ldots$ yields
\[
\frac{S_t}{N_t}\longrightarrow0
\qquad
\mathbf P\text{-a.s.}
\]
Equivalently,
\[
\frac1{N_t}
\sum_{\tau\in H_t}\zeta_\tau
\longrightarrow0
\qquad
\mathbf P\text{-a.s.},
\]
which proves the selected-sample convergence required in this step.

\paragraph{Step 2: Uniform exponential-moment and continuity bounds over the compact action space.}

EPY 2021 assumes a finite action space, so the moment and continuity bounds used in their argument are automatically uniform over actions. With a compact action space, we establish directly the two uniform bounds needed here. First, we derive a conditional exponential-moment bound for the centered local likelihood envelope used in Step 1. Second, we establish a uniform oscillation bound in the parameter.

\medskip
\noindent
\emph{Claim 1: Uniform conditional exponential-moment bound.}

Fix $\theta\in\Theta$ and $\varepsilon>0$, and recall that
\[
Z_\tau
:=
\sup_{\theta'\in O(\theta,\varepsilon)}
g(\theta',y_\tau,a_\tau)
\]
and
\[
\zeta_\tau
:=
Z_\tau-\mathbf E[Z_\tau\mid\mathcal G_\tau].
\]
There exists a constant $C\in(0,\infty)$, independent of $\tau$, such that
\[
\mathbf E
\left[
e^{\lambda\zeta_\tau}
\mid
\mathcal G_\tau
\right]
\leq
e^{C\lambda^2}
\qquad
\mathbf P\text{-a.s.}
\]
for every $|\lambda|\leq 1/2$.

To prove the claim, let
\[
R
:=
\int_{\mathbb Y}
M(y)r(y)\,\nu(dy)
<\infty.
\]
The likelihood-ratio bound in Assumption~\ref{ass:game}(iii) implies that, for every $\theta'\in\Theta$, every $a\in\mathbb A$, and $\nu$-almost every $y$,
\[
|g(\theta',y,a)|
=
\left|
\log
\frac{q(y\mid a)}
     {q_{\theta'}(y\mid a)}
\right|
\leq
\log M(y).
\]
Therefore,
\[
|Z_\tau|
\leq
\log M(y_\tau).
\]

Since, conditional on $\mathcal G_\tau$, the consequence $y_\tau$ is distributed according to $Q(\cdot\mid a_\tau)$,
\begin{align*}
\mathbf E
\left[
e^{|Z_\tau|}
\mid
\mathcal G_\tau
\right]
&\leq
\int_{\mathbb Y}
M(y)q(y\mid a_\tau)\,\nu(dy)
\\
&\leq
\int_{\mathbb Y}
M(y)r(y)\,\nu(dy)
\\
&=
R.
\end{align*}
Thus the exponential moment of $Z_\tau$ is bounded uniformly in the realized action $a_\tau$.

Similarly, since $\log M(y)\leq M(y)$ for $M(y)\geq1$,
\begin{align*}
\mathbf E
\left[
|Z_\tau|
\mid
\mathcal G_\tau
\right]
&\leq
\int_{\mathbb Y}
\log M(y)q(y\mid a_\tau)\,\nu(dy)
\\
&\leq
\int_{\mathbb Y}
M(y)r(y)\,\nu(dy)
\\
&=
R.
\end{align*}

Now, using the triangle inequality,
\[
|\zeta_\tau|
=
\left|
Z_\tau-\mathbf E[Z_\tau\mid\mathcal G_\tau]
\right|
\leq
|Z_\tau|
+
\left|
\mathbf E[Z_\tau\mid\mathcal G_\tau]
\right|.
\]
By conditional Jensen's inequality,
\[
\left|
\mathbf E[Z_\tau\mid\mathcal G_\tau]
\right|
\leq
\mathbf E[|Z_\tau|\mid\mathcal G_\tau]
\leq
R.
\]
Hence
\[
|\zeta_\tau|
\leq
|Z_\tau|+R.
\]
It follows that
\begin{align*}
\mathbf E
\left[
e^{|\zeta_\tau|}
\mid
\mathcal G_\tau
\right]
&\leq
e^R
\mathbf E
\left[
e^{|Z_\tau|}
\mid
\mathcal G_\tau
\right]
\\
&\leq
e^R R.
\end{align*}
Let
\[
K:=e^R R.
\]

For every integer $k\geq2$, the inequality
\[
|x|^k\leq k!e^{|x|}
\]
implies
\[
\mathbf E
\left[
|\zeta_\tau|^k
\mid
\mathcal G_\tau
\right]
\leq
Kk!.
\]

Using the Taylor expansion of the exponential and the fact that
\[
\mathbf E[\zeta_\tau\mid\mathcal G_\tau]=0,
\]
we obtain, for $|\lambda|<1$,
\begin{align*}
\mathbf E
\left[
e^{\lambda\zeta_\tau}
\mid
\mathcal G_\tau
\right]
&=
1+
\sum_{k=2}^{\infty}
\frac{\lambda^k}{k!}
\mathbf E
\left[
\zeta_\tau^k
\mid
\mathcal G_\tau
\right]
\\
&\leq
1+
\sum_{k=2}^{\infty}
\frac{|\lambda|^k}{k!}
\mathbf E
\left[
|\zeta_\tau|^k
\mid
\mathcal G_\tau
\right]
\\
&\leq
1+
K
\sum_{k=2}^{\infty}
|\lambda|^k.
\end{align*}

For $|\lambda|\leq1/2$,
\[
\sum_{k=2}^{\infty}
|\lambda|^k
=
\frac{\lambda^2}{1-|\lambda|}
\leq
2\lambda^2.
\]
Therefore,
\[
\mathbf E
\left[
e^{\lambda\zeta_\tau}
\mid
\mathcal G_\tau
\right]
\leq
1+2K\lambda^2.
\]
Using $1+x\leq e^x$ for $x\geq0$,
\[
\mathbf E
\left[
e^{\lambda\zeta_\tau}
\mid
\mathcal G_\tau
\right]
\leq
e^{2K\lambda^2}.
\]
Thus the claim holds with
\[
C:=2K=2e^R R\in(0,\infty).
\]

\medskip
\noindent
\emph{Claim 2: Uniform oscillation bound.}
For every $\theta\in\Theta$ and $\varepsilon>0$, there exists
$\delta(\theta,\varepsilon)>0$ such that
\[
\sup_{a\in\mathbb A}
\mathbf E_{Q(\cdot\mid a)}
\left[
\sup_{\theta'\in O(\theta,\delta(\theta,\varepsilon))}
|g(\theta',Y,a)-g(\theta,Y,a)|
\right]
<
\frac{\varepsilon}{4}.
\]

For $s>0$, define
\[
H_s(y)
:=
\sup_{\substack{\|\theta'-\theta\|\leq s\\ a\in\mathbb A}}
|g(\theta',y,a)-g(\theta,y,a)|.
\]

For $\nu$-almost every $y$, the map
\[
(\theta',a)\mapsto g(\theta',y,a)
\]
is continuous on the compact set $\Theta\times\mathbb A$. Hence it is
uniformly continuous. Therefore, for every such $y$,
\[
H_s(y)\to0
\qquad
\text{as }s\downarrow0.
\]

Moreover, by the likelihood-ratio bound in
Assumption~\ref{ass:game}(iii),
\[
|g(\theta',y,a)|
\leq
\log M(y)
\]
for every $\theta'\in\Theta$, every $a\in\mathbb A$, and
$\nu$-almost every $y$. Hence
\[
H_s(y)
\leq
2\log M(y)
\leq
2M(y)
\]
for $\nu$-almost every $y$.

Therefore, for every $a\in\mathbb A$,
\begin{align*}
&\mathbf E_{Q(\cdot\mid a)}
\left[
\sup_{\theta'\in O(\theta,s)}
|g(\theta',Y,a)-g(\theta,Y,a)|
\right]
\\
&\qquad\leq
\int_{\mathbb Y}
H_s(y)q(y\mid a)\,\nu(dy)
\\
&\qquad\leq
\int_{\mathbb Y}
H_s(y)r(y)\,\nu(dy).
\end{align*}

Since
\[
H_s(y)r(y)
\leq
2M(y)r(y)
\]
and
\[
\int_{\mathbb Y}
2M(y)r(y)\,\nu(dy)
<\infty,
\]
the dominated convergence theorem implies
\[
\int_{\mathbb Y}
H_s(y)r(y)\,\nu(dy)
\to0
\qquad
\text{as }s\downarrow0.
\]

The bound on the right-hand side does not depend on $a$. Hence
\[
\sup_{a\in\mathbb A}
\mathbf E_{Q(\cdot\mid a)}
\left[
\sup_{\theta'\in O(\theta,s)}
|g(\theta',Y,a)-g(\theta,Y,a)|
\right]
\to0
\qquad
\text{as }s\downarrow0.
\]

Choosing
\[
s=\delta(\theta,\varepsilon)
\]
sufficiently small proves the claim.

\paragraph{Step 3: Uniform convergence over the parameter space.}

We now adapt the finite-cover argument in EPY 2021 to the selected sample. Fix $\varepsilon>0$. By Claim 2, for every $\theta\in\Theta$ there exists $\delta(\theta,\varepsilon)>0$ such that
\[
\sup_{a\in\mathbb A}
\mathbf E_{Q(\cdot\mid a)}
\left[
\sup_{\theta'\in O(\theta,\delta(\theta,\varepsilon))}
|g(\theta',Y,a)-g(\theta,Y,a)|
\right]
<
\frac{\varepsilon}{4}.
\]
By compactness of $\Theta$, there exist $\theta_1,\ldots,\theta_J\in\Theta$ such that
\[
\Theta\subseteq\bigcup_{j=1}^J O(\theta_j,\delta_j),
\qquad
\delta_j:=\delta(\theta_j,\varepsilon).
\]

For every date $t$ such that $N_t>0$, define
\[
D_t(\theta)
:=
\frac{1}{N_t}
\sum_{\tau\in H_t}
\left[
g(\theta,y_\tau,a_\tau)-K(\theta,a_\tau)
\right].
\]
Recall that
\[
K(\theta,a)
=
\mathbf E_{Q(\cdot\mid a)}
\left[
g(\theta,Y,a)
\right].
\]

For each element of the finite cover, define
\[
U_{t,j}
:=
\frac{1}{N_t}
\sum_{\tau\in H_t}
\left\{
\sup_{\theta\in O(\theta_j,\delta_j)}
g(\theta,y_\tau,a_\tau)
-
\mathbf E_{Q(\cdot\mid a_\tau)}
\left[
\sup_{\theta\in O(\theta_j,\delta_j)}
g(\theta,Y,a_\tau)
\right]
\right\},
\]
and
\[
L_{t,j}
:=
\frac{1}{N_t}
\sum_{\tau\in H_t}
\left\{
\inf_{\theta\in O(\theta_j,\delta_j)}
g(\theta,y_\tau,a_\tau)
-
\mathbf E_{Q(\cdot\mid a_\tau)}
\left[
\inf_{\theta\in O(\theta_j,\delta_j)}
g(\theta,Y,a_\tau)
\right]
\right\}.
\]

For each fixed $j$, Step 1 applied with center $\theta_j$ and radius $\delta_j$ implies
\[
U_{t,j}\to0
\qquad
\mathbf P\text{-a.s.}
\]
To obtain the analogous result for $L_{t,j}$, apply Step 1 to the function $-g$. Indeed,
\[
\inf_{\theta\in O(\theta_j,\delta_j)}
g(\theta,y,a)
=
-
\sup_{\theta\in O(\theta_j,\delta_j)}
[-g(\theta,y,a)].
\]
Thus the centered local infimum for $g$ is the negative of the centered local supremum for $-g$. Hence
\[
L_{t,j}\to0
\qquad
\mathbf P\text{-a.s.}
\]

Since the cover contains only finitely many sets, these convergences hold simultaneously for all $j=1,\ldots,J$ on a probability-one event. Therefore,
\[
\max_{1\leq j\leq J}
\left\{
|U_{t,j}|,|L_{t,j}|
\right\}
\to0
\qquad
\mathbf P\text{-a.s.}
\]

We next use the local upper and lower envelopes to control $D_t(\theta)$. Fix $j$ and let $\theta\in O(\theta_j,\delta_j)$. For every $(y,a)$,
\[
g(\theta,y,a)
\leq
\sup_{\theta'\in O(\theta_j,\delta_j)}
g(\theta',y,a).
\]
Moreover,
\[
K(\theta,a)
=
\mathbf E_{Q(\cdot\mid a)}
[g(\theta,Y,a)]
\geq
\mathbf E_{Q(\cdot\mid a)}
\left[
\inf_{\theta'\in O(\theta_j,\delta_j)}
g(\theta',Y,a)
\right].
\]
Subtracting the second inequality from the first and averaging over the selected observations gives
\[
D_t(\theta)\leq U_{t,j}+B_{t,j},
\]
where
\[
B_{t,j}
:=
\frac{1}{N_t}
\sum_{\tau\in H_t}
\mathbf E_{Q(\cdot\mid a_\tau)}
\left[
\sup_{\theta\in O(\theta_j,\delta_j)}
g(\theta,Y,a_\tau)
-
\inf_{\theta\in O(\theta_j,\delta_j)}
g(\theta,Y,a_\tau)
\right].
\]

Similarly, since
\[
g(\theta,y,a)
\geq
\inf_{\theta'\in O(\theta_j,\delta_j)}
g(\theta',y,a)
\]
and
\[
K(\theta,a)
\leq
\mathbf E_{Q(\cdot\mid a)}
\left[
\sup_{\theta'\in O(\theta_j,\delta_j)}
g(\theta',Y,a)
\right],
\]
we obtain
\[
D_t(\theta)\geq L_{t,j}-B_{t,j}.
\]
Combining the upper and lower bounds yields
\[
|D_t(\theta)|
\leq
\max\{|U_{t,j}|,|L_{t,j}|\}
+
B_{t,j}.
\]
Since this holds for every $\theta\in O(\theta_j,\delta_j)$,
\[
\sup_{\theta\in O(\theta_j,\delta_j)}
|D_t(\theta)|
\leq
\max\{|U_{t,j}|,|L_{t,j}|\}
+
B_{t,j}.
\]

It remains to bound the oscillation term $B_{t,j}$. For every $(Y,a)$,
\[
\sup_{\theta\in O(\theta_j,\delta_j)}
g(\theta,Y,a)
-
\inf_{\theta\in O(\theta_j,\delta_j)}
g(\theta,Y,a)
\leq
2
\sup_{\theta\in O(\theta_j,\delta_j)}
|g(\theta,Y,a)-g(\theta_j,Y,a)|.
\]
Indeed, every value of $g(\theta,Y,a)$ in the neighborhood differs from $g(\theta_j,Y,a)$ by at most the latter supremum, so the distance between the largest and smallest values is at most twice that quantity.

By the choice of $\delta_j$ in Claim 2,
\[
\sup_{a\in\mathbb A}
\mathbf E_{Q(\cdot\mid a)}
\left[
\sup_{\theta\in O(\theta_j,\delta_j)}
|g(\theta,Y,a)-g(\theta_j,Y,a)|
\right]
<
\frac{\varepsilon}{4}.
\]
Consequently, for every selected action $a_\tau$,
\[
\mathbf E_{Q(\cdot\mid a_\tau)}
\left[
\sup_{\theta\in O(\theta_j,\delta_j)}
g(\theta,Y,a_\tau)
-
\inf_{\theta\in O(\theta_j,\delta_j)}
g(\theta,Y,a_\tau)
\right]
<
\frac{\varepsilon}{2}.
\]
Averaging over $\tau\in H_t$ therefore gives
\[
B_{t,j}
<
\frac{\varepsilon}{2}
\]
for every $t$ with $N_t>0$ and every $j$.

Since the sets $O(\theta_j,\delta_j)$ cover $\Theta$, it follows that
\[
\sup_{\theta\in\Theta}
|D_t(\theta)|
\leq
\max_{1\leq j\leq J}
\left\{
|U_{t,j}|,|L_{t,j}|
\right\}
+
\frac{\varepsilon}{2}.
\]
Hence, on the probability-one event on which all the finitely many $U_{t,j}$ and $L_{t,j}$ converge to zero,
\[
\limsup_{t\to\infty}
\sup_{\theta\in\Theta}
|D_t(\theta)|
\leq
\frac{\varepsilon}{2}.
\]

To conclude almost-sure convergence, apply the preceding argument to the countable sequence $\varepsilon_n:=1/n$, $n=1,2,\ldots$, and intersect the corresponding probability-one events. On this common probability-one event,
\[
\limsup_{t\to\infty}
\sup_{\theta\in\Theta}
|D_t(\theta)|
\leq
\frac{1}{2n}
\]
for every $n\in\mathbb N$. Therefore,
\[
\sup_{\theta\in\Theta}
|D_t(\theta)|
\to0
\qquad
\mathbf P\text{-a.s.}
\]
Equivalently,
\[
\sup_{\theta\in\Theta}
\left|
\frac{1}{N_t}
\sum_{\tau\in H_t}
\left[
g(\theta,y_\tau,a_\tau)
-
K(\theta,a_\tau)
\right]
\right|
\to0
\qquad
\mathbf P\text{-a.s.}
\]

Finally,
\[
L_t(\theta)
=
\frac{1}{N_t}
\sum_{\tau\in H_t}
g(\theta,y_\tau,a_\tau),
\]
while
\[
\int_{\mathbb A}
K(\theta,a)\,\hat\sigma_t(da)
=
\frac{1}{N_t}
\sum_{\tau\in H_t}
K(\theta,a_\tau).
\]
Thus
\[
D_t(\theta)
=
L_t(\theta)
-
\int_{\mathbb A}
K(\theta,a)\,\hat\sigma_t(da),
\]
and hence
\[
\sup_{\theta\in\Theta}
\left|
L_t(\theta)
-
\int_{\mathbb A}
K(\theta,a)\,\hat\sigma_t(da)
\right|
\to0
\qquad
\mathbf P\text{-a.s.},
\]
which is the desired uniform SLLN.

The proof separates the two extensions of EPY 2021. General selected sampling affects Step 1: non-anticipation preserves the centered conditional structure of selected likelihood contributions, while the conditional exponential-moment bound and the sample-growth condition yield almost-sure convergence through an exponential tail bound and a Borel--Cantelli argument. The extension from a finite to a compact action space affects Step 2, where the required exponential-moment and oscillation bounds are made uniform over actions. Given these two ingredients, Step 3 upgrades the local selected-sample convergence to uniform convergence over the parameter space by compactness.

\section{Learning foundation for the support refinement}

Section \ref{subsec:supportref} introduced the support refinements $\Gamma^+$ and $\Gamma^{CS+}$. This appendix develops their learning foundation by formalizing the idea that recurrence of an action is synchronized with the empirical distribution that justifies it.

For a nonempty set $S\subseteq\mathbb A$, define
\[
\Gamma^{+}(S)
=
\left\{
a\in\mathbb A:
\begin{array}{l}
\text{for every }i\in I\text{ there exist }\sigma^i\in\Delta S
\text{ and }\mu^i\in\Delta\Theta^{m,i}(\sigma^i)\\
\text{such that }
a\in\operatorname{supp}(\sigma^i)
\text{ and }
a^i\in F^i(\mu^i,(\sigma^i)^{-i})
\end{array}
\right\}.
\]
The common-sample refinement $\Gamma^{CS+}$ is defined analogously by requiring a single $\sigma\in\Delta S$, with $a\in\operatorname{supp}(\sigma)$, to justify all players. We call $S$ \emph{$+$-self-justified} if $S\subseteq\Gamma^+(S)$, and \emph{common-sample $+$-self-justified} if $S\subseteq\Gamma^{CS+}(S)$.

Because $a\in\Gamma^+(S)$ requires $a\in\operatorname{supp}(\sigma^i)\subseteq S$, we have $\Gamma^+(S)\subseteq S$. Hence $+$-self-justified sets are exactly the fixed points of $\Gamma^+$. Since $\Gamma^+$ is monotone, the union of all fixed points is itself a fixed point, so a largest fixed point always exists. The same holds for $\Gamma^{CS+}$. Unlike the baseline operator $\Gamma$, however, these refined operators need not be nonempty-valued on nonempty sets, so their largest fixed points may be empty. When $\mathbb A$ is finite, the decreasing sequence of iterates stabilizes after finitely many steps and yields the usual iterative characterization.\footnote{In infinite action spaces, the iterative characterization need not hold because the condition $a\in\operatorname{supp}(\sigma)$ is not preserved under weak convergence.}

We next provide a learning foundation for these refinements. A nonempty set $S\subseteq\mathbb A$ is \emph{self-synchronized} along a realized path if, for every $a\in S$, there exists a subsequence $(t_k)_k$ and distributions $\sigma^i\in\Delta S$, $i\in I$, such that
\[
a_{t_k}\to a,
\qquad
\hat\sigma_{t_k}^i\Rightarrow\sigma^i,
\qquad
a\in\operatorname{supp}(\sigma^i)
\quad\text{for every }i\in I.
\]
Thus self-synchronization strengthens recurrence by requiring the limiting empirical distributions along the same subsequence to continue to represent the recurrent action. It excludes actions that are revisited only along exceptional dates while the samples used for learning converge to a different empirical regime. This also suggests a robustness interpretation: actions supported only by an asymptotically negligible set of dates may disappear under small perturbations of the realized history, whereas synchronized actions remain represented in the limiting empirical distribution. We leave a fuller exploration of this notion to future work. In finite action spaces, however, self-synchronization has an immediate frequency interpretation.

\begin{proposition}[Frequency interpretation]\label{prop:synchronization-frequency}
Suppose that $\mathbb A$ is finite and all players use the same selected sample. Then the largest self-synchronized set is
\[
\left\{
a\in\mathbb A:
\limsup_{t\to\infty}\hat\sigma_t(a)>0
\right\},
\]
where $\hat\sigma_t$ is the empirical distribution of the common selected sample. In particular, under full-history learning, this is the set of action profiles played with nonvanishing empirical frequency.
\end{proposition}

\begin{proof}
Let
\[
P:=\left\{a\in\mathbb A:\limsup_{t\to\infty}\hat\sigma_t(a)>0\right\}.
\]
We first show that $P$ is self-synchronized. Fix $a\in P$ and let
$\ell=\limsup_t\hat\sigma_t(a)>0$. Since $|H_t|\to\infty$, the change in
$\hat\sigma_t(a)$ generated by adding one selected observation converges to
zero. Moreover, the frequency of $a$ can increase only when an observation
with $a_t=a$ is selected. It follows that there are infinitely many selected
dates $t_k$ with $a_{t_k}=a$ such that
$\liminf_k\hat\sigma_{t_k}(a)>0$. Indeed, otherwise every sufficiently late
increase in the frequency of $a$ would start from a value bounded away from
$\ell$, and the vanishing size of a single update would contradict
$\limsup_t\hat\sigma_t(a)=\ell$.

Since $\Delta\mathbb A$ is compact, after passing to a subsequence we have
$\hat\sigma_{t_k}\to\sigma$ for some $\sigma\in\Delta\mathbb A$. Then
$\sigma(a)>0$, so $a\in\operatorname{supp}(\sigma)$. If
$b\in\operatorname{supp}(\sigma)$, then
$\limsup_t\hat\sigma_t(b)\geq\sigma(b)>0$, and hence $b\in P$. Thus
$a\in\operatorname{supp}(\sigma)\subseteq P$, so $P$ is self-synchronized.

Conversely, let $S$ be any self-synchronized set and fix $a\in S$. There
exist a subsequence $(t_k)_k$ and $\sigma\in\Delta S$ such that
$a_{t_k}\to a$, $\hat\sigma_{t_k}\to\sigma$, and
$a\in\operatorname{supp}(\sigma)$. Since $\mathbb A$ is finite,
$\sigma(a)>0$, and therefore
$\limsup_t\hat\sigma_t(a)\geq\sigma(a)>0$. Hence $a\in P$. Since $a\in S$
was arbitrary, $S\subseteq P$. Therefore $P$ is the largest
self-synchronized set. Under full-history learning, $\hat\sigma_t$ is the
usual empirical distribution of play, which gives the final claim.
\end{proof}
 
Because self-synchronization focuses on a stronger notion of long-run persistence, we can correspondingly weaken the behavioral requirement. We say that behavior is \emph{synchronized asymptotically myopic} if, whenever for some player $i$ and some subsequence $(t_k)_k$,
\[
a_{t_k}\to a,
\qquad
\hat\sigma_{t_k}^i\Rightarrow\sigma^i,
\qquad
a\in\operatorname{supp}(\sigma^i),
\]
there exists $\varepsilon_k\downarrow0$ such that
$a_{t_k}^i\in F_{\varepsilon_k}^i(\mu_{t_k}^i,\tilde\sigma_{t_k}^{-i})$
for all $k$. The asymptotically myopic behavior imposed in Section \ref{sec:foundations} clearly implies synchronized asymptotic myopia.

The forward-looking interpretation from the preceding subsection extends to this weaker requirement under correspondingly weaker identification. In particular, under the weak-identification condition of EP16, Appendix C, the same argument implies that forward-looking Bayesian behavior is synchronized asymptotically myopic: along a synchronized subsequence, best-fitting models agree on the consequence distribution generated by the limiting action, so dynamic optimality implies myopic optimality in the limit.

\begin{theorem}[Self-synchronized behavior is \(+\)-self-justified]
\label{theo:self-synchronized}
Consider a game with Bayesian players, selected samples, consistent forecasts, and synchronized asymptotically myopic behavior. With probability one, every self-synchronized set $S$ satisfies
$S\subseteq\Gamma^+(S)$. If the selected sample is common across players, then every self-synchronized set satisfies
$S\subseteq\Gamma^{CS+}(S)$.
\end{theorem}

\begin{proof}
Fix a realization in the $\mathbf P$-probability-one event on which Lemma \ref{lemm:Berk-selected} holds simultaneously for every player, and let $S$ be self-synchronized. Pick $a\in S$. By self-synchronization, there exists a subsequence $(t_k)_k$ and, for every player $i$, a distribution $\sigma^i\in\Delta S$ such that $a_{t_k}\to a$, $\hat\sigma_{t_k}^i\Rightarrow\sigma^i$, and $a\in\operatorname{supp}(\sigma^i)$. Passing to a further subsequence if necessary, let $\mu_{t_k}^i\Rightarrow\mu^i$ for every $i$. Lemma \ref{lemm:Berk-selected} then implies $\mu^i\in\Delta\Theta^{m,i}(\sigma^i)$.

By forecast consistency, $\tilde\sigma_{t_k}^{-i}\Rightarrow(\sigma^i)^{-i}$. Since $a\in\operatorname{supp}(\sigma^i)$, synchronized asymptotic myopia applies along this subsequence, so there exists $\varepsilon_k^i\downarrow0$ such that $a_{t_k}^i\in F_{\varepsilon_k^i}^i(\mu_{t_k}^i,\tilde\sigma_{t_k}^{-i})$. The closed-graph property of $F_\varepsilon^i$ therefore yields $a^i\in F^i(\mu^i,(\sigma^i)^{-i})$.

Thus, for every player $i$, $\sigma^i\in\Delta S$, $a\in\operatorname{supp}(\sigma^i)$, $\mu^i\in\Delta\Theta^{m,i}(\sigma^i)$, and $a^i\in F^i(\mu^i,(\sigma^i)^{-i})$. Hence $a\in\Gamma^+(S)$. Since $a\in S$ was arbitrary, $S\subseteq\Gamma^+(S)$.

If the selected sample is common across players, the synchronizing subsequence has a common limiting distribution $\sigma\in\Delta S$, and the same argument gives $S\subseteq\Gamma^{CS+}(S)$.
\end{proof}

In particular, when $\mathbb A$ is finite, Proposition \ref{prop:synchronization-frequency} and Theorem \ref{theo:self-synchronized} imply the existence of a nonempty $+$-self-justified set; under a common sample they likewise imply the existence of a nonempty common-sample $+$-self-justified set. Hence the largest fixed points of $\Gamma^+$ and $\Gamma^{CS+}$ are nonempty in the finite-action case.

\end{document}